\documentclass[prd,twocolumn,superscriptaddress,nofootinbib,aps,11pt]{revtex4-1}
\usepackage{natbib}
\usepackage{graphicx}
\usepackage{bm}
\usepackage{amssymb}
\usepackage{color}
\usepackage{amsmath}
\usepackage{mathrsfs} 
\usepackage{amstext}
\usepackage[english]{babel}
\usepackage{latexsym}
\usepackage{array}             
\usepackage{multirow}         
\usepackage[usenames,dvipsnames]{xcolor}
\usepackage[colorlinks=true,citecolor=Blue,linkcolor=RubineRed,urlcolor=Blue]{hyperref}
\usepackage{subfigure}
\usepackage[version=4]{mhchem}
\usepackage{changes}

\newcommand{\beq}{\begin{equation}}
\newcommand{\eeq}{\end{equation}}
\newcommand{\beqa}{\begin{eqnarray}}
\newcommand{\eeqa}{\end{eqnarray}}

\begin{document}

\title{Counterflow and coflow instabilities in miscible binary superfluids}
\author{\textbf{Yu-Ping An}}%
 \email{anyuping@itp.ac.cn}
 \affiliation{Institute of Theoretical Physics,
Chinese Academy of Sciences, Beijing 100190, China}
\affiliation{School of Physical Sciences, University of Chinese Academy of Sciences, Beijing 100049, China}
 \author{\textbf{Blaise Gout\'eraux}}
 \email{blaise.gouteraux@polytechnique.edu}
 \affiliation{CPHT, CNRS, \'Ecole polytechnique, Institut Polytechnique de Paris, 91120 Palaiseau, France}
 \author{\textbf{Li Li}}
\email{liliphy@itp.ac.cn}
 \affiliation{Institute of Theoretical Physics,
Chinese Academy of Sciences, Beijing 100190, China}
\affiliation{School of Physical Sciences, University of Chinese Academy of Sciences, Beijing 100049, China}
\affiliation{School of Fundamental Physics and Mathematical Sciences, Hangzhou Institute for Advanced Study, University of Chinese Academy of Sciences, Hangzhou 310024, China}

 \vspace{1cm}

\begin{abstract}
We explore instabilities in binary superfluids with a nonvanishing relative superflow, particularly focusing on counterflow and coflow instabilities. We extend recent results on the thermodynamic origin of finite superflow instabilities in single-component superfluids to binary systems and derive a criterion for the onset of instability through a hydrodynamic analysis, which applies to interacting many-body systems at finite temperature. We find that the onset of these instabilities is signaled by the determinant of the Hessian of the thermal free energy diverging and changing sign.  We verify this hydrodynamic prediction in a holographic binary superfluid modeled with gauge/gravity duality, which naturally incorporates strong coupling, finite temperature, and dissipation.  We also compare to results obtained using the Gross-Pitaevskii equation for weakly interacting Bose-Einstein condensates and find that the same criterion continues to apply at zero temperature, where it reduces to evaluating derivatives of the supercurrents with respect to the superfluid velocities. 
We observe that the critical velocities of these instabilities follow a general scaling law related to the interaction strength between superfluid components. Finally, the nonlinear stages of the instabilities are studied by full time evolution using gauge/gravity duality, where vortex annihilation leads to a decrease of superfluid velocity back to a value where the binary superfluid phase is stable.

\end{abstract}

\maketitle

\tableofcontents

\section{Introduction and summary of results}
In both classical fluid systems and quantum superfluids with nonzero velocity, instabilities, such as the Kelvin-Helmholtz instability~\cite{KH1,KH2,QKH2,QKH3,QKH4,QKH5,QKH6}, and the two-stream instability~\cite{counterflow1,counterflow2,counterflow3,counterflow4,two-stream1,two-stream2,two-stream3,two-stream4}, appear very generally. Such instabilities typically develop nonlinearly into turbulence. Understanding the origin and final fate of these instabilities is crucial to understanding turbulence. Binary superfluids are an appealing platform to study these instabilities since there is no friction between the two components and one can study equilibrium states with nonzero relative velocity. This is called a counterflow instability. The counterflow instability of weakly interacting Bose-Einstein condensates (BECs) has been extensively studied theoretically, \emph{e.g.}~\cite{khalatnikov1973sound,Nepomnyashchii1974,nepomnyashchil1976microscopic,Yukalov1980-03-01,counterflow1,counterflow2,counterflow3,counterflow4}, and experimentally, \emph{e.g.}~\cite{PhysRevLett.106.065302,PhysRevLett.110.025301,PhysRevLett.119.185302}, see \emph{e.g.} \cite{pethick2008bose,pitaevskii2003bose} for textbooks on this topic. In these studies, the Gross-Pitaevskii equation (GPE) plays a prominent role, and  describes the dynamics of a gas of weakly-interacting bosons at zero temperature. 

In this work, we study the coflow and couterflow instabilities of dissipative binary superfluids at finite temperature employing hydrodynamics and gauge/gravity duality.  These approaches probe a regime different from that of the GPE. Namely, they are both applicable in the presence of strong interactions, and so do not assume the existence of underlying long-lived quasiparticles. They also allow to model thermal dissipative effects. We also comment on how our results relate to the analyses using the GPE. 

\subsection{Non-quasiparticle approaches to superfluids}

Hydrodynamics \cite{landaubookfluids,chaikinlubensky1995,Kovtun:2012rj,forster2018hydrodynamic}, which we review in greater detail in Appendix \ref{subsec:hydro}, is a celebrated and appealing framework for capturing the late-time collective dynamics of interacting systems at finite temperature, such as electron flow in Graphene \cite{Lucas:2017idv,fritz2024hydrodynamic} or the Quark-Gluon plasma formed in heavy ion collisions \cite{Luzum:2008cw}. The late-time and finite temperature assumptions are crucial as they allow us to postulate the existence of a local thermodynamic equilibrium state, which then relaxes to global equilibrium. Since microscopic details of the system enter only through the thermodynamic equation of state and the so-called transport coefficients characterizing out-of-equilibrium dynamics, hydrodynamics allows one to derive universal results applying to widely different microscopic systems, which is the philosophy we adopt in this work. Specifically, the results we derive using hydrodynamics will be seen to correctly reproduce the onset of instabilities using either the GPE or gauge-gravity duality.

Gauge-gravity duality \cite{Ammon:2015wua, RN499,holoreview} is a microscopic framework which maps the dynamics of a certain class of strongly-coupled systems (specifically, strongly-coupled non-Abelian gauge theories with an infinite-dimensional gauge group) to gravity theories defined on spacetimes with a negative global Ricci curvature, called anti-de Sitter spacetimes.  See Appendix \ref{subsec:holo} for more details. Originally, this duality was discovered \cite{Maldacena:1997re,Witten:1998qj,Gubser:1998bc} in the context of String Theory, a theory of quantum gravity. Charged black holes in the gravity theory map to thermal, finite density states in the dual field theory. The area of the event horizon is dual to the entropy of the state, and its temperature is given by the Hawking temperature of the black hole. The late-time dynamics of these black holes is found to be governed by hydrodynamics, as expected on general grounds for interacting systems (whether strongly or weakly coupled). Gauge-gravity duality then offers a solvable framework to investigate and confirm hydrodynamic predictions in a class of models which do not rely on the assumption of weakly-coupled quasiparticles, and is complementary to such approaches.   While the gravitational model we use in this work could in principle be connected to a precise microscopic theory by embedding it into String Theory, it suffices for our purposes that it matches hydrodynamic predictions in the infrared, irrespective of specific microscopic UV completions. In fact, this serves to emphasize the main point of this work, that the onset of the instabilities we study may be understood without referring to a specific microscopic model.

Noteworthy results of the dialogue between gauge-gravity duality and hydrodynamics are the prediction of a lower bound on the shear viscosity to entropy density ratio of strongly-coupled quantum fluids, \cite{Kovtun:2004de}, verified by the Quark-Gluon-Plasma and trapped cold atoms gases \cite{Schafer:2009dj}; and on the diffusion of quantum fluids \cite{Hartnoll:2014lpa}, such as those found in cold atoms systems \cite{Brown:2019vef} or in the strange metal phase of high $T_c$ cuprate superconductors \cite{Zhang:2016ofh}.

Both approaches, hydrodynamics and gauge-gravity duality, can be extended to superfluids. Superfluid hydrodynamics was devised by Landau in the context of Helium 4 \cite{Landau:1941}, followed by many subsequent developments (see \cite{Schmitt:2014eka} for a review of relativistic superfluid hydrodynamics, \cite{chaikinlubensky1995} for Galilean superfluids). In the context of cold atoms, zero-temperature superfluids are described by ideal superfluid hydrodynamics, where temperature effects, and thus dissipation, are neglected \cite{PitStringbook,Sonin_2016}. This hydrodynamic theory should be viewed as a low-energy effective theory, and can be derived more microscopically starting from the GPE. The underlying reason why this can be done is the emergence of two global symmetries in the infrared, a global U(1) symmetry corresponding to charge conservation, and the conservation of the winding of the phase of the superfluid order parameter, which in the absence of vortices gives rise to conserved supercurrents. These global symmetries constrain the effective theory in a way that reduces it to ideal, zero-temperature superfluid hydrodynamics, even though the notion of local thermodynamic equilibrium which underlies hydrodynamics as an effective theory is missing.  

In gauge-gravity duality, a superfluid is modeled by constructing a black hole endowed with scalar `hair', meaning it has a `cloud' of a charged, complex scalar condensate hovering outside its horizon \cite{Gubser:2008px,Hartnoll:2008kx,Hartnoll:2008vx}. This is dual to the condensate (order parameter) in the dual field theory. Correspondingly, the late-time dynamics of these hairy black holes maps to superfluid hydrodynamics \cite{Amado:2009ts,Herzog:2009md,Sonner:2010yx,Herzog:2011ec,Bhattacharya:2011eea,Gouteraux:2019kuy,Gouteraux:2020asq,Arean:2021tks,thermoinstability,Arean:2023nnn}. In previous works by some of us~\cite{An:2024ebg,An:2024dkn}, the dynamical Kelvin-Helmholtz instability at the interface of strongly interacting immiscible binary superfluids was also explored using gauge-gravity duality.

\subsection{Thermodynamic origin of the Landau instability of single-component superfluids}
In single-component superfluids, a well-known instability occurs for states with sufficiently large superflow. An elegant intuitive explanation was given by Landau, making use of the quasi-particle description of weakly-coupled Galilean superfluids such as Helium 4. The superfluid phonons are bosonic quasiparticles and their quasiparticle energy may become negative at large enough superflow, using a Galilean transformation to boost back to the superfluid rest frame \cite{landau1980course9,Schmitt:2014eka}. In \cite{thermoinstability,Arean:2023nnn}, it was shown that the instability can equally well be characterized by the divergence and change of sign of one of the eigenvalues of the Hessian of the free energy density $f$, which occurs when
\begin{equation}
\label{intro:instabcritss}
    \left.\frac{\partial(n_s v_s)}{\partial v_s}\right|_{T,\mu}=0\,,\quad n_s v_s =\left.\frac{\partial f}{\partial v_s}\right|_{T,\mu}\,.
\end{equation}
Here, $(n_s,v_s)$ are the superfluid density and the norm of the superfluid velocity, respectively, which give the superfluid contribution to the charge current. This gives a concrete connection between the microscopic description of the instability (quasiparticles are unstable) and a macroscopic one: the system becomes locally thermodynamically unstable, with a precise identification of the sector where the instability occurs. Beside Helium 4, this criterion also matches the critical supercurrent condition for BCS superconductors~\cite{RevModPhys.34.667} 

This instability is also dynamical, in the sense that time-dependent plane-wave perturbations $\sim e^{-i\omega t+i k x}$ of the state grow exponentially with time, $\textrm{Im}\,\omega>0$. This is manifested by one of the collective hydrodynamic excitations moving into the upper complex frequency plane when \eqref{intro:instabcritss} is crossed. 

This macroscopic argument can easily be extended to other systems than weakly-coupled, Galilean superfluids, and indeed it correctly predicts the onset of the instability in strongly-coupled superfluids with a gravitational dual~\cite{thermoinstability,Arean:2023nnn} as well as many other systems~\cite{Gouteraux:2024adm}. In~\cite{Gouteraux:2024adm}, a general recipe was given to construct hydrodynamic theories with stable linear dynamical evolution provided that the static susceptibility matrix is positive-definite and the local second law of thermodynamics holds. 

One of the main results of the present work is to extend the instability criterion \eqref{intro:instabcritss} to binary superfluids at finite temperature and check that it correctly predicts the onset of coflow and counterflow instabilities in a gauge-gravity duality model of binary superfluids.

\subsection{Summary of results \label{subsec:summary}}

The main aim of this work is to show that dynamical instabilities which
occur in binary superfluids can be predicted by computing
the appropriate static susceptibility. The precise criterion is given in \eqref{criterion0} below. We now summarize our main results.

We first write the hydrodynamics of dissipative binary superfluids, based on the standard hydrodynamic theory of superfluids~\cite{chaikinlubensky1995,PitStringbook,Sonin_2016} and going beyond previous work on binary superfluids~\cite{khalatnikov1957hydrodynamics,khalatnikov1973sound, nepomnyashchil1976microscopic,andreev1975three} by including dissipative terms to first order in derivatives of the hydrodynamic fields and by relaxing the requirement of boost invariance.

Doing so allows us to demonstrate that the onset of these instabilities is predicted by one of the eigenvalues of the Hessian of the free energy diverging and changing sign provided that the local second law of thermodynamics holds, as for single-component superfluids. Conversely, this establishes that the hydrodynamic theory of binary superfluids is stable under linear perturbations provided the local thermodynamic equilibrium is stable and the divergence of the local entropy current is positive.

The exact form of criterion for instabilities is given in \eqref{criterion0} below. In the limit where we freeze the velocity of the normal component of the superfluid as well as temperature fluctuations, this is controlled by the dependence of the superfluid densities on the superfluid velocities.

We carefully verify that this is precisely reproduced in our holographic model of binary superfluids, by analysing the spectrum of quasi-normal modes (QNMs) on top of stationary black hole states. In brief, this means solving the Maxwell's equations coupled to a complex scalar on the black hole background spacetime. Indeed, these QNMs directly correspond to the poles  of superfluid hydrodynamics in the complex frequency plane through the holographic dictionary. In the unstable region predicted by \eqref{criterion0}, we both find that the Hessian is no longer positive-definite and that the spectrum includes a collective mode with $\textrm{Im}\,\omega>0$.

The dynamics of a weakly-interacting Bose gas described by the GPE can be mapped to perfect superfluid hydrodynamics, setting the temperature to zero and neglecting the role of dissipation~\cite{PitStringbook,Sonin_2016}. Since temperature and normal velocity effects are unimportant to leading order for capturing the coflow and counterflow instabilities hydrodynamics, we expect our criterion to hold as well, when modeling the dynamics using the GPE. For the GPE, the critical velocity for counterflow instability has been computed from the microscopic Bogoliubov-de Gennes theory in previous literature~\cite{counterflow1,counterflow2,counterflow3,counterflow4} and we show that this expression exactly matches the one derived from perfect binary superfluid hydrodynamics when applying our criterion \eqref{criterion0}. The reason behind this is clear: since in the perfect superfluid limit, the second law is automatically satisfied, it suffices to examine the appropriate `thermodynamic' derivatives. Even at strictly zero temperature, these derivatives are well defined (we can define the charge and superfluid densities and the superfluid velocity from the microscopic quantum operators), and so our criterion continues to hold.

Thus the instability criterion we establish using the collective hydrodynamic description is found to work in widely different kinds of binary superfluids: an ideal, zero temperature superfluid with weakly-coupled Bogoliubov-de Gennes quasiparticles on one hand, and strongly-coupled binary superfluids without quasiparticles modeled using gauge-gravity duality.

Further, both in GPE and in the holographic model, we find that the critical superfluid velocities where these instabilities occur present a general scaling law with respect to the interaction strength between the two components. 

Finally, by using a full nonlinear time evolution scheme, we find that at the endpoint of the instability the superfluid velocity is lowered back to a stable value through vortex formation and annihilation.

This paper is organized as follows. In section \ref{sec_hydro}, we present the theory of homogeneous binary superfluids and establish the instability criterion, see equation \eqref{criterion0} below. In section \ref{sec:GPE}, we study the counterflow instability with GPE and check its onset is predicted by \eqref{criterion0}. In section \ref{sec:Holo}, we turn to holographic binary superfluids, study their linear instabilities and check that their onset matches the criterion \eqref{criterion0}. Finally, we determine their nonlinear time evolution. We discuss the universal scaling law of the critical velocity in section \ref{sec4.2} and end with some conclusions \ref{sec:conclusion}. Finally, more details and discussion are given in a set of appendices , such as reviews of hydrodynamics (Appendix \ref{subsec:hydro}) and gauge/gravity duality (Appendix \ref{subsec:holo}), as well as details of our holographic setup (Appendix \ref{app:hsetup}).

\section{Counterflow and coflow instabilities from binary superfluid hydrodynamics}

 In this section, we first review how to set up the hydrodynamics of binary superfluids in subsection \ref{sec_hydro}. While we start our treatment very generally, in section \ref{subsec_instab_hydro} we freeze the fluctuations of the normal fluid and of the temperature in order to facilitate the analysis. Indeed, our purpose is not to formulate the hydrodynamic theory in full generality, but rather to analyze how the counterflow and coflow instabilities arise within hydrodynamics, and demonstrate that they can be predicted by a simple condition on static susceptibilities, \eqref{criterion0}.

\subsection{Hydrodynamics of homogeneous binary superfluids}
\label{sec_hydro}
 In this subsection, we review the formulation of binary superfluid hydrodynamics. This differs from ordinary superfluid hydrodynamics (see eg \cite{chaikinlubensky1995} for a standard treatment), as there are generally two conserved $U(1)$ charges and two independent Goldstone degrees of freedom. At ideal order and with Galilean symmetry, this is well known \cite{andreev1975three}. Our goal is to formulate the theory for non-Galilean invariant states (in particular so we can connect to the holographic treatment, which is Lorentz-invariant), and to account for thermal dissipation, along the lines of non-ideal superfluid hydrodynamics \cite{chaikinlubensky1995}.

The conservation equations take the form of generalized continuity equations~\footnote{While we find it convenient to write the conservation equations and constitutive relations for the spatial fluxes non-covariantly, for the purposes of this work we always have in mind systems ultimately invariant under Galilean or Lorentz boost, which set $g^i=j^i_1+j^i_2$ and $g^i=j_\epsilon^i$, respectively. At ideal order, boost invariance does not lead to major differences, but the number of derivative corrections proliferates rapidly without boost invariance \cite{Novak:2019wqg,deBoer:2020xlc,Armas:2020mpr}.}
\begin{equation}
    \label{conservation}
\begin{aligned}
        &\partial_t \epsilon+\partial_ij_\epsilon^i=0,  \quad \partial_tg^i+\partial_i\tau^{ji}=0, \\ &\partial_t n_I+\partial_ij_I^i=0, \quad(I=1,2)\,,
\end{aligned}
\end{equation}
where $\epsilon$, $g^i$ and $n_I$ are the energy, momentum and charge densities, $j_\epsilon^i$ and $j_I^i$ are the energy and charge currents, and $\tau^{ji}$ is the spatial stress tensor. The former two equations follow from invariance under time and space translations, respectively, and the last one is from $U(1)$ global transformations. Denoting $\bm{v^n}$ to be the normal fluid velocity and $\bm{v^s_I}$ the superfluid velocities, one has the Josephson relations:
\begin{equation}
    \label{Josephson}
    \partial_t\varphi_I+\bm{v^n}\cdot \bm{\mathrm{\partial}}\varphi_I+\mu_I=0\,,
\end{equation}
where $\mu_I$ is the chemical potential (including possible derivative corrections, see below) and $\varphi_I$ the Goldstone field for the $I$-th superfluid component. The latter is closely related to the superfluid velocity, $\bm{v^s_I}=\nabla\varphi_I$.

In this paper we work in the grand-canonical ensemble where the thermodynamic variables are the temperature $T$, the chemical potential $\mu$ and the norm of the superfluid velocities $v_I^s=|\bm{v^s_I}|$. The local first law of thermodynamics reads
\begin{equation}
    \label{1stlaw0}
 \begin{aligned}
d\epsilon=&Tds+\mu^Idn_I+\bm{v^n}\cdot d\bm{g}\\&+\bm{h^I}\cdot d\bm{v^s_I}+Nd(\bm{v^s_1}\cdot \bm{v^s_2})\,,
 \end{aligned}
\end{equation}
where $s$ is the entropy density, $\bm{v^n}$ is the normal fluid velocity, $\bm{h_I}=n^s_I(\bm{v^s_I}-\bm{v^n})$ is the conjugate source to the superfluid velocity and $n^s_I$ is the superfluid density of the $I^{\textrm{th}}$ component (here, $I$ is not summed over). In contrast to the single component case~\cite{chaikinlubensky1995}, there is an additional cross term which accounts for the energy change due to relative motion between the two components of superfluids.\footnote{In principle in absence of boost invariance there is a relative coefficient between $\bm{v^s_I}$ and $\bm{v^n}$ in $\bm{h_I}$~\cite{Armas:2023ouk}. As it plays no role in our analysis, we have set it to unity.}
By introducing $\bm{\bar{h}_I}=\bm{h_I}+N\sigma^{IJ}\bm{v^s_J}$ with $\sigma^{IJ}=\left(\begin{smallmatrix} 0&1\\1&0 \end{smallmatrix}\right)$, this cross term can be absorbed and one has
\begin{equation}
    \label{1stlaw}
 \begin{aligned}
    d\epsilon  =Tds+\mu^Idn_I+\bm{v^n}d\bm{g}+\bm{\bar{h}^I}d\bm{v^s_I}\,.
 \end{aligned}
\end{equation}
Unless otherwise stated, repeated capital Latin flavour indices $I,J,\dots$ are summed over, just like the usual small case Latin spatial indices $i,j,\dots$ (\emph{i.e.} the Einstein summation convention).

We now consider non-ideal corrections to the ideal order constitutive relations, labeled by tildes. One central hypothesis of hydrodynamics is that of local equilibrium, ie we coarse-grain the system over scales larger than the local equilibration scales and define local equilibrium thermodynamic quantities, $\epsilon(t,x^i)$, $n(t,x^i)$, etc. The global version of the second law of thermodynamics can then be rewritten locally as the divergence of an entropy current, with $s(t,x^i)$ and $T(t,x^i)$ the non-equilibrium entropy density and temperature:

\begin{equation}
    \label{2ndlaw}
    T\partial_ts+T\partial_i j^i_s\equiv \Delta\ge0\,.
\end{equation}
$\Delta\geq0$ is the local version of positivity of entropy production , with $j^i_s$ the entropy current. 

The first law of the thermodynamics can also be promoted to a local version, and used to express $\partial_t s$ and $\partial_i s$ in terms of the other quantities in \eqref{1stlaw}.
After using the evolution equations \eqref{conservation}, \eqref{Josephson} and doing some integration by parts \cite{chaikinlubensky1995}, imposing the absence of entropy production at ideal level gives the constitutive relations for the currents in equilibrium state compatible with parity and time-reversal invariance:
\begin{equation}
\label{constitutive}
    \begin{aligned}
        &j_I^i=n_Iv^{ni}+\bar{h}_I^i+\tilde{j}_I^i\,,  \\ &j_\epsilon^i=(\epsilon+p)v^{ni}+\partial_t\varphi^I\bar{h}_I^i+\tilde{j_\epsilon^i}\,,\\
        &\tau^{ji}=p\delta^{ij}+v^{ni}g^j+\bar{h}_I^jv_s^{Ii}+\tilde{\tau}^{ji}\,, \\
        & j^i_s=sv^{ni}+\tilde{j}^i_s\,,
    \end{aligned}
\end{equation}
where $p=-\epsilon+sT+n_I\mu_I+v^{ni}g_i$, while tilded quantities are higher-order in derivatives corrections to the ideal constitutive relations. \footnote{While not manifest from \eqref{constitutive}, the stress tensor $\tau^{ij}$ is symmetric for isotropic states. This can be verified by inverting the relation between $\bm{\bar h_I^i}$ and $\bm{v_s^I}$, working out the static susceptibilities $\chi_{v_{s}^{Ii}g^j}$, using Onsager $\chi_{v_{s}^{Ii}g^j}=\chi_{g^i v_{s}^{Ij}}$ and deducing the expression for $g^i$. Substituting in \eqref{constitutive}, $\tau^{ij}=\tau^{ji}$. } Then, the local second law \eqref{2ndlaw} becomes 
\begin{equation}
\label{Deltanonideal}
    \Delta=-\tilde{j}_s^i\partial_iT/T-\tilde{j}_I^i\partial_i\mu^I-\tilde{\tau}^{ji}\partial_iv^n_j-\tilde{\mu^I}\partial_i\bar{h}_I^i\,,
\end{equation}
where $\tilde{\mu^I}$ are derivative corrections in the Josephson equations.  We can now use the second law constraint $\Delta\geq0$ to determine the most general form of the non-ideal constitutive relations for $\tilde j^i_I$, $\tilde j^i_s$, $\tilde \mu_I$ and $\tilde \tau^{ij}$ compatible with positivity of entropy production. To proceed, a choice of frame must be made, ie a choice of out-of-equilibrium definition of temperature, chemical potentials and velocities, \cite{Kovtun:2012rj}. We choose a special `thermodynamic' frame \cite{Jensen:2012jh, Armas:2020mpr} where only spatial fluxes $\tilde j^i_I$, $\tilde j^i_s$, $\tilde \mu_I$ and $\tilde \tau^{ij}$, not the densities $O_A$, receive derivative corrections, and time derivatives are removed by using the ideal EoMs.

\subsection{Linear analysis and the instabilities \label{subsec_instab_hydro}}

Next, we turn to the dynamical instability from hydrodynamics. One can linearize the EoMs around equilibrium state by perturbing the thermodynamic quantities $O_A=O_{A0}+e^{-i\omega t+i\bm{k}\cdot \bm{x}}\delta O_A$ and their sources $s_A=s_{A0}+e^{-i\omega t+i\bm{k}\cdot \bm{x}}\delta s_A$, where $O_A=(s,\bm{g}, n_I, \bm{v^s_I})$ and $s_A=(T,\bm{v^n}, \mu_I, \bm{\bar{h}_I})$. They are related by the static susceptibility matrix $\bar\chi_{AB}=\delta O_A/\delta s_B=-\delta^2 \bar F/\delta s_A\delta s_B$, where $\bar F=-T\mathrm{ln}Z-\int d^d x \,\bm{v^s_I}\cdot \bm{\bar{h}_I}$ is the Legendre-transformed thermal Gibbs free energy and $Z$ static partition function.  Given our choice of frame, the linearized EoMs in Fourier space are
\begin{equation}
\label{eom}
\begin{aligned}
        &(-i\omega+i\bm{k}\cdot \bm{v^n})\delta O_A+M_{AB}(\bm{k})\delta s_B=0\\ \Rightarrow\;& (-i\tilde{\omega}+M(\bm{k})\cdot \bar\chi^{-1})\delta O=0\,,
\end{aligned}
\end{equation}
where $\tilde{\omega}=\omega-\bm{k}\cdot \bm{v^n}$.  The fact that the matrix $M$ only depends on ${\bf k}$ follows from our choice of frame. We show this explicitly in equation \eqref{MAB} below.

Then the spectrum of collective modes can be obtained by solving $\det(-i\tilde{\omega}+M\cdot \bar\chi^{-1})=0$. The onset of a dynamical instability occurs when the imaginary part of $\tilde{\omega}$ becomes positive. In~\cite{Gouteraux:2024adm}, a general proof is given that provided positivity of entropy is imposed $\Delta\geq0$, this can only occur when an eigenvalue of $\bar\chi$, and so  $\det(\bar\chi)$ also, diverges and changes sign, revealing the thermodynamic origin of these instabilities. In this work, we restrict to their study in the `probe' limit where normal velocity and temperature fluctuations are set to zero. Thus, only the charge conservations equations and Josephson equations need to be solved. We further simplify the analysis by restricting to choices of the superfluid velocities relevant for the coflow and counterflow instabilities, meaning that we take both velocities in the same direction with equal (coflow) or opposite (counterflow) orientations. Then, also taking the wavenumber ${\bf k}$ in the same direction as the superfluid velocities, the modes can be solved for analytically and the onset of the instabilities can be studied by direct inspection, without needing to resort to the general proof in \cite{Gouteraux:2024adm}.\footnote{The proof in \cite{Gouteraux:2024adm} is presented for single-component superfluids, and gives rigorous justification for earlier results obtained in \cite{thermoinstability,Arean:2023nnn}. Its extension to binary superfluids is straightforward.}

We now consider a frame where the normal fluid is at rest, $\bm{v^n}=0$. This is the case for holographic superfluids in the probe limit, and also for the superfluids modeled by GPE. For simplicity, we shall consider the case where the chemical potentials and charge densities are the same for both superfluid components.\footnote{This is equivalent to setting the imbalance charge density $n_-=n_1-n_2$ identically to zero. This can straightforwardly be relaxed if needed.} Then the thermodynamic quantities and sources we need to consider are just $\{n_I=(n,n),\,\bm{v^s_I}\}$ and $\{\mu_I=(\mu,\mu),\,\bm{\bar{h}_I}\}$. In this case, the EoMs are
\begin{equation}\label{sfeoms}
\begin{aligned}
        &\partial_t n+\partial_i(j^i+\tilde{j^i})=0, 
 \\ &\partial_tv_I^{si}+\partial_i(\mu_I+\tilde{\mu_I})=0\,.
\end{aligned}
\end{equation}

Positivity of entropy production $\Delta\geq0$ leads us to consider non-ideal constitutive relations in the form

\begin{equation}
\label{constrels}
\left(\begin{array}{c}
    \tilde j^i  \\
      \tilde\mu_I
\end{array}\right)=-\tilde\Sigma\cdot\left(\begin{array}{c}
     \partial_j\mu  \\
    \partial_j \bar{h}_J^j 
\end{array}\right)
\end{equation}
so that
\begin{equation}
    \Delta = \left(  \partial_i\mu  ,\,
    \partial_j \bar{h}_I^j \right)\cdot\tilde\Sigma\cdot\left(\begin{array}{c}
     \partial_j\mu  \\
    \partial_j \bar{h}_J^j 
\end{array}\right)\geq0
\end{equation}
means imposing that the matrix $\tilde \Sigma$ is positive-definite.
It suffices then to parametrize $\tilde\Sigma$ in the most general way compatible with the index structures and with time reversal symmetry (Onsager reciprocity).\footnote{Under time-reversal, $\left\{n,\mu,{\bf v_s^I},{\bf \bar h^I}\right\}\mapsto\left\{n,\mu,-\bf v_s^I,-{\bf \bar h^I}\right\}$. Onsager symmetry then implies $G^R_{AB}(\omega,{\bf k},{\bf v_s^I})=\eta_A\eta_B G^R_{AB}(\omega,-{\bf k},-{\bf v_s^I})$, with $\eta_A$ the eigenvalue under time reversal of operator $A$. These constraints must also be obeyed by the static susceptibility matrix $\bar\chi_{AB}$ as well as by the matrix $\tilde\Sigma$. This implies in particular that the same coefficient $\beta$ appears in the off-diagonal block of this matrix.} This gives:
\begin{equation}
\label{Sigmaij} \tilde\Sigma=\left(\begin{array}{cc}
    \sigma_1 \delta^{ij}+\sigma_2\frac{v_J^{si}v_J^{sj}}{|v_J^s|^2} & \beta v_J^{si} \\
    \beta v_I^{sj} & \delta_1 \delta_{IJ}+\delta_2 v_I^{si}v_{Ji}^s 
\end{array}\right)
\end{equation}
where $(\sigma_1, \sigma_2,\beta,\delta_1,\delta_2)$ are five independent transport coefficients.\footnote{In principle, $\tilde{j^i}$ can be different for each component, \emph{i.e.} $\Delta\tilde{j^i}=\tilde{j^i_1}-\tilde{j^i_2}\ne 0$. Nevertheless, from~\eqref{conservation} one has $\partial_i\Delta\tilde{j^i}=0$ since $n_1=n_2$. On the other hand, only divergence of $\tilde{j^i}$ appears in the EoMs~\eqref{sfeoms}. Therefore, $\Delta\tilde{j^i}$ has no effects and one can simply take $\tilde{j^i_1}=\tilde{j^i_2}=\tilde{j^i}$.} 
 We can now write $\Delta$ as: 
 \begin{align}
 \Delta =&\, \sigma_1\left(\partial^i\mu+\frac{\beta}{4\sigma_1}v^{si}_I\partial^k \bar h_{kI}\right)^2\nonumber\\
 &+\left(\delta_2-\frac{\beta^2}{2\sigma_1}\right)v^{si}_I\partial^k \bar h_{kI}v^s_{iJ}    \partial^j\bar h_{jJ}\nonumber\\
 &+\sigma_2\left(v^{si}_I\partial_i\mu+\frac{\beta|v^s_J|^2}{2\sigma_2}\partial_k \bar h^k_I\right)^2\nonumber\\
 &+\left(\delta_1-\frac{\beta^2|v^s_J|^2}{4\sigma_2}\right)\partial_k \bar h^k_I\partial^j \bar h^I_j\,,
 \end{align}
     and imposing that it is a positive-definite quadratic form for any configuration of sources $\mu$, $\bar{h}_I^i$ results in the following constraints:
 \begin{equation}
 \label{constraintsDelta}
    \begin{aligned}
            & \sigma_{1,2}\geq0\,,\quad \beta^2\leq 4 \sigma_1\delta_2\,,\quad \beta^2\leq\frac{4\sigma_2\delta_1}{|v^s_J|^2}\,.
    \end{aligned}
 \end{equation}

The case with general superfluid velocities is still very complicated. We focus on the collinear case where $\bm{v^s_1}// \bm{v^s_2} // \bm{k}$.\footnote{There is no conceptual obstacle generalising to other cases. But the resulting expressions are very lengthy.}  For a single species superfluid, this is the sector where the dominant instability is found \cite{Amado:2013aea,Schmitt:2014eka}. Evidence that this continues to be true for binary superfluids was presented in \cite{counterflow3}. For later convenience, we change variables to
\begin{equation}\label{pmv}
\begin{aligned}
 & v^s_+=\frac{v^s_1+v^s_2}{2},\quad v^s_-=\frac{v^s_1-v^s_2}{2}\,,\\
 &\bar{h}_+=\bar{h}_1+\bar{h}_2,\quad \bar{h}_-=\bar{h}_1-\bar{h}_2\,.
\end{aligned}   
\end{equation}
By linearizing the EoMs~\eqref{sfeoms},  inserting the constitutive relations \eqref{constrels} and \eqref{Sigmaij}  we find the matrix $M_{AB}$ to be 
\begin{widetext}
\begin{equation}\label{MAB}
  M=\left(
    \begin{array}{ccc}
        \sigma k^2   & ik+\beta v^s_+k^2/2 & \beta v^s_-k^2/2\\
        ik+\beta v^s_+k^2/2 & (\delta_1+\delta_2 (v^s_+)^2)k^2/4 & \delta_2 v^s_+v^s_-k^2/4\\
        \beta v^s_-k^2/2 & \delta_2 v^s_+v^s_-k^2/4 & (\delta_1+\delta_2 (v^s_-)^2)k^2/4 
    \end{array}
    \right),
\end{equation} 
\end{widetext}
with $\sigma=\sigma_1+2\sigma_2$  (we present the linearized equations in the appendix \ref{app:eig}). In the equation above, we have treated $\bar{h}_I$ as a source as it makes linearizing the EoMs easier. In practice, it is more convenient to treat $v^s_I$ as a source (as this is easier to control the value of this parameter). Let us denote $(\delta n, \delta v^s_I)^T=\bar{\chi}\cdot(\delta \mu,\delta \bar{h}_I)^T$ and $(\delta n,\delta \bar{h}_I)^T=\chi\cdot(\delta \mu,\delta v^s_I)^T$. Then we can express $\bar{\chi}$ in terms of $\chi$:
\begin{widetext}
\begin{equation}\label{chiAB}
    \bar{\chi}=\left(
    \begin{array}{ccc}
      \chi_{nn}+\chi_{n+}\frac{\chi_{n+}\chi_{--}-\chi_{n-}\chi_{-+}}{\chi_{++}\chi_{--}-\chi_{+-}\chi_{-+}}+\chi_{n-}\frac{-\chi_{n+}\chi_{+-}+\chi_{n-}\chi_{++}}{\chi_{++}\chi_{--}-\chi_{+-}\chi_{-+}}   & \frac{\chi_{n+}\chi_{--}-\chi_{n-}\chi_{-+}}{\chi_{++}\chi_{--}-\chi_{+-}\chi_{-+}} & \frac{-\chi_{n+}\chi_{+-}+\chi_{n-}\chi_{++}}{\chi_{++}\chi_{--}-\chi_{+-}\chi_{-+}} \\
        \frac{\chi_{n+}\chi_{--}-\chi_{n-}\chi_{-+}}{\chi_{++}\chi_{--}-\chi_{+-}\chi_{-+}}  & \frac{\chi_{--}}{\chi_{++}\chi_{--}-\chi_{+-}\chi_{-+}} & \frac{-\chi_{+-}}{\chi_{++}\chi_{--}-\chi_{+-}\chi_{-+}} \\
        \frac{-\chi_{n+}\chi_{+-}+\chi_{n-}\chi_{++}}{\chi_{++}\chi_{--}-\chi_{+-}\chi_{-+}} & \frac{-\chi_{-+}}{\chi_{++}\chi_{--}-\chi_{+-}\chi_{-+}}  & \frac{\chi_{++}}{\chi_{++}\chi_{--}-\chi_{+-}\chi_{-+}}
    \end{array}
    \right),
\end{equation}
\end{widetext}
where $\chi_{nn}=\partial n/\partial \mu|_{v^s_I}$, $\chi_{nI}=\partial n/\partial v^s_I|_\mu$ and $\chi_{IJ}=\partial \bar{h}_I/\partial v^s_J|_\mu$, We also have $\chi_{+-}=\chi_{-+}$, Onsager reciprocity and as is clear from $\bar\chi_{AB}\equiv-\delta^2\bar F/\delta s_A\delta s_B$. By solving (\ref{eom}) with~\eqref{MAB} and~\eqref{chiAB}, we find two sound modes and one diffusive mode:\footnote{ In the more general case where the two U(1) densities are kept different, there are instead two pairs of superfluid sound modes \cite{andreev1975three}.}
\begin{equation}
    \label{modes}
\begin{split}
    \omega_\pm&=v_\pm k+i\Gamma_\pm k^2+\mathcal{O}(k^3)\,, \\
    \omega_0&=i\Gamma_0 k^2+\mathcal{O}(k^3)\,,
\end{split}
\end{equation}
where
\begin{equation}
    \label{soundspeed}
\begin{aligned}
        v_\pm&=\frac{-\chi_{n+}\pm\sqrt{\chi_{n+}^2+\chi_{++}\chi_{nn}}}{\chi_{nn}}, \\ 
    \Gamma_0&=-\frac{(\chi_{++}\chi_{--}-\chi_{+-}^2)(\delta_2(v^s_-)^2+\delta_1)}{4\chi_{++}}\,,
\end{aligned}
\end{equation}

\begin{widetext}
    \begin{equation}
        \begin{aligned}
\Gamma_\pm=&\frac {1} {4(\chi_{++}-v_\pm(4\chi_{n+}+3v_\pm \chi_{nn}))} \\&(
\delta_1( v_\pm^2\chi_{nn}(\chi_{++}+\chi_{--})+(v_\pm\chi_{n-}-\chi_{+-})^2+v_\pm\chi_{n+}(2\chi_{--}+v_\pm\chi_{n+})-\chi_{++}\chi_{--}) \\& 
+\delta_2 (
v_\pm^2(v_+^s)^2(\chi_{n+}^2+\chi_{++}\chi_{nn})+(v_-^s)^2(
v_\pm^2\chi_{--}\chi_{nn}+(v_\pm\chi_{n-}-\chi_{+-})^2+(2v_\pm\chi_{n+}-\chi_{++})\chi_{--})\\& +2v_\pm v_+^s v_-^s(\chi_{n-}(v_\pm\chi_{n+}-\chi_{++})+\chi_{+-}(v_\pm\chi_{nn}+\chi_{n+}))
) 
+4\sigma v_\pm^2 \\&
+4\beta (v_+^s+v_-^s) v_\pm (v_+^s(v_\pm\chi_{n+}-\chi_{++})+v_-^s(v_\pm\chi_{n-}-\chi_{+-})))\,.\\
        \end{aligned}
        \end{equation}
\end{widetext}
Checking that $\Gamma_\pm <0$ provided the constraints \eqref{constraintsDelta} hold and that $\bar\chi$ is positive-definite is tedious, highlighting the interest of a more general derivation, \cite{Gouteraux:2024adm}.

The criterion for the onset of instability \cite{Gouteraux:2024adm} is given by  $\mathrm{det}(\bar{\chi})=-\chi_{nn}/(\chi_{++}\chi_{--}-\chi_{+-}^2)$ diverging and changing sign. It gives\footnote{In principle, it may also be that $\chi_{nn}$ changes sign. But this is not what we observe in the two microscropic examples based on the GPE and gauge/gravity duality.} 
\begin{equation}\label{criterion0}
    \chi_{++}\chi_{--}-\chi_{+-}^2=0\,,
\end{equation}
at which both the sound and diffusive modes of~\eqref{modes} could begin to develop a positive imaginary frequency, causing perturbations to grow exponentially.
One can directly verify that the criterion~\eqref{criterion0} is also the same for non-identical binary superfluids for which $n_1\neq n_2$. The calculation is straightforward but lengthier. 

In the counterflow and coflow cases, the expressions for the modes become simpler and the analysis can be performed by direct inspection more easily.

In the counterflow case, $v^s_+=0$, we have $\chi_{+-}=0$, since $\chi_{+-}=\delta^2 F\left[(v^s_+)^2,(v^s_-)^2\right]/(\delta v^s_+\delta v^s_-)\sim v^s_+ v^s_-$. For a similar reason, we also have  $\chi_{n+}=0$. Then, the expressions for the modes become
\begin{align}
\omega_\pm=&\pm\sqrt{\frac{\chi_{++}}{\chi_{nn}}}k-\frac{ik^2}{2\chi_{nn}}\left[\sigma-v_-\beta\chi_{n-}\right.\nonumber\\
&\left.+\frac{\chi_{n-}^2}{4}(\delta_1+v_-^2\delta_2)+\frac{\chi_{nn}\chi_{++}}{4}\delta_1\right],\label{soundcounterflow}\\
\omega_0=&-\frac{i}4k^2(\delta_1+v_-^2\delta_2)\chi_{--}\,.
\end{align}
It is clear from both expressions that an instability occurs if $\chi_{++}$ becomes negative, since then the sound speed becomes purely imaginary, with $\omega_-$ moving into the upper half plane. This is precisely what we observe in the next sections in the GPE \cite{counterflow4} and in holographic superfluids. In principle, $\chi_{--}$ changing sign would also cause an instability by sending $\omega_0$ into the upper half plane, but this is not what we find happens in the GPE and in holographic superfluids.

For the coflow case $v_-=0$, we have $\chi_{+-}=\chi_{n-}=0$. The modes  simplify to
\begin{align}
    \omega_0=&-i\frac{\chi_{--}}{4}\delta_1 k^2 \label{diffcoflow}\,,\\
    \omega_\pm=&v_\pm k+i\Gamma_\pm k^2\,,\\
    \Gamma_\pm=&-\frac12 v^s_+v_\pm\beta-\frac{(v_+^s)^2}8(\chi_{++}-\chi_{n+}v_\pm)\delta_2\nonumber\\
    &\mp v_\pm\frac{16\sigma+\chi_{nn}^2(v_+-v_-)^2\delta_1}{16\chi_{nn}(v_+-v_-)}\,.\label{soundcoflow}
\end{align}
The expression for $v_\pm$ is given in \eqref{soundspeed}.

We also see two possible instabilities in this case, when either $\chi_{--}$ or $\chi_{++}$ change sign. In holographic superfluids, the former is realized first as $v_+^s$ is increased and leads to $\omega_0$ crossing to the upper half plane. A secondary instability occurs when $\chi_{++}$ changes sign, upon which both $v_+$ and $\Gamma_+$ vanish and then change sign, moving $\omega_+$ to the upper half plane.  To highlight this, we expand the relevant dispersion relation around $\chi_{++}=0$, assuming $\chi_{n+}>0$:\footnote{If $\chi_{n+}<0$, then $\omega_-$ becomes unstable instead.}
\begin{equation}
\label{coflowunstablesound}
    \begin{aligned}
         \omega_+=\,&\frac{\delta\chi_{++}}{2\chi_{n+}}k-ik^2\frac{\delta\chi_{++}}{16}\times \\& \left( \delta_1  +\delta_2(v_+^s)^2+\frac{4\beta v_+^s}{\chi_{n+}}+\frac{4\sigma}{\chi_{n+}^2} \right).
    \end{aligned}
\end{equation}
When $\delta\chi_{++}<0$, the mode crosses to the upper half plane and becomes unstable.

\section{Counterflow instability from the GPE}
\label{sec:GPE}
The development of quantum turbulence from two counter-propagating superfluids of miscible BECs has been studied experimentally and by using the mean-field GPE~\cite{counterflow1,counterflow2,counterflow3,counterflow4,beattie2013persistent,yakimenko2013stability,abad2014persistent}. In the mean field approximation, the condensate wave functions are described by $\Psi_J(\mathbf{r},t)=\sqrt{n^s_J(\mathbf{r},t)}e^{i\theta_J}(\mathbf{r},t)$ where $n^s_J$ and $\theta_J$ are the particle density and phase of the $J$-th component of the binary superfluids. The latter is nothing but the Goldstone field $\varphi_J$ for the $J$-th superfluid component. The coupled GPEs are given as (here, capital Latin indices are not summed over) 
\begin{equation}
    \label{GP}
    \begin{aligned}
        i\partial_t \Psi_I=&\left(-\frac{1}{2m_I}\nabla^2-\mu_I+g_I|\Psi_I|^2+g_{IJ}|\Psi_J|^2\right)\Psi_I,\\&\quad  (I,J=1,2, \quad I\ne J)\,.     
    \end{aligned}
\end{equation}
They describe binary superfluids without dissipation or a normal fluid component. Here $m_I$ is the mass of the $I$-th component, and $g_1, g_2$ and $g_{12}$ are the coupling
constants. For more about BEC and GPE, readers can refer to the textbooks~\cite{pethick2008bose,pitaevskii2003bose}. To obtain stable miscible condensates, we consider the conditions  $g_{12}<\sqrt{g_1 g_2}$ and $g_{I}>0$. 
It was shown that the countersuperflow becomes unstable and quantized vortices are nucleated as the relative velocity exceeds a critical value, leading to isotropic quantum turbulence consisting of two superflows. 

We show that the counterflow instability of the two-component BECs via~\eqref{GP} continues to be given by the divergence of $\det(\bar{\chi})$. Since the derivatives involved in the criterion we derived assuming the existence of a hydrodynamic regime involve quantities that are well-defined even at zero temperature (superfluid velocity, charge densities), it is meaningful to evaluate them at zero temperature with the GPE. Further, perfect superfluid hydrodynamics can be derived from the GPE \cite{PitStringbook,Sonin_2016}, at zero temperature, neglecting dissipation and in the absence of a normal component.
For the case where there are two $U(1)$ conserved charges, one can find $\det(\bar{\chi})$ diverges when (\ref{criterion0}) is satisfied. To be more specific, it is given by
\begin{equation}
    \chi_{++}=n^s_++v^s_-\partial_{v^s_+}n^s_-+N=0\,,
\end{equation}
with $n^s_+=n^s_1+n^s_2$ and $n^s_-=n^s_1-n^s_2$.
For simplicity, let's consider the case where $m_1=m_2=m$, $g_1=g_2=g$ and $\mu_1=\mu_2=\mu$. From (\ref{GP}), the superfluid densities for stationary and uniform state can be found to be 
\begin{equation}
\begin{aligned}
        n^s_I=\frac{g(\mu-\frac{m}{2}(v^s_I)^2)-g_{12}(\mu-\frac{m}{2}(v^s_J)^2)}{g^2-g_{12}^2},& \\ I,J=1,2,  \quad I\ne J,&
\end{aligned}
\end{equation}
with $n^s_I=|\Psi_I|^2$.
In terms of $v^s_+$ and $v^s_-$ of~\eqref{pmv}, the corresponding densities are
\begin{equation}
\begin{aligned}
        n^s_+&=\frac{2\mu-m((v^s_+)^2+(v^s_-)^2)}{g+g_{12}}, \\ n^s_-&=-\frac{2mv^s_+v^s_-}{g-g_{12}}\,.
\end{aligned}
\end{equation}
We can obtain $N$ by noting that
\begin{equation}
    \chi_{+-} - \chi_{-+}=v^s_-\partial_{v^s_+}N-\frac{4v^s_+v^s_-g_{12}}{g^2-g_{12}^2}=0\,.
\end{equation}
Therefore, we have $N=2(v^s_+)^2g_{12}/(g^2-g_{12}^2)$, which is 0 for $v^s_+=0$, \emph{i.e.} the counterflow case. Then the criterion for the onset of instability is given by 
\begin{equation}
    \chi_{++}|_{v^s_+=0,v^s_-=v_c}=\frac{2\mu-mv_c^2}{g+g_{12}}-\frac{2mv_c^2}{g-g_{12}}=0\,,
\end{equation}
which gives
\begin{equation}
 v_c=\sqrt{(g-g_{12})n^s_+/2m}\,.   
\end{equation}
This is exactly the critical velocity derived from Bogoliubov–de Gennes analysis~\cite{counterflow1,counterflow2,counterflow3,counterflow4}, with $n^s_+=2n$. Therefore, we see that the microscopic Bogoliubov-de Gennes theory offers a microscopic realization of the macroscopic description of the dynamical counterflow instability that we established in \eqref{criterion0}. As we shall see now, the macroscopic criterion \eqref{criterion0} applies to a completely different class of superfluids, which are those that have a gravitational dual.

\section{Holographic binary superfluids}
\label{sec:Holo}

We now move to the gravity description of a two-component superfluid in two spatial dimensions. More details about the holographic model are given in appendix \ref{app:hsetup}. The field theory system involves two complex scalar operators $\mathcal{\hat{O}}_I$, which carry the same charge $e$ under a global $U(1)$ symmetry.\footnote{Thus there will be only one conserved density, which is a special limit of the hydrodynamic and GPE analyses of the previous sections.} Below a critical temperature $T_c$, the scalar operators develop a nonzero expectation value, which breaks the $U(1)$ symmetry spontaneously and drives the system into a superfluid phase. The dual (3+1)-dimensional gravity system (the {\em bulk}) is defined on an asymptotically AdS spacetime. It includes a bulk dynamical $U(1)$ gauge field $A_\mu$ and two bulk complex scalar fields $(\Psi_1, \Psi_2)$ carrying charge $e$ under $A_\mu$, dual to the field theory scalar operators $\mathcal{\hat{O}}_I$. For definiteness, we will choose two identical scalar fields with the potential $V(\Psi_1, \Psi_2)=m^2(|\Psi_1|^2+|\Psi_2|^2)+\frac{\nu}{2}|\Psi_1|^2|\Psi_2|^2$.\footnote{Though the scalar fields are only directly coupled through the density-density coupling term, they are also indirectly coupled through interactions with gauge field. Therefore, current-current entrainment is naturally incorporated in our model. To see this explicitly, one can refer to Figure~\ref{thermo} and notice that $2N|_{v_+=v_-=0}=(\chi_{++}-\chi_{--})|_{v_+=v_-=0}$.} We pick the value $m^2L^2=-2$, so that the dual scalar operators have the scaling dimension $\Delta_\Psi=2$. We do not expect our results to differ qualitatively
for other choices of the mass. There is an inter-component coupling $\nu$,  which characterizes the interaction between the two superfluid components and determines whether the components are miscible~\cite{An:2024ebg}. For the homogeneous case with $\nu>0$, only one of the superfluid components can condense below the critical temperature $T_c$ (immiscible binary superfluids), while $\nu<0$ gives miscible binary superfluids, which is the focus of this work. 

At finite temperature, in general each of the two species
contains both superfluid and normal components. However, the essential physics of the counterflow and coflow instabilities we are after can be accessed by freezing out the normal velocity and temperature fluctuations. This implies that the dynamics of momentum and energy are decoupled from the charge and superfluid sectors in the field theory. This corresponds to the so-called probe limit in the gravity description, where the backreaction of the matter sector on the metric is neglected. More precisely, the background is given by the Schwarzschild AdS black brane (see Appendix \ref{app:hsetup} for details on the bulk action and the metric we use). After solving the bulk EoMs, we can read off all the relevant observables thanks to the standard holographic dictionary, such as the temperature $T$ or the superfluid velocity $\bm{v}_I^s\equiv(v_{Ix}^s, v_{Iy}^s)$ for each superfluid condensate $\mathcal{O}_I\equiv \langle \mathcal{\hat{O}}_I \rangle$.
This gravity dual provides a first-principle description of superfluid dynamics.

Phase-separated binary holographic superfluids have been studied recently in~\cite{An:2024ebg,An:2024dkn,An:2024mbx}, and rich dynamics was found. In this work, we are interested in miscible binary holographic superfluids, which provide a good platform for checking the hydrodynamic framework developed in Section~\ref{sec_hydro} and the onset of instabilities. The construction of the stationary configuration for miscible binary superfluids is provided in Appendix~\ref{app:eom}.

\subsection{Dynamical instability from linear analysis}
\label{sec:eig}
In this section we study the dynamical  counterflow and coflow instabilities in miscible holographic binary superfluids using linear response theory. To do so, we turn on small perturbations $\sim e^{-i(\omega t-\bm{k\cdot x})}$ over the stationary background geometry, thanks to the translation invariance of the background along the time and spatial directions. Due to thermal dissipation by the event horizon of the  black brane background geometry, the frequency $\omega$ typically takes a complex value , ie poles of retarded Green's functions lie in the complex plane. A dynamical instability is triggered if the imaginary part of a pole at some value of the wave number $\bm{k}$ becomes positive, Im$\omega>0$. The poles of the retarded Green's functions of the dual field theory are determined from the quasinormal modes (QNMs) of gravito-electric waves in the gravitational theory.\footnote{Both the hydrodynamic and non-hydrodynamic modes of the dual field theory can be obtained by computing the QNMs of the dual black hole.}

Without loss of generality, we assume that the superfluid velocity $v_y$ is aligned along the $y$ direction. More precisely, the counterflow case corresponds to $\bm{v}_1^s=-\bm{v}_2^s=(0, v_y)$, and the coflow case to $\bm{v}_1^s=\bm{v}_2^s=(0, v_y)$. Since the instability will typically develop fastest along the flow of the superfluids, we restrict our analysis to wavevectors parallel to the superfluid velocity, \emph{i.e.} $\bm{k\cdot x}=k\,y$. 

\begin{figure}[htpb]
    \centering
        \includegraphics[width=0.9\linewidth]{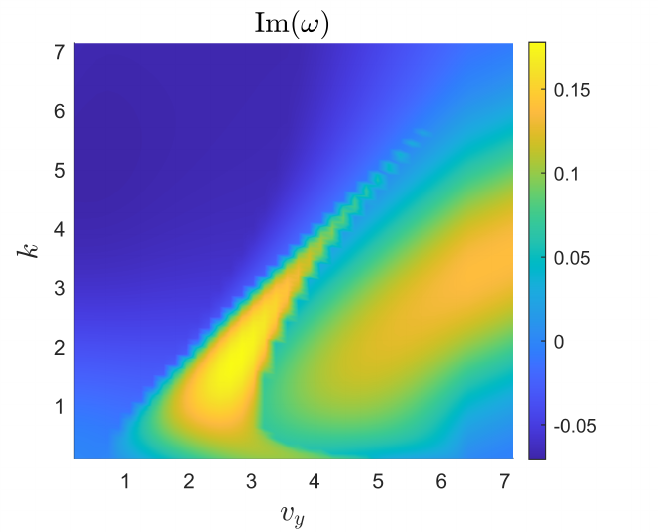}
        \includegraphics[width=0.9\linewidth]{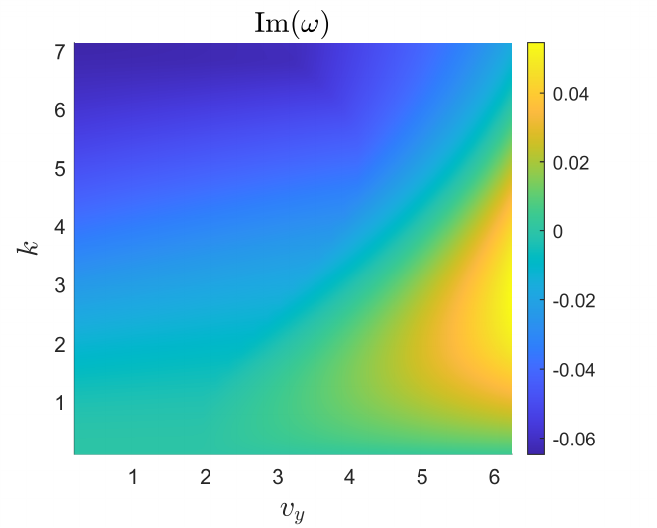}
        \caption{The QNMs spectrum with respect to $k$ and $v_y$ for the counterflow case (top) and the coflow cases (bottom). The stationary configuration is dynamical unstable whenever $\mathrm{Im}\omega_k>0$. We have considered holographic miscible binary superfluids at $T/T_c=0.677$ and $\nu=-0.2$.}
    \label{omega}
\end{figure}

The full set of linearized EoMs can be found in Appendix~\ref{app:eig}, which results in a generalized eigenvalue problem that can be solved numerically. In Figure~\ref{omega}, we give a representative example for the counterflow instability (upper panel) and the coflow instability (lower panel), where we show the dominant QNMs with respect to the wave number $k$. While the spectral patterns are quite different, one finds that a dynamical instability will develop beyond a critical velocity for both the counterflow and coflow cases.

\begin{figure}[htpb]
    \centering
        \includegraphics[width=0.98\linewidth]{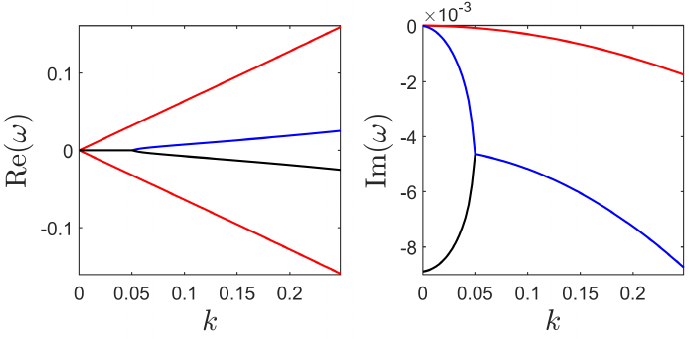}
        \caption{The QNM spectrum for holographic miscible binary superfluids with $T/T_c=0.677$, $\nu=-0.1$ and $v_y=0$. The left panel shows real part of QNMs while the right panel shows imaginary part. The red lines are the superfluid sound modes. The blue and black lines correspond to the type $\mathrm{\uppercase\expandafter{\romannumeral2}}$ Goldstone boson when $\nu=0$. At nonzero $\nu$, we expect one branch of this mode to otain an imaginary gap, which is indeed what we observe.}
    \label{example1}
\end{figure}

The spectrum of QNMs for the case with $\nu=0$ and $v_y=0$ was studied in~\cite{type2G} where the scalar fields transform as a doublet, \emph{i.e.} there is a global $SU(2)$ symmetry (systems with such a symmetry have been studied in the context of binary superfluids for some time \cite{manakov1974theory}). In~\cite{type2G}, the holographic analysis found that there exist two kinds of gapless modes. The first are the superfluid sound modes, and their dispersion relations is linear in the hydrodynamic regime $\omega_{\mathrm{\uppercase\expandafter{\romannumeral1}}}=\pm (v_sk+\Bar{b} k^2)-i\Gamma k^2$. These modes are classified as type $\mathrm{\uppercase\expandafter{\romannumeral1}}$  Goldstone boson~\cite{countingrule} and due to the spontaneous breaking of the $U(1)$ symmetry. The second type of gapless modes presents a quadratic dispersion relation at small $k$, $\omega_{\mathrm{\uppercase\expandafter{\romannumeral2}}}=\pm b k^2-ick^2$, and are classified as type $\mathrm{\uppercase\expandafter{\romannumeral2}}$ Goldstone bosons. These modes appear because  the bulk global $SU(2)$ symmetry is spontaneously broken. Instead, when $\nu\ne 0$, the $SU(2)$ symmetry is explicitly broken, leaving a residual $U(1)\times U(1)$ symmetry. Therefore, one branch of type $\mathrm{\uppercase\expandafter{\romannumeral2}}$ Goldstone boson becomes gapped, while the other branch remains gapless but becomes a purely dissipative, quadratic mode. This is clearly shown in Figure~\ref{example1}, which shows there are two modes with real linear dispersion relations at small $k$ (red lines), one purely imaginary gapless mode with quadratic dispersion relation (blue line) and a purely imaginary gapped mode (black line).
\begin{equation}\label{HSmodes}
\begin{aligned}
        &\omega_\pm=\pm v_sk-i\Gamma k^2+\mathcal{O}(k^3)\,, \\ &\omega_0=-i\Gamma_0 k^2+\mathcal{O}(k^3)\,, \\ &\omega_{gap}=-i\Gamma_{gap}+\mathcal{O}(k)\,.
\end{aligned}
\end{equation}
We find that this pattern is true for all parameters of our holographic binary superfluids, as long as $\nu<0$. This is also consistent with the dispersion relations (\ref{modes}) derived from hydrodynamics of binary  miscible superfluids (which does not incorporate the dynamics of the gapped mode).

When the superfluid velocity $v_y$ becomes larger than a system-dependent critical value $v_c$, one of these modes~\eqref{HSmodes} crosses to the upper half plane (Im$\omega>0$) and therefore signals a dynamical instability. While this occurs both for the counterflow and the coflow cases, we will show that the dominant unstable modes are different. 

\subsubsection{Dispersion relation for the counterflow case}
\begin{figure}[htpb]
    \centering
        \includegraphics[width=0.98\linewidth]{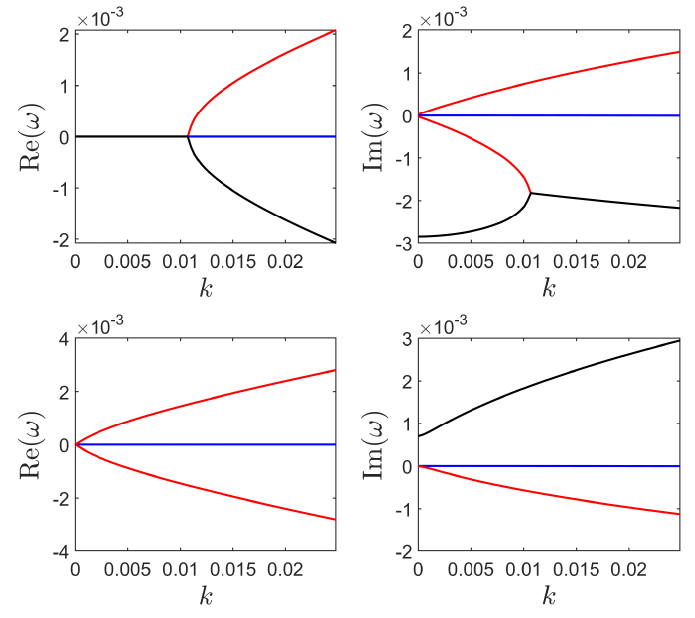}
        \caption{QNM spectrum at low momentum for the counterflow case. \textbf{Top}: QNM spectrum for superfluid velocity $v_{y1}=0.283$, so that $v_{c1}<v_{y1}<v_{c2}$. \textbf{Bottom}: QNM spectrum for superfluid velocity $v_{c2}<v_{y2}=0.314$. The red lines are the sound modes, the blue line is the diffusive mode and the black line is the gapped  mode. The relevant parameter values are $T/T_c=0.677$, $\nu=-0.2$. The unstable mode for $v_{y1}$ is the sound mode, while for $v_{y2}$, the unstable mode is the gapped  mode. For reference, the two critical velocities at $T/T_c=0.677$ and $\nu=-0.2$ are numerically found to be $v_{c1}=0.251$, $v_{c2}=0.308$.}
    \label{v1v2}
\end{figure}

In the counterflow case $\bm{v}_1^s=-\bm{v}_2^s=(0, v_y)$, the instability appears first in one of the sound modes of~\eqref{HSmodes}. 
More precisely, above the critical velocity $v_{c1}$, the speed of sound of one of the modes becomes purely imaginary ($v_s^2<0$), and this mode obtains a positive imaginary part, see the red lines in the upper panel of Figure~\ref{v1v2}.  
There is another critical velocity $v_{c2}$, above which the unstable mode becomes gapped at $k=0$, as shown by the black line in the lower panel of Figure~\ref{v1v2}. 

\begin{figure}[htpb]
    \centering
        \includegraphics[width=0.98\linewidth]{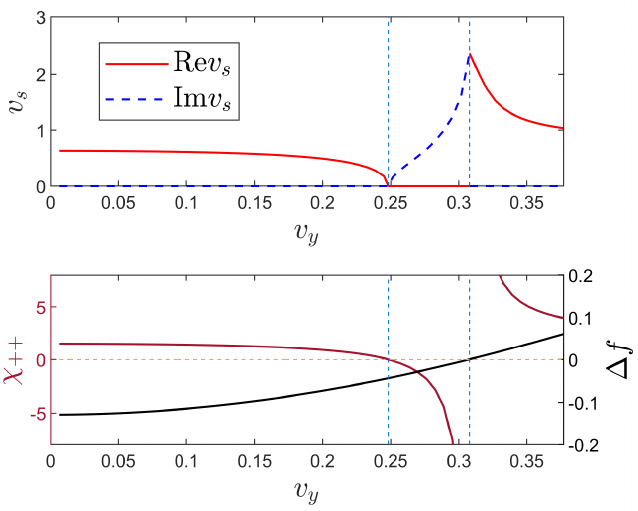}
        \caption{\textbf{Top}:Velocity $v_s$ of the sound mode for different superfluid velocities $v_y$ in the counterflow case. \textbf{Bottom}: Susceptibility $\chi_{++}$ and free energy density difference $\Delta f=f_{\mathrm{binary}}-f_{\mathrm{single}}$. The vertical dashed lines denote the critical velocities $v_{c1}$ and $v_{c2}$. Above $v_{c1}$, $\chi_{++}$ changes sign, the sound velocity becomes purely imaginary, and one sound mode becomes unstable. Above $v_{c2}$, $\Delta f$ becomes positive. When $\Delta f>0$, the binary superfluid phase is globally thermodynamically unstable, and the gapped mode becomes unstable with a positive imaginary part at $k=0$. Relevant parameters are $T/T_c=0.677$, $\nu=-0.2$. }
    \label{vs_F}
\end{figure}

As predicted by the hydrodynamics of binary superfluids in Section~\ref{sec_hydro}, the onset of the dynamical instability is given by~\eqref{criterion0}. Since $\chi_{+-}$=0 for both the counterflow and coflow cases, the criterion reduces to $\chi_{++}=0$ or $\chi_{--}=0$. We find that the former one $\chi_{++}=0$ explains precisely the dynamical instability we observe in the counterflow case, while $\chi_{--}$ does not change sign. According to the holographic dictionary, the free energy $F$ at fixed superfluid velocity is computed from the on-shell bulk action with Euclidean signature times temperature. 
Then we can extract $\chi_{++}$ by calculating the free energy density for different $v^s_+$ while keeping $v^s_-$ fixed. See Appendix~\ref{app:fe} for more details. In Figure~\ref{vs_F}, we plot the sound velocity $v_s$ of the unstable mode and $\chi_{++}$ for different superfluid velocities $v_y$.\footnote{In Figure~\ref{vs_F}, the sound velocity is extracted from the QNMs directly by $v_s=\lim_{k\rightarrow 0}\omega/k$. One can also compute the sound velocity from the susceptibilities according to (\ref{soundspeed}). We have verified that the results from these two methods agree within numerical error, and the discrepancy is given in Appendix~\ref{app:sound}.} The two critical velocities $v_{c1}$ and $v_{c2}$ are denoted by vertical dashed lines.
The first critical velocity occurs exactly when $\chi_{++}=0$. Then from \eqref{soundcounterflow},
we expect the sound velocity to become purely imaginary when $\chi_{++}$ changes sign, which is exactly what happens above the first critical velocity (see the dashed blue line of Figure~\ref{vs_F}). 

We point out that the fastest unstable mode develops at a finite wave number. Interestingly, as $v_y$ is increased, $\chi_{++}$ becomes positive again. In this case, the hydrodynamics of Section~\ref{sec_hydro} predicts no instability. Nevertheless, we find that, beyond the second critical velocity $v_{c2}$, the system does become unstable at $k=0$ despite a positive $\chi_{++}$. This is in fact a global instability associated with  the gapped mode becoming unstable, due to order competition between the two components.\,\footnote{Early studies on competing holographic orders can be found \emph{e.g.} in~\cite{Basu:2010fa,Cai:2013wma}.} Note that there exist two kinds of broken phases: the single superfluid phase and the binary superfluid phase. When $v_y>v_{c2}$, the binary superfluid phase becomes less thermodynamically favored, and order competing between the two components triggers a global instability. We can verify this by comparing the free energy of the two phases. By calculating the free energy density difference between the two superfluid phases $\Delta f=f_{\mathrm{binary}}-f_{\mathrm{single}}$, we find that $\Delta f$ becomes positive exactly when $v_y>v_{c2}$, see the black line in the bottom panel of Figure~\ref{vs_F}. Therefore, there will be a first-order phase transition from the binary superfluid phase to the single superfluid phase. Moreover, as shown in Figure~\ref{vs_F}, $\chi_{++}$ diverges at $v_{c2}$.
This instability is also of thermodynamic origin but beyond the scope of our hydrodynamic description, as it is a global rather than a local instability.

\subsubsection{Dispersion relations for the coflow case}
We now consider the coflow case, $\bm{v}_1^s=\bm{v}_2^s=(0, v_y)$. We first recall that the boost symmetry of the GPEs~\eqref{GP} removes the coflow instability for binary superfluids without relative velocity. Indeed, at zero temperature where the GPE applies, there is no normal component and any nonzero superfluid velocity can be boosted back using a Galilean transformation to a state where the superfluid is at rest. 

Instead, the holographic setup enjoys relativistic symmetry and is invariant under Lorentz boosts. However, at finite temperature, states with a nonzero superfluid velocity cannot be boosted back to a state where all components are at rest, since there is a preferred rest frame for the normal component, defined by setting the normal velocity to be $u^\mu=(1,0,0)$. 
This means that the interface coflow instability for immiscible binary superfluids, which is absent in the GPE, can be observed in holographic superfluids~\cite{An:2024dkn}. 

In the coflow case, we find two critical velocities denoted as $v'_{c1}$ and  $v'_{c2}$. Comparing with the counterflow case, we find that in the coflow case the binary superfluid phase always has the lowest free energy, and thus there is no global instability towards a single superfluid phase. Thus, we expect both critical velocities should follow from the thermodynamic criterion \eqref{criterion0}.

\begin{figure}[htpb]
    \centering
        \includegraphics[width=0.98\linewidth]{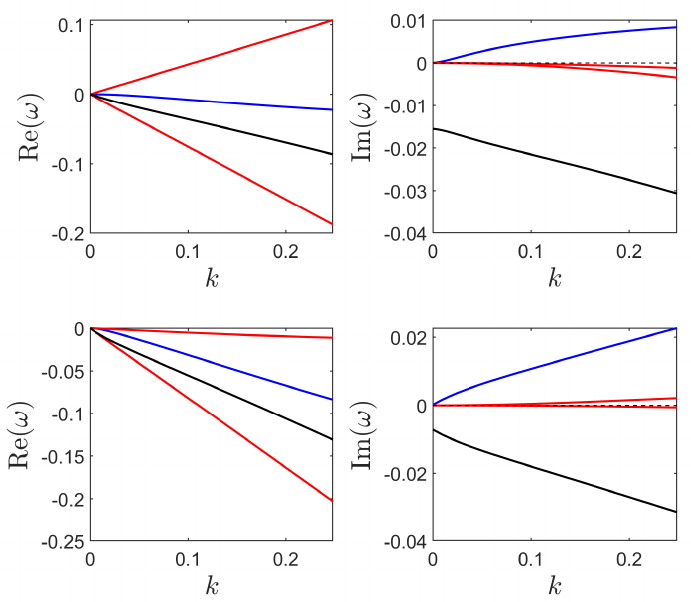}
        \caption{QNM spectrum for the coflow case. \textbf{Top}: QNM spectrum for a superfluid velocity $v'_{y1}=1.257$, such that $v'_{c1}<v'_{y1}<v'_{c2}$. \textbf{Bottom}: QNM spectrum for a superfluid velocity $v'_{y2}=2.199$, such that $v'_{y2}>v'_{c2}$. The red lines are the superfluid sound modes, the blue line is the diffusive mode and the black line is the gapped mode. Relevant parameters are $T/T_c=0.677$, $\nu=-0.2$. At $v'_{y1}$, only the diffusive mode is unstable, while at $v'_{y2}$, one of the sound mode also becomes unstable. For reference, the two critical velocities at $T/T_c=0.677$ and $\nu=-0.2$ are numerically found to be $v'_{c1}=0.691$, $v'_{c2}=2.136$.}
    \label{v1v2_2}
\end{figure}

As depicted in the upper panel of Figure~\ref{v1v2_2}, the diffusive mode is the first to become unstable when the superfluid velocity crosses over $v'_{c1}$. Meanwhile, the sound modes and the gapped mode remain stable. As the superfluid velocity increases further to above $v'_{c2}$, the other instability shows up from one of the sound modes, see the red lines in the lower panel of Figure~\ref{v1v2_2}. We find that these two critical velocities exactly correspond to $\chi_{--}=0$ and $\chi_{++}=0$, respectively. For the coflow case case with $v^s_-=0$, the attenuation is $\Gamma_0=-\chi_{--}\delta_1/4$,  \eqref{diffcoflow}. Therefore, when $\chi_{--}$ changes sign, the diffusive mode becomes unstable. Moreover, when $\chi_{++}$ changes sign, the velocity of one of the sound modes \eqref{soundcoflow} becomes negative \eqref{coflowunstablesound}, and its attenuation becomes positive. A
negative sound speed signals modes with negative energy are excited, and thus implies an energetic instability, which is also accompanied by a dynamical instability due to the positive imaginary part.

\begin{figure}[htpb]
    \centering       \includegraphics[width=0.98\linewidth]{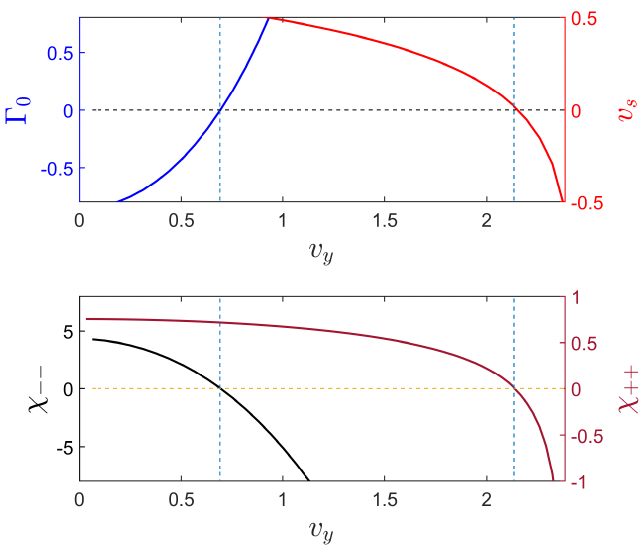}
        \caption{\textbf{Top}: Sound velocity $v_s$ of the unstable sound mode and attenuation $\Gamma_0$ of the diffusive mode for different superfluid velocities $v_y$ in the coflow case. \textbf{Bottom}: susceptibilities $\chi_{++}$ and $\chi_{--}$. The vertical dashed lines denote the critical velocities $v'_{c1}$ and $v'_{c2}$, given by $\chi_{--}=0$ and $\chi_{++}=0$ respectively. Above $v'_{c1}$, the diffusivity changes sign and the diffusive mode becomes unstable. Above $v'_{c2}$, the sound mode also becomes unstable, with its velocity and attenuation both changing sign. Relevant parameters are $T/T_c=0.677$, $\nu=-0.2$.}
    \label{thermo}
\end{figure}

As shown in Figure~\ref{thermo}, the values for the critical velocities found from the QNM analysis exactly match the thermodynamic prediction from hydrodynamics. At $v_{c1}'$, both $\chi_{--}$ and $\Gamma_0$ vanish and change sign. At the larger critical velocity $v_{c2}'$, $\chi_{++}$, $v_s$ and $\Gamma$ change sign. Note that $\chi_{--}$ is always negative when $v_y>v_{c1}'$. As we explain in Appendix~\ref{app:eig}, in the coflow case, there are two sets of solutions  $(u_1,v_1)=\pm(u_2,v_2)$ to the eigenvalue problem governing the QNMs. By checking the eigenfunctions numerically, we find that the sound modes correspond to $(u_1,v_1)=(u_2,v_2)$, while the quadratic mode and the gapped mode correspond to $(u_1,v_1)=-(u_2,v_2)$. Sincec $(u_I,v_I)$ are the fluctuations of the condensates at the boundary, This makes the physical meaning of the two kinds of coflow instability clear. For the first instability, the two condensates are fluctuating with opposite phases, \emph{i.e.} the two components have relative motion. For the second instability, the condensates are fluctuating in phase, \emph{i.e.} the two components move as a whole. In this sense, the second instability can be interpreted as the generalization of the Landau instability to binary superfluids, which is due to both superfluids moving relative to the normal component. The Landau instability in single-species superfluids was studied in \cite{Amado:2013aea,Landauinstability,thermoinstability,Arean:2023nnn}.

\subsection{Nonlinear time evolution of the counterflow and coflow instabilities}
\label{sec:tevol}
\begin{figure*}[htpb]
    \centering
        \includegraphics[width=0.85\linewidth]{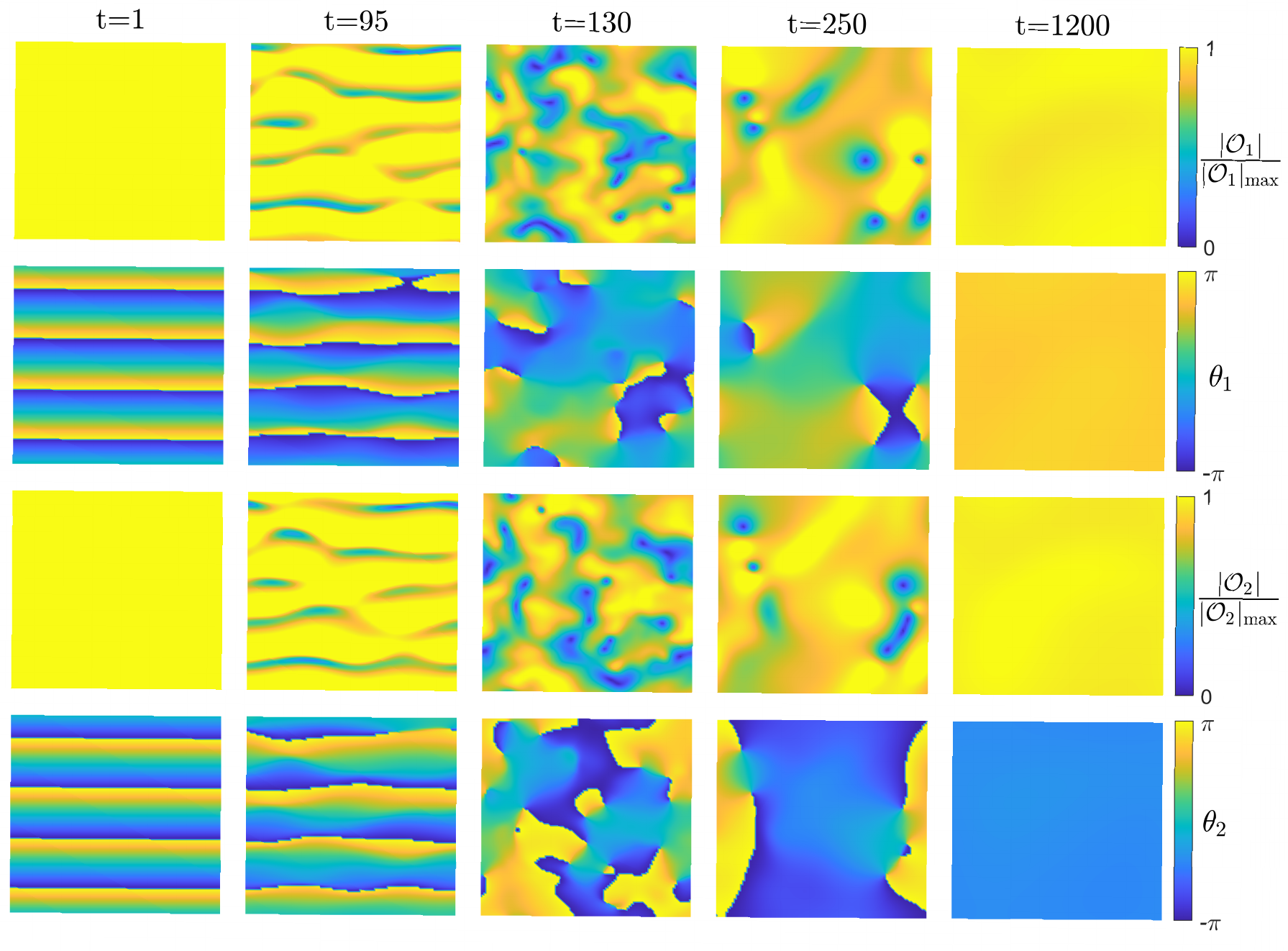}
        \caption{Nonlinear evolution of the counterflow instability at $T/T_c=0.677$, $\nu=-0.2$. The initial superfluid velocity is $v_y=0.628$. The region plotted is [0,40]$\times$[0,40]. From top to bottom, the values plotted are $|\mathcal{O}_1|/|\mathcal{O}_1|_{\mathrm{max}}$, $\theta_1$, $|\mathcal{O}_2|/|\mathcal{O}_2|_{\mathrm{max}}$ and $\theta_2$. As time goes by, the initial small perturbations grows exponentially. Then a dark soliton forms and vortex nucleation occurs. Due to strong dissipation within the system, vortices annihilate with anti-vortices. During this process, the kinetic energy of the superfluid components is also dissipated and the final state is a homogeneous binary superfluid with a velocity below the critical velocity $v_{c1}$. The region plotted is [0,40]$\times$[0,40].}
    \label{counterflow}
\end{figure*}
\begin{figure*}[htpb]
    \centering
        \includegraphics[width=0.85\linewidth]{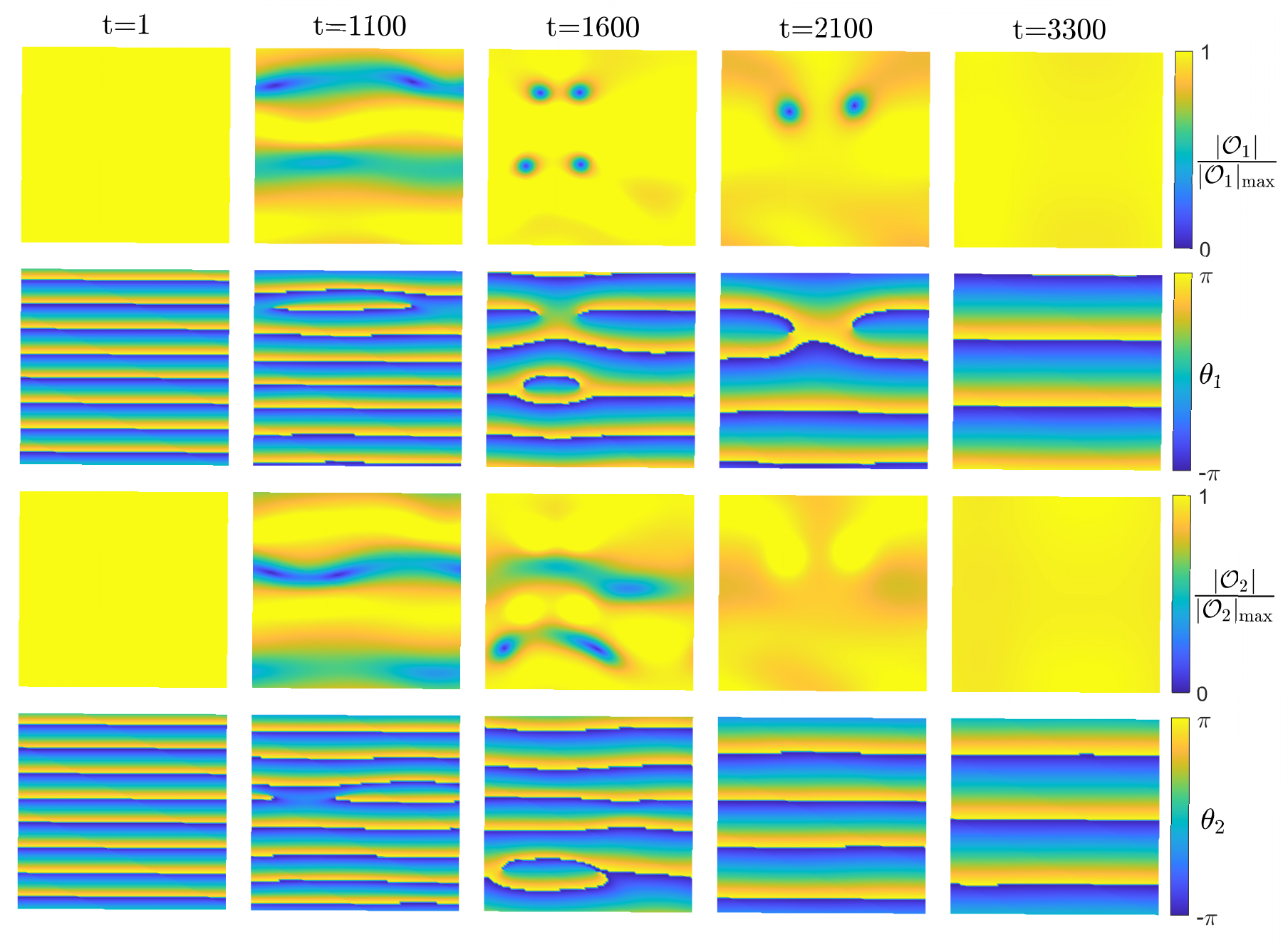}
        \caption{Nonlinear evolution of the coflow instability at $T/T_c=0.677$, $\nu=-0.2$. The initial superfluid velocity is $v_y=1.257$. The same values and regions are plotted as those for the counterflow case. This dynamics is similar to that of counterflow case, except that the coflow instability develops much more slowly and mildly. The final state is a homogeneous binary superfluid with a velocity below the critical velocity $v'_{c1}$.}
    \label{coflow}
\end{figure*}

From our linear analysis, we showed that miscible binary superfluids with superflow velocity generically present dynamical instabilities that follow from thermodynamic instabilities. In this section we explore the nonlinear stage of the counterflow and coflow instabilities by evolving the system forward in time using the full set of nonlinear EoMs.  The explicit form of the EoMs and time evolution scheme can be found in Appendix~\ref{app:eom}. 

In Figure~\ref{counterflow} and Figure~\ref{coflow}, we show typical examples of such nonlinear time evolution for the counterflow and coflow instabilities, respectively. The dynamics for both cases looks similar, except that the coflow instability develops much more slowly and mildly. Since the system is dynamically unstable, small  inhomogeneous perturbations initially grow exponentially over time and eventually drive the system into a highly inhomogeneous state. Then nonlinear dynamics takes over and a dark soliton forms. Due to the instability of the dark soliton, vortex nucleation begins to occur.\footnote{ Vortices only nucleate if the initial perturbation breaks translations in both directions.} As the number of vortices increases, the system transitions into a quantum turbulent stage. In the next stage, due to strong dissipation in holographic superfluids, vortices and anti-vortices annihilate with each other and the total number of vortex/anti-vortex pairs gradually decreases. During this process, the superfluid kinetic energy is also dissipated and the overall superfluid velocity is lowered. After a long time, when all vortices are annihilated, the final state of system is again a homogeneous binary superfluid, but with a superfluid velocity below the critical velocity, respectively below $v_{c1}$ for the counterflow case and $v'_{c1}$ for the coflow case. Such slowing down mechanism might be universal for dissipative superfluids. In single-component superfluids, the superfluid velocity is also lowered to below the Landau critical velocity by vortex nucleation and annihilation~\cite{Landauinstability}.

\section{Universal scaling law of critical velocity}\label{sec4.2}
While we have shown some key differences between the GPEs and the holographic description for binary superfluids, both share a similar scaling law of critical velocity. 

\begin{figure}[htpb]
    \centering
        \includegraphics[width=0.99\linewidth]{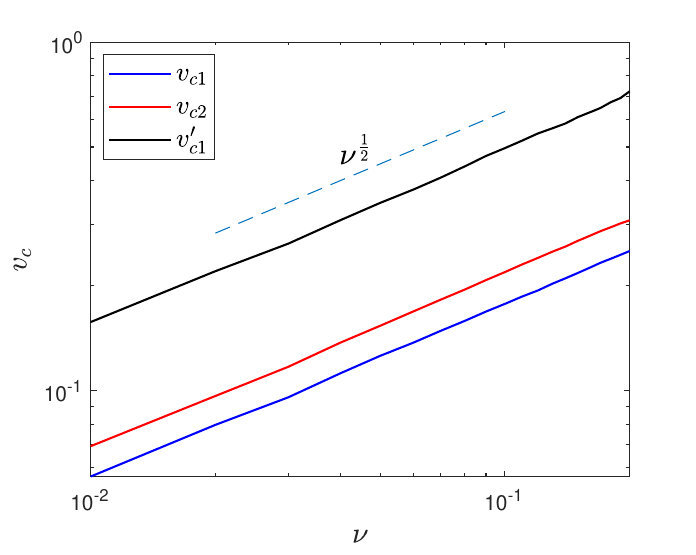}
        \caption{Two critical velocities $v_{c1}$, $v_{c2}$ for counterflow case and the first critical velocity $v'_{c1}$ for the coflow case versus coupling strength $\nu$ at $T/T_c=0.677$. All critical velocities scale as $\nu^{1/2}$. This result also holds at other temperatures.}
    \label{vc}
\end{figure}

When the two components are identical, as is the case in the present work,  the critical velocity of counterflow instability from the coupled GPEs~\eqref{GP} was found to be~\cite{counterflow1,counterflow2,counterflow3,counterflow4}
\begin{equation}
    \label{scale}
    v_c\sim \sqrt{1-\Delta^2},\quad \Delta=g_{12}/\sqrt{g_1g_2}\,.
\end{equation}
In the holographic theory, we have found four critical velocities: $v_{c1}$, $v_{c2}$ for the counterflow case and $v_{c1}'$, $v_{c2}'$ for the coflow case. Interestingly, we find that three of the four critical velocities scale as $\nu^{1/2}$, see Figure~\ref{vc}. This scaling behavior persists for all temperatures. The only velocity which does not follow this scaling behavior is the critical velocity $v'_{c2}$. $v'_{c2}$ is nonzero even when $\nu$ vanishes and changes little as $\nu$ decreases. This is because $v'_{c2}$ corresponds to the Landau critical velocity, due to the superfluid moving as a whole relative to the normal component, and thus persists even when the two components do not interact. 

Such similar scaling behavior between the two models prompts us to identify $\nu$ with $1-\Delta^2$. In the recent work on interface instability of immiscible binary superfluids by some of us~\cite{An:2024ebg}, similar scaling behavior for the interface width between the two models was found for small $\nu$ by identifying $\nu$ with $1-\Delta$. In fact, the two identifications are equivalent for small $\nu$. Taking $\nu=1-\Delta^2$, for small $\nu$ we have
\begin{equation}
    1-\Delta=1-\sqrt{1-\nu}=\frac{1}{2}\nu+\mathcal{O}(\nu^2)\,.
\end{equation}
The coefficient $1/2$ has no effect on the scaling laws. It is nontrivial to find the same identification and scaling law between weakly and strongly interacting binary superfluids independently in the analysis of dynamical instabilities of miscible binary superfluids and in the stationary state of immiscible binary superfluids. This result suggests that the scaling law for critical velocity might be universal, whether the superfluid is weakly or strongly interacting.

\section{Conclusion and discussion}
\label{sec:conclusion}

We have studied the counterflow and coflow instabilities in miscible
binary superfluids. The former is generic in two-component fluid systems, while the latter is only present in non-Galilean invariant systems, such as dissipative systems. We first developed a hydrodynamic framework for binary superfluids in Section~\ref{sec_hydro}. Based on the local thermodynamic stability and positivity of entropy production, we have shown that the hydrodynamic theory of binary superfluids becomes linearly dynamically unstable whenever the matrix of static susceptibility $\chi_{AB}$ is no longer positive definite.

We have demonstrated that the dynamical counterflow instability found from GPEs is reproduced by our criterion, which provides a macroscopic origin for it. In the framework of GPE, there is no coflow instability in an isolated uniform system because of Galilean invariance. In contrast to GPE, the holographic superfluids describe strongly-coupled, dissipative superfluids at finite temperature. We found holographic miscible binary superfluids feature rich dynamics. 

Linear analysis reveals that for both the counterflow and coflow cases, a dynamical instability only occurs above some critical superfluid velocity, see \emph{e.g.} Figures~\ref{v1v2} and~\ref{v1v2_2}. For the counterflow case, it first arises as a sound mode crosses to the upper half complex frequency plane, while for the coflow case it is a diffusive mode. As the superfluid velocity is further increased, another critical velocity shows for each case. For the counterflow case, this is due to phase competition. Above the second critical velocity, free energy of the counterflow binary superfluid phase is higher than that of single component phase, therefore the binary superfluid phase becomes less thermodynamically favored. In accordance, the unstable mode above the second critical velocity becomes the gapped mode, which is unstable even at zero wave number, signaling a global instability. For the coflow case, the second critical velocity corresponds to the Landau critical velocity in single component superfluids. Above this critical velocity, one of the sound modes also becomes unstable. Nevertheless, the unstable diffusive mode always dominates the instability. 

By formulating hydrodynamics of binary superfluids, we have found that except $v_{c2}$, which is due to order competing, the other three instabilities originate in the divergences of susceptibility matrix $\chi_{AB}$. 

Then we have explored the nonlinear stage of these dynamical instabilities by evolving the systems forward in time. We found the phenomena for two kinds of instabilities are similar, see Figures~\ref{counterflow} and~\ref{coflow}. In linear stage, small perturbations grow exponentially and drive the system into inhomogeneous state and then nonlinearity takes over. Firstly, a dark soliton forms. Due to the instability of dark soliton, vortices are nucleated. Then the overall velocity of the system is reduced due to vortex annihilation and dissipation. The final state of the evolution is a homogeneous binary superfluid with a velocity below the critical velocity. Such mechanism seems to be general in dissipative superfluids. 

We have limited ourselves to superfluids with identical components under the simplifying assumption of colinearity. Nevertheless, there should be no conceptual obstacle relaxing this assumption. Moreover, it is desirable to understand the range of application of the hydrodynamic framework. In particular, one could expand the hydrodynamic expansion to higher orders to compare with the non-linear stage of the system. In future work, we shall also explore the properties of vortices in strongly interacting miscible binary superfluids (see~\cite{An:2025gid}) and compare them with those of weakly interacting binary superfluids~\cite{vortex1,vortex2,vortex3}.
While we have uncovered the macroscopic origin of counterflow and coflow instabilities in miscible binary superfluids, it would be interesting to generalize the discussion to immiscible binary superfluids, for which one has to deal with a phase-separated configuration. The tension from the interface could play an important role for the instability of the system~\cite{Armas:2016xxg}.

Finally, it would be interesting to investigate whether phonon-phonon Beliaev scattering at zero temperature produces a linearly dynamical instability with the GPE when the critical velocity is crossed. This would be the zero temperature counterpart of the finite temperature dynamical instabilities we found using dissipative hydrodynamics and gauge/gravity duality.

\acknowledgments
B.~G.~ thanks Laurent Sanchez-Palencia for helpful comments on the draft. 
This work was supported by the National Natural Science Foundation of China Grants No. 12525503, No. 12075298,
No. 12122513, and No. 12447101.
We thank Jilin University for hospitality during the School of Frontiers in Hydrodynamics, Effective Field Theory and Holographic Duality. We acknowledge the use of the High Performance Cluster at Institute of Theoretical Physics, Chinese Academy of Sciences.

\appendix

\section{Review of fluid hydrodynamics \label{subsec:hydro}}

Hydrodynamics \cite{landaubookfluids,chaikinlubensky1995,Kovtun:2012rj,forster2018hydrodynamic} posits that the collective dynamics of an interacting system on scales long compared to the local equilibrium scales (\emph{e.g.} the mean collision time or the mean free path for a system composed of weakly-interacting particles) is captured by the conservation equations of the densities of conserved operators (charge $n$, energy $\epsilon$, momentum $g^i$...) following from the global symmetries of the system ($U(1)$, time and space translations...). These equations are generalized continuity equations, eg
\begin{equation}
    \partial_t n+\partial j^i=0
\end{equation}
for charge conservation, where $n(t,x^i)$ is the charge density and $j^i(t,x^i)$ the spatial charge current.
The expectation values of the spatial fluxes (charge $j^i$ and energy $j_\epsilon^i$ currents, spatial stress tensor $\tau^{ij}$...) in the thermal ensemble can be expanded order by order in gradients of the densities, with each term featuring a so-called transport coefficient (the shear and bulk viscosities, thermal conductivity, etc.). For instance, the charge current is
\begin{equation}
    j^i=n v^i-\sigma_0\partial^i\mu-\alpha_0 \partial^i T+O(\partial^2)\,.
\end{equation}
$v^i$ is the fluid velocity and $\sigma_0$, $\alpha_0$ are electrothermal conductivities.\footnote{In a Galilean system, we would have $\sigma_0=\alpha_0=0$ so that the charge current is full convective, but in a non-Galilean system, such as Graphene \cite{Crossno:2016fvs, Lucas:2015sya,Lucas:2017idv} near the charge neutrality point, other processes can generate a charge current.} 

Together with the first law of thermodynamics $d\epsilon=Tds+\mu dn+v_i dg^i$, this allows to solve the continuity equations and to compute the hydrodynamic spectrum.

In thermodynamics, global equilibrium is characterized by a few macroscopic variables, such as the charge and the energy. The first law of thermodynamics $dE=T dS+\mu dN$ relates variations of the energy $E$, and charge number $N$ to variations of the entropy $S$ through state variables such as the temperature and the chemical potential $(T,\mu)$. The second law states that the entropy is maximized in global equilibrium; it never decreases in any spontaneous process $\Delta S\geq0$. This means that the entropy is a Lyapunov function and the Lyapunov stability theorems \cite{LiapunovBook} guarantee that global equilibrium is stable: small perturbations away from it decay over time. 

In its modern incarnation, hydrodynamics is viewed as the local, out-of-equilibrium extension of global thermodynamic equilibrium~\cite{chaikinlubensky1995,Kovtun:2012rj}. That is, it is postulated that at finite temperature and for an interacting system, there exists a local equilibration time and length over which the system can be coarse-grained and microscopic dynamics (of \emph{e.g.} particles) are recast into the conservation equations of the densities of conserved operators (charge $n(t,x^i)$, energy $\epsilon(t,x^i)$, momentum $g^i(t,x^i)$ densities...) following from the global symmetries of the system (U(1), time and space translations...). This hypothesis can be verified explicitly by a Boltzmann equation treatment of the dynamics of a weakly-interacting gas of particles, or in strongly-interacting systems with a gravitational dual by gauge-gravity duality (see Appendix~\ref{subsec:holo}). The hydrodynamic regime is enhanced by the strength of interactions, since elementary constituents will equilibrate faster, and strongly-interacting systems exhibit a larger window of scales where hydrodynamics applies. 

We thus assume that each `mesoscopic' volume element of the system is in local thermodynamic equilibrium. An immediate consequence of this is that local versions of the first and second laws of thermodynamics can be formulated~\cite{Nicolis1979,chaikinlubensky1995}
 \begin{equation}
 \label{firstlaw}
     d\epsilon = Tds + \mu dn+v_i dg^i\,,
 \end{equation}
 \begin{equation}
 \label{localsecondlaw}
     T\partial_t s+T\partial_i j_s^i\equiv \Delta\geq0\,,
 \end{equation}
where $T(t,x^i)$, $\mu(t,x^i)$, $v^i(t,x^i)$ are the local temperature, chemical potential and fluid velocity, and $s(t,x^i)$, $j_s^i(t,x^i)$ the local entropy density and entropy current, respectively. We use the Einstein summation convention for which repeated indices should be summed, and Latin indices $i$ run over spatial coordinates.

The conservation equations take the form of generalized continuity equations
\begin{equation}
\label{hydroeomsintro}
    \partial_t \langle O^A\rangle+\partial_i \langle j^{iA}\rangle=0\,,
\end{equation}
for the set of local expectation values of the densities of conserved operators  $O^A=\{\epsilon,n,g^i\}$. The angular brackets denote an expectation value $\langle O^A(t,x^i)\rangle=Z^{-1}\textrm{Tr}\left[O^A(t,x^i) e^{\beta H}\right]$ in a thermal ensemble with inverse temperature $\beta$ and Hamiltonian $H$. From now on, we drop the angular brackets for brevity. To be closed, the equations require constitutive relations for the spatial fluxes $j^{iA}(t,x^i)=\{j^i_\epsilon,j^i_n,\tau^{ij}\}$ (energy current, charge current and spatial stress tensor). Writing these is in general a very complicated task but is made tremendously easier by the hypothesis of local thermodynamic equilibrium: as $j^{iA}$ are not conserved, their expectation values in local thermal equilibrium can be written in an expansion in gradients of $O^A$: 
\begin{equation}
j^{iA}=\sum_{n\geq0}D^{AB}_{(n)i} O_B    
\end{equation}
where $D^{AB}_{(n)i}$ is a differential operator of order $n$ weighted by the appropriate power of the local equilibration length. Simply put, we write $j^{iA}$ as a series order by order in gradients. Each subsequent derivative term suppressed compared to the lower orders so that spatial variations about local equilibrium are small (but not restricted to linear perturbations, as manifest in the Navier-Stokes equations). Each independent gradient term in the differential operator $D^{AB}_{(n)i}$ comes with an independent transport coefficient, and our task now becomes to classify and constrain them. 

In simple systems, there are only a handful of such terms at first order in gradients. In Galilean fluids, they are the shear and bulk viscosities and the thermal diffusivity ~\cite{landaubookfluids,chaikinlubensky1995}. They appear in the constitutive relations as:
\begin{equation}
\label{constitutivefluid}
    \begin{aligned}
        g^i=&
        m n v^{i}\,,  \quad        j_s^i=sv^i-\frac{\kappa}T\partial^i T\,,\\ 
        \tau^{ji}=& p\delta^{ij}+v^{i}g^j
        -\eta \left(\partial^i v^j+\partial^j v^i\right)\\
        &-\left(\frac2d\eta+\zeta\right)\delta^{ij} \partial\cdot v\,.
    \end{aligned}
\end{equation}
Here $p=Ts+\mu n+v_i g^i-\epsilon$ is the pressure and $d$ the spatial dimension of the system.

In more complicated systems with a lower amount of symmetry, their number quickly proliferates. For instance, there are fourteen first-order coefficients in a $(3+1)$-dimensional relativistic (or Galilean) superfluid where the conservation of the particle number is spontaneously broken~\cite{Bhattacharya:2011eea}. Classifying them implies enumerating all possible tensor structures, eliminating redundancies using the equations of motion (EoMs) at lower orders, and choosing a frame. The choice of frame amounts to a choice of an out-of-equilibrium definition for the temperature, chemical potential etc.~\cite{Kovtun:2012rj}, which are only unambiguously defined in equilibrium. 

Once all of these are done, an essential step remains: the local version of the second law of thermodynamics~\eqref{localsecondlaw} must be obeyed on all fluid configurations, which sets some combination of transport coefficients to zero and sets inequality constraints on others. Finally, the Onsager reciprocity relations can be imposed depending on the set of discrete symmetries of the system (such as invariance under parity or time reversal).

In Galilean fluids \eqref{constitutivefluid}, the positivity of entropy production amounts to imposing that the quadratic form
\begin{equation}
\begin{split}
   \Delta=&\zeta(\partial\cdot v)^2+\frac{\kappa}T(\partial_iT)^2\\
    &+\eta\left(\partial^i v^j+\partial^j v^i-\frac2d\delta^{ij}\partial\cdot v\right)^2
    \end{split}
\end{equation}
is positive for all fluid configurations, resulting in the simple set of constraints $\eta,\zeta,\kappa\geq0$. 

Once the hydrodynamic theory is formulated, we Fourier transform in space and time, such that after linearizing around local equilibrium $O^A(t, x^i)=O^A_0+\delta O^A e^{-i\omega t+i k_i x^i}$, the equations \eqref{hydroeomsintro} take the form:\footnote{The matrix $M$ only depends on $k^i$ and not $\omega$ provided the hydrodynamic frame, ie the out-of-equilibrium definition of temperature, chemical potential and fluid velocity \cite{Kovtun:2012rj}, is chosen appropriately. }
\begin{equation}
\label{lineomintro}
\begin{aligned}
        &(-i\omega+M(k^i)\cdot \chi^{-1})\delta O=0\,,
\end{aligned}
\end{equation}
where $\chi_{AB}\equiv\delta O_A/\delta s_B=\delta^{(2)}p/(\delta s_A \delta s_B)$ is the matrix of static susceptibilities. Here $\delta s_B=(\delta T,T\delta(\mu/T),\delta v^i)$ is the set of sources conjugate to $\delta O^A=(\delta\epsilon,\delta n,\delta g^i)$. The static susceptibilities are just thermodynamic derivatives, for instance $\chi_{\epsilon\epsilon}=\left.\partial \epsilon/\partial T\right|_{(\mu/T)}$.    

The spectrum of collective excitations $\omega(k)$ in the complex frequency plane ($k=\sqrt{k_i k^i}$) is determined by finding the $\omega(k) $  such  that $\det(i\omega(k)-M\cdot\chi^{-1})=0$. Hydrodynamics only reliably captures poles such that $\omega(k\to0)\to0$, \emph{i.e.} these poles can be expanded as $\omega(k)=\sum_{n>0}\omega_n k^n$. Since we expanded in plane waves $e^{-i\omega t+i k_i x^i}$, any any hydrodynamic pole in the upper half complex frequency plane $\textrm{Im}\, \omega>0$ leads to exponential growth of perturbations and signals the dynamical instability. 

Returning to Galilean fluids~\eqref{constitutivefluid}, the hydrodynamic modes  (thermal diffusion, transverse momentum diffusion and longitudinal sound) take relatively simple expressions~\cite{chaikinlubensky1995} and it is straightforward to show by direct inspection of their closed form expressions that the constraints from positivity of entropy production, together with local thermodynamic stability (the matrix $\chi_{AB}$ is positive-definite) implies linear dynamical stability: all the hydrodynamic poles are in the lower complex frequency plane, $\textrm{Im} \,\omega(k)\leq0$. On the other hand, if one of the eigenvalues of $\chi_{AB}$ changes sign, then a linear dynamical instability is found.
 
In a sense, this result is expected from the property that the entropy is a Lyapunov function and from the Lyapunov stability theorem. However, showing in practice that the modes are all linearly stable and that an instability occurs whenever $\chi_{AB}$ is no longer positive definite can be technically complicated by several factors: if many modes couple; the number of transport coefficients is not small;  the system has a reduced amount of symmetry; or the frame choice is such that the linearized equations are not of the form \eqref{lineomintro} and spurious non-hydrodynamic unstable modes are present in the spectrum.

\section{Review of gauge/gravity duality \label{subsec:holo}}

Gauge/gravity duality, also known as holography or AdS/CFT (Anti-de Sitter/Conformal Field Theory) correspondence, has been widely used to study strongly-correlated condensed matter systems without quasiparticles (see~\cite{RN499} for a pedagogical textbook, \cite{holoreview} for more advanced applications to quantum matter and~\cite{Ammon:2015wua} for a more string-theoretic exposition). Gauge/gravity duality originates from string theory~\cite{Maldacena:1997re,Witten:1998qj,Gubser:1998bc}, which is a theory of quantum gravity, and states that certain conformal  gauge field theories are dual to a classical gravity theory in the `t Hooft limit when the rank of the gauge group is taken to infinity and the coupling constant is large. This is both a strong/weak coupling and a quantum/classical duality, which makes it particularly appealing. The gravity theory is defined on asymptotically AdS spacetimes (the `bulk'). Anti de Sitter is a maximally symmetric solution to Einstein's equations with a negative cosmological constant. It has a negative global Riemann curvature and so is hyperbolic, which endows it with a conformal timelike boundary \cite{Witten:1998qj}. This is at the heart of this holographic duality as, loosely speaking, the energy scale of the dual field theory is geometrized as the radial coordinate interpolating between the center of the AdS spacetime and its boundary. The dual field theory has a lower dimensionality (that of the conformal boundary of the bulk spacetime) than the gravity theory. In this work, we consider only $(3+1)$-dimensional gravity theories, dual to $(2+1)$-dimensional field theories.

In practice, the conformal symmetry is broken, either by placing the theory at finite density or turning on various other deformations of the dual field theory.
Finite temperature, finite density states in the dual field theory are modeled by black holes in the bulk charged under a global U(1) symmetry. They are mathematical solutions to the classical equations of the Einstein-Maxwell theory with a negative cosmological constant. For zero Maxwell field, the solution is the Schwarzschild black hole, while for nonzero Maxwell field it is the Reissner-Nordstr\"om black hole. The temperature in the dual field theory is given by the Hawking temperature of the black hole event horizon, while the charge is given through a generalized Gauss's law by the electric flux due to the bulk Maxwell field coming from the black hole horizon.

Since the gauge theory is in the strong coupling regime, we use its correlation functions to characterize it. This is done by identifying the field theory generating functional $Z$ with the on-shell, renormalized, Euclidean gravity path integral according to the well-established holographic dictionary~\cite{holoreview,RN499}. Computing correlation functions on the field theory side amounts to solving the linearized Einstein and Maxwell equations and working out their responses to electromagnetic and gravitational perturbations. The bulk fields are `projected' on the conformal boundary, and the sources and responses are read off from their asymptotic expansions. As an example, using $z$ to denote the radial coordinate interpolating between the boundary at $z=0$ and the horizon at $z=z_h$, the bulk Maxwell field is found to have the following near-boundary expansion:
\begin{equation}
    A_\mu(z,t,x,y) = a_\mu(t,x,y)+b_\mu(t,x,y)z+O(z^2)\,.
\end{equation}
The indices $\mu=(z,t,x,y)$ run over the holographic radial coordinate $z$, the time coordinate $t$ and spatial coordinates $(x,y)$. The dual field theory has coordinates $(t,x,y)$. $a_\mu$ and $b_\mu$ are integration functions which are fixed by boundary conditions at $z=0$ and $z_h$. There are two since Maxwell's equations are second-order PDEs. Subleading coefficients are entirely fixed by $a_\mu$ and $b_\mu$. $a_t$ is proportional to the chemical potential of the state, and $b_t$ to its charge density. More generally, the integration function leading in $z$ is the source (here, $a_\mu$), and the subleading one the response (here, $b_\mu$). Thus the current-current retarded Green's functions is $G^R_{J^\mu J^\nu}\propto\delta b_\mu/\delta a^\nu$.

Since holographic systems are dual to strongly-coupled gauge theories, it is natural to expect that their late time, long distance collective dynamics are captured by hydrodynamics. This was indeed verified for systems dual to relativistic fluids first at linear level~\cite{Policastro:2001yc,Policastro:2002se,Policastro:2002tn,Herzog:2003ke}, and then at nonlinear level~\cite{Bhattacharyya:2007vjd,Banerjee:2008th,Erdmenger:2008rm}. At linear level, the spectrum of collective hydrodynamic modes maps onto the quasinormal modes (QNMs) of the black hole perturbations~\cite{Kovtun:2005ev}. These electromagnetic and gravitational waves (in asymptotically AdS spacetimes rather than Minkowsky as for astrophysical gravitational waves) are damped and dissipation is encoded by imposing that in the classical gravity approximation, they can only be absorbed by the black hole horizon~\cite{Son:2002sd}.

How does one model a superfluid using Gauge-Gravity duality? This was answered in a series of papers~\cite{Gubser:2008px, Hartnoll:2008vx, Hartnoll:2008kx}, where a charged, complex scalar field is added to the Einstein-Maxwell theory, which now has an extra classical Klein-Gordon equation for the complex scalar. This complex scalar field in the bulk is dual to a complex scalar operator in the boundary field theory, with a scaling dimension set by the mass of the bulk scalar. Below a certain critical temperature $T_c$, two solutions compete. One is the Reissner-Nordstr\"om black hole, with the scalar field vanishing everywhere in the bulk; the other has a non-vanishing bulk scalar field. Since the Klein-Gordon equation is second-order, the solution also comes with two free integration constants, one interpreted as the source of the scalar operator and the other as the scalar expectation value. Any nontrivial scalar profile breaks the global $U(1)$ symmetry explicitly for nonzero source. The insight in~\cite{Gubser:2008px, Hartnoll:2008vx, Hartnoll:2008kx} was that a solution can be found where the source is zero and the $U(1)$ symmetry is then broken spontaneously, giving rise to a superfluid phase in the dual field theory.

The match between the superfluid hydrodynamics and the collective thermal dynamics of these holographic superfluids was thoroughly established over a series of works~\cite{Amado:2009ts,Herzog:2009md,Sonner:2010yx,Herzog:2011ec,Bhattacharya:2011eea,Gouteraux:2019kuy,Gouteraux:2020asq,Arean:2021tks,thermoinstability,Arean:2023nnn}.

\section{Holographic setup}
\label{app:hsetup}

The holographic $2+1$-dimensional binary superfluid model reads~\cite{An:2024ebg}
    \begin{equation}\label{lagrangian}
        \begin{aligned}
             \mathcal{L}=&-(\mathcal{D}_\mu\Psi_1)^*
            \mathcal{D}^\mu\Psi_1-m_1^2|\Psi_1|^2-(\mathcal{D}_\mu\Psi_2)^*
            \mathcal{D}^\mu\Psi_2\\&-m_2^2|\Psi_2|^2-\frac{\nu}{2}|\Psi_1|^2|\Psi_2|^2-\frac{1}{4}F^{\mu\nu}F_{\mu\nu},
          \end{aligned}  
    \end{equation}
where $\mathcal{D}_\mu\Psi_I=(\nabla_\mu-ie_IA_\mu)\Psi_I$, $A_\mu$ is the $U(1)$ gauge field, $F_{\mu\nu}$ is the field strength
 and $\Psi_I$s are two complex scalar fields coupled with each other. As is mentioned in previous work by some of us \cite{An:2024ebg}, this model corresponds to immiscible binary superfluid for $\nu>0$ and miscible binary superfluid for $\nu<0$. 
In terms of the background, we utilize a $3+1$-dimensional planar AdS black hole described in Eddington-Finkelstein coordinates:
\begin{equation}
\label{backg}
ds^2=\frac{L^2}{z^2}(-(1-(z/z_h)^3)dt^2-2dtdz+dx^2+dy^2).
\end{equation}
The dual field theory coordinates are $(t,x,y)$ while $z$ is the holographic radial coordinate, with the boundary located at $z=0$ and the black hole horizon at $z=z_h$.
This represents a heat bath at the boundary with temperature $T=3/(4\pi z_h)$. For simplicity, we set $L=z_h=1$, $m_1^2=m_2^2=-2$, $e_1=e_2=1$, and adopt the radial gauge $A_z=0$. 

Consequently, the asymptotic expansions for $A_\mu$ and $\Psi_I$ near the AdS boundary are:
\begin{equation}
\begin{aligned}
&A_\mu=a_\mu+b_\mu z+\mathcal{O}(z^2),\ \\
&\Psi_I=(\Psi_I)_0 z+(\Psi_I)_1 z^2+\mathcal{O}(z^3).
\end{aligned}
\end{equation}
In the context of gauge-gravity duality, the coefficients $a_t$, $a_i$ $(i = x,y)$, and $(\Psi_I)_0$ can be recognized as the chemical potential $\mu$, vector potential, and scalar operator source at the boundary, respectively. $(\Psi_I)_1$ corresponds to the expectation value of the order parameter $\langle \mathcal{\hat{O}}_I \rangle$. In our study, we work within the grand-canonical ensemble, \emph{i.e.} by fixing $\mu$, without considering alternative quantization. It is important to note that $T$ and $\mu$ are not independent quantities due to the scaling symmetry of the system. After setting $z_h=1$, $\mu$ becomes the sole free parameter. A second-order phase transition occurs in this model when $\mu\ge \mu_c\simeq 4.064$, which also determines the ratio $T/T_c=\mu_c/\mu$.

To investigate holographic superfluids with spontaneously broken symmetry, we turn off electromagnetic fields and scalar sources at the boundary, meaning $a_x=a_y=(\Psi_I)_0=0$. Consequently, the charge conservation equation at the boundary is given by $-\partial_t b_t=b_i$, where $-b_t=n$ represents the charge density and $b_i=j_i$ is the charge current. The superfluid velocity is determined by $\bm{v}^s=\nabla\theta|_{z=0}$, where $\theta$ is the phase (the Goldstone field) of the order parameter $\langle \mathcal{\hat{O}} \rangle$.

\section{Equations of motion and time evolution scheme}\label{app:eom}
  The explicit form of EoMs for $\Psi_I$ and $A_\mu$ are 
\begin{widetext}
    \begin{equation}
        \begin{aligned}
            \label{phi}
            2\partial_t\partial_z\Phi_I-[2i A_t\partial_z\Phi_I+i \partial_zA_t\Phi_I+\partial_z(f\partial_z\Phi_I)-z\Phi_I
            +\partial_x^2\Phi_I+\partial_y^2\Phi_I
            -i (\partial_xA_x+\partial_yA_y)\Phi_I&\\
            -(A_x^2+A_y^2)\Phi_I-2i (A_x\partial_x\Phi_I+A_y\partial_y\Phi_I)
            -\frac{\nu}{2}|\Phi_j|^2\Phi_I]=0, \qquad(i,j=1,2,\quad i\ne j)&
        \end{aligned}
    \end{equation}
    \begin{equation}
        \label{At}
        \begin{aligned}
            \partial_t\partial_zA_t-[\partial_x^2A_t+\partial_y^2A_t+f\partial_z(\partial_xA_x+\partial_yA_y)-\partial_t(\partial_xA_x+\partial_yA_y)
            -2A_t\sum_I|\Phi_I|^2&\\
            -2f\mathrm{Im}(\sum_I\Phi_I^*\partial_z\Phi_I)+2\mathrm{Im}(\sum_I\Phi_I^*\partial_t\Phi_I)]=0,&
        \end{aligned}
    \end{equation}
    \begin{equation}
        \label{Ax}
        \begin{aligned}
            2\partial_t\partial_zA_x-[\partial_z(\partial_xA_t+f\partial_zA_x)+\partial_y(\partial_yA_x-\partial_xA_y)-2A_x\sum_I|\Phi_I|^2
            +2\mathrm{Im}(\sum_I\Phi_I^*\partial_x\Phi_I)]=0,
        \end{aligned}
    \end{equation}
    \begin{equation}
        \label{Ay}
        \begin{aligned}           
        2\partial_t\partial_zA_y-[\partial_z(\partial_yA_t+f\partial_zA_y)+\partial_x(\partial_xA_y-\partial_yA_x)-2A_y\sum_I|\Phi_I|^2
            +2\mathrm{Im}(\sum_I\Phi_I^*\partial_y\Phi_I)]=0,
        \end{aligned}
    \end{equation}
        \begin{equation}
        \label{constraint}
        \begin{aligned}     
        \partial_z(\partial_xA_x+\partial_yA_y-\partial_zA_t)-2\mathrm{Im}(\sum_I\Phi_I^*\partial_z\Phi_I)=0,
        \end{aligned}
    \end{equation}
\end{widetext}
    where $\Phi_I=\Psi_I/z$ and $\mathrm{Im}$ represents imaginary part. Notice the last equation is a constraint with no time derivative. These equations are not independent. They obey the following constraint equation
    \begin{equation}
    \label{relation}
    \begin{aligned}
       & -\partial_t\mathrm{Eq.}(\ref{constraint})-\partial_z\mathrm{Eq.}(\ref{At})+\partial_x\mathrm{Eq.}(\ref{Ax})\\&+\partial_y\mathrm{Eq.}(\ref{Ay})=2\mathrm{Im}(\sum_I\mathrm{Eq.}(\ref{phi})\times\Phi_{0I}^*).
    \end{aligned}
    \end{equation}

The initial condition is set to be stationary mixture perturbed by some random noise:
    \begin{equation}
    \Phi_I=\Phi_{0I}e^{i(v_I)_yy}(1+\alpha n(x,y))\,,
    \end{equation}
where $\alpha$ is a small number set to be $10^{-3}$, and $n(x,y)$ is the random noise superimposed on the stationary background.

For time evolution, we use fourth order Runge-Kutta method and the following scheme: First, we use \eqref{phi}, \eqref{Ax} and \eqref{Ay} to evolve $\Phi$, $A_x$ and $A_y$ with boundary conditions $\Phi(z=0)=A_x(z=0)=A_y(z=0)=0$. Then we use \eqref{At} to evolve $\partial_zA_t$ on the boundary. Note that $-\partial_zA_t(z=0)$ is just the charge density or number density $\rho$ of the dual field theory. Finally, we use \eqref{constraint} to solve $A_t$ by evolved $\Phi$, $A_x$, $A_y$ and boundary conditions $\partial_zA_t(z=0)=-\rho$ and $A_t(z=0)=\mu$. Such scheme keeps $\mu$, the chemical potential, unchanged, so we are in fact working in grand canonical ensemble. In $z$ direction, we use Chebyshev pseudo spectral method. In $x$ and $y$ direction we use Fourier pseudo spectral method and periodic boundary condition.

\section{Stationary miscible binary superfluids}
\label{app:stationary}
Stationary configuration for miscible binary superfluids is homogeneous in spatial directions. We choose the following ansatz around the bulk geometry~\eqref{backg}:
\begin{equation}
\label{ansatzps}
\begin{split}
&\Psi_I=z\phi_I(z)e^{i\Theta_I(z,x,y)}, \\&  A_t=A_t(z), \\ &  V_{Ix}(z)\equiv\partial_x\Theta_I-A_x\,,\\
&V_{Iy}(z)\equiv\partial_y\Theta_I-A_y,\\&
(I=1,2)\,,\\
\end{split}
\end{equation}
together with the gauge $\partial_z\Theta_I=-{A_t}/{f}$. The consistence of above ansatz requires $\partial_z(A_y+V_{Iy})=0$ and $\partial_z(A_x+V_{Ix})=0$. Note that the phase $\Theta_I$ of the condensation $\mathcal{O}_i$ corresponds to $\Theta_I|_{z=0}$. According to the holographic dictionary, we have $\bm{V_I}|_{z=0}=\bm{v^s_I}$, where $\bm{v^s_I}$ is the superfluld velocity of $I$-th component. If $\bm{v_I}$ does not depend on spatial and radial cordinates, as is the case in this work, we would have $\bm{V_I}=\bm{v_I-A}$ and $\Theta_I(z,x,y)=\Theta(z)+(v^s_I)_xx+(v^s_I)_yy$. 

The resulting bulk EoMs are given as
\begin{equation}\label{app:Seom}
\begin{split}
    \partial_z(f\partial_z\phi_I)-(V_{Ix}^2+V_{Iy}^2)\phi_I+\frac{A_t^2}{f}\phi_I\\-z\phi_I-\frac{\nu}{2}\phi_J^2\phi_I=0, \, (I\neq J)\,,\\
        f\partial_z^2A_t-2A_t\sum_I\phi_I^2=0\,,\\
        \partial_z(f\partial_zV_{Ix})-2\sum_I V_{ix}\,\phi_I^2=0\,, \\
\partial_z(f\partial_zV_{Iy})-2\sum_I V_{Iy}\,\phi_I^2=0\,,     
\end{split}
\end{equation}
which involves 7 coupled ordinary differential equation for $(\phi_I, V_{Ix}, V_{Iy}, A_t)$ that depend on the variables $z$ only.  The boundary conditions at the AdS boundary $z=0$ read
\begin{equation}
\begin{aligned}
        &\phi_I(z=0)=0,\quad A_t(z=0)=\mu,,\\ &V_{Ix}(z=0)=(v^s_I)_x\,,\quad V_{Iy}(z=0)=(v^s_I)_y\,.
\end{aligned}
\end{equation}
We consider the regular condition at the event horizon $z=1$. Then, we solve the above equations numerically by adopting the Newton-Raphson method together with the Chebyshev pseudo-spectral method. In this work, we focus on the cases where $\bm{v^s_1}//\bm{v^s_2}$. Without loss of generality, we assume that they are along the $y$ direction.

Apparently, there exists two kinds of solution. One is single component superfluid solution $\phi_I\ne 0$, $\phi_J=0$, and the other is binary superfluid solution, which is the focus of this work. For fixed temperature, binary superfluid solution only exists for $\nu>\nu_c$~\cite{0order}. Below this critical value, superfluid would collapse due to strong attraction between the two components. 
For the counterflow case $v_+=0$ and the coflow case $v_-=0$, there is a additional symmetry $\phi_1=\phi_2=\phi$ between the two components. Utilizing the symmetry, the EoMs~\eqref{app:Seom} can be slightly further simplified. For counterflow binary superfluids $(v_1)_y=-(v_2)_y=v_y$, the difference of the two equations for $\phi_I$ gives $v_yA_y\phi=0$, which means $A_y=0$ for nonzero $\phi$. Then, the equation of motion for $V_{Iy}$ are eliminated, and the full EoMs become
\begin{equation}
    \begin{split}
    \partial_z(f\partial_z\phi)-v_y^2\phi+\frac{A_t^2}{f}\phi-z\phi-\frac{\nu}{2}\phi^3=0\,,\\
        f\partial_z^2A_t-4A_t\phi^2=0\,.  
\end{split}
\end{equation}
Note that these are not identical to EoMs of the single component holographic superfluid, since equation for $A_y$ is absent. For coflow binary superfluids $(v_1)_y=(v_2)_y=v_y$, the two components are completely identical. We would also have $V_{1y}=V_{2y}=V_y$. In this case, the EoMs are 
\begin{equation}
    \begin{split}
    \partial_z(f\partial_z\phi)-V_y^2\phi+\frac{A_t^2}{f}\phi-z\phi-\frac{\nu}{2}\phi^3=0\,,\\
        f\partial_z^2A_t-4A_t\phi^2=0\,,\\
    \partial_z(f\partial_zV_{y})-4 V_{y}\,\phi_I^2=0\,,     
\end{split}
\end{equation}
which are identical to the EoMs of the single component superfluid with extra self interaction $\nu'=\nu/2$ and the superfluid velocity $v_y$ if we rescale the scalar field $\phi\rightarrow \phi/\sqrt{2}$. Nevertheless, in the main text we have seen that the presence of the second component introduces additional instability into the system beside the Landau instability of single component superfluids.

\subsection{Free energy density and susceptibilities}\label{app:fe}    
Given the stationary solutions, we can proceed to calculate the free energy density and susceptibilities. 
According to the holographic dictionary, the free energy is simply identified with on-shell bulk action with Euclidean signature times temperature. In the probe limit with sources of condensates set to 0, we can neglect the gravity part of the action, the Euclidean action is finite and no counter term is needed. Therefore the free energy density can easily be calculated as\footnote{Due to our boundary conditions, it is $F$ where $v^s_\pm$ are held fixed which we calculate, and not $\bar F$ where ${\bf \bar h_\pm}$ are held fixed.}
\begin{equation}
    f\equiv\frac{F}{Vol}=-\int dz \sqrt{-g} \mathcal{L}_{\mathrm{on-shell}}\,.
\end{equation}
where $Vol$ is the volume of the spatial subspace on the boundary, $\mathcal{L}$ is given by~\eqref{lagrangian} and $T=3/(4\pi z_h)$. This is essentially the free energy density difference with respect to pure AdS planar black hole without matter fields. To get $\chi_{AB}$, 
we need to calculate the free energy density for different values of the  parameters $(\mu, v^s_+, v^s_-)$. Once we obtain $f(\mu, v^s_+, v^s_-)$, we can use the finite difference method to extract the value of $\chi_{AB}$ numerically. For instance, if we want to calculate $\chi_{++}$, we need to calculate the free energy density for different $v^s_+$ while keeping $(\mu,v^s_-)$ fixed to $(\mu_0,v^s_{-0})$, and then calculate $\chi_{++}=-\partial_{v^s_+}^2f(\mu_0, v^s_+, v^s_{-0})$ numerically.

\section{Linearized equations of motion}
\label{app:eig}

\subsection{Linearized EoMs of Hydrodynamics}
 After inserting the constitutive relations \eqref{constrels} and \eqref{Sigmaij}, the explicit form of EoMs~\eqref{sfeoms} is
\begin{equation}
\begin{aligned}
    &\partial_t n+\partial_i\sum_I\bar{h}_I^i\\&=\partial_i[(\sigma_1 \delta^{ij}+\sigma_2\frac{v_J^{si}v_J^{sj}}{|v_J^s|^2})\partial_j\mu + \beta v_J^{si}\partial_j \bar{h}_J^j]\,, 
 \\ &\partial_tv_I^{si}+\partial_i\mu\\&=\partial_i[\beta v_I^{sj}\partial_j\mu + (\delta_1 \delta_{IJ}+\delta_2 v_I^{si}v_{Ji}^s)\partial_j \bar{h}_J^j)]\,.\\
\end{aligned}
\end{equation}
By perturbing the expectation values $O_A=O_{A0}+e^{-i\omega t+i\bm{k}\cdot \bm{x}}\delta O_A$ and their sources $s_A=s_{A0}+e^{-i\omega t+i\bm{k}\cdot \bm{x}}\delta s_A$, where $O_A=(n, \bm{v^s_I})$ and $s_A=(\mu, \bm{\bar{h}_I})$, we can obtain the linearized EoMs
\begin{equation}
\begin{aligned}
    &i\omega\delta n+ik\sum_I\delta\bar{h}_I\\&=-(\sigma_1 +2\sigma_2)k^2\delta\mu - \beta v_J^{s}k^2 \delta\bar{h}_J\,, 
 \\ &i\omega \delta v_I^{s}+ik\delta \mu\\&=-\beta v_I^{sj}k^2\delta\mu - (\delta_1 \delta_{IJ}+\delta_2 v_I^{s}v_{J}^s) k^2\delta\bar{h}_J\,.\\
\end{aligned}
\end{equation}
Changing variables to \eqref{pmv} we get
\begin{equation}
\begin{aligned}
\left(\begin{array}{c}
    \delta n \\
      \delta v^{s}_+\\
      \delta v^{s}_-
\end{array}\right)=M\cdot\left(\begin{array}{c}
     \delta \mu  \\
    \delta \bar{h}_+\\ 
    \delta \bar{h}_-
\end{array}\right)\,,
\end{aligned}
\end{equation}
with matrix $M$ given by \eqref{MAB} in the main text.

\subsection{Linearized EoMs of Holographic binary superfluids}
In this section we consider dynamical counterflow and coflow instabilities in miscible holographic binary superfluids using linear response theory. We turn on small perturbations on the stationary background
\begin{equation}
\begin{aligned}
     &\Phi_I=(\Phi_{I0}+\delta\Phi_i)e^{i(v_I)_yy},\\ &A_t=A_{t0}+\delta A_t, \\ &A_y=A_{y0}+\delta A_y\,, 
\end{aligned}
\end{equation}
and linearize the EoMs~\eqref{phi}-\eqref{constraint}. Here $\Phi_{0I}=\phi_Ie^{i\Theta_I}$, $A_{t0}$, $A_{x0}$ and $A_{y0}$ are stationary solutions solved in last section. Taking into account the translation invariance of the background along the time and spatial directions, we express the bulk perturbation fields as
    \begin{equation}
    \begin{aligned}
        &\delta\Phi_I=u_i(z)e^{-i(\omega t-\bm{k\cdot x})}, \quad \delta\Phi_I^*=v_i(z)e^{-i(\omega t-\bm{k\cdot x})},\\&\delta A_t=a_t(z)e^{-i(\omega t-\bm{k\cdot x})},\quad\delta A_y=a_y(z)e^{-i(\omega t-\bm{k\cdot x})} .
    \end{aligned}
    \end{equation}
For simplicity, in this work we only consider wave vectors parallel to the superfluid velocity, \emph{i.e.} $\bm{k\cdot x}=ky$. 

The full set of linear perturbation equations are given explicitly as
\begin{widetext}
   \begin{equation}
    \begin{aligned}
     \label{u}        &2iA_{t0}\partial_zu_I+2ia_t\partial_z\Phi_{0I}+i\partial_zA_{t0}u_I+i\partial_za_t\Phi_{0I}
        +\partial_z(f\partial_zu_I)-zu_I \\&-(k+(v_I)_y)^2u_I
        +k\Phi_{0I}a_y-A_{y0}^2u_I-2A_{y0}\Phi_{0I}a_y
        +2(k+(v_I)_y)A_{y0}u_I\\&+2a_y(v_I)_y\Phi_{0I}-\frac{\nu}{2}|\Phi_{0j}|^2u_I
        -\frac{\nu}{2} \Phi_{0j}^*\Phi_{0I}u_j-\frac{\nu}{2}  \Phi_{0j}\Phi_{0I}v_j\\&=-2i\omega\partial_zu_I, \qquad(I,J=1,2\quad I\ne J)\,,\\
    \end{aligned}
    \end{equation}
    \begin{equation}
    \begin{aligned}
    \label{v}
        &-2iA_{t0}\partial_zv_I-2ia_t\partial_z\Phi_{0I}^*-i\partial_zA_{t0}v_I-i\partial_za_t\Phi_{0I}^*
        +\partial_z(f\partial_zv_I)-zv_I \\&-(k-(v_I)_y)^2v_I
        -k\Phi_{0I}^*a_y-A_{y0}^2v_I-2A_{y0}\Phi_{0I}^*a_y
        -2(k-(v_I)_y)A_{y0}v_I\\&
        +2a_y(v_I)_y\Phi_{0I}^*-\frac{\nu}{2}|\Phi_{0j}|^2v_I
        -\frac{\nu}{2} \Phi_{0j}\Phi_{0I}^*v_j-\frac{\nu}{2}  \Phi_{0j}^*\Phi_{0I}^*u_j\\&=-2i\omega\partial_zv_I, \qquad(I,J=1,2\quad I\ne J)\,,\\
    \end{aligned}
    \end{equation}
    \begin{equation}
    \begin{aligned}
    \label{at}
        &-k^2a_t+ikf\partial_za_y-2a_t\sum_I|\Phi_{0I}|^2-2A_{t0}\sum_I(\Phi_{0I}^*u_I+\Phi_{0I}v_I)+if\sum_I(\Phi_{0I}^*\partial_zu_I\\
        &-\Phi_{0I}\partial_zv_I+v_I\partial_z\Phi_{0I}-u_I\partial_z\Phi_{0I}^*)=-i\omega(\partial_za_t+ika_y)+\omega\sum_I(\Phi_{0I}^* u_I-\Phi_{0I} v_I)\,,\\
    \end{aligned}
    \end{equation}
    \begin{equation}
    \begin{aligned}
    \label{ay}
        &ik\partial_za_t+\partial_z(f\partial_za_y)-2a_y\sum_I|\Phi_{0I}|^2-2A_{y0}\sum_I(\Phi_{0I}^*u_I+\Phi_{0I}v_I)\\
        &+\sum_I((k+(v_I)_y)\Phi_{0I}^*u_I-(k-(v_I)_y)\Phi_{0I}v_I+(v_I)_yv_I\Phi_{0I}+(v_I)_yu_I\Phi_{0I}^*)\\&=-2i\omega\partial_za_y\,.
    \end{aligned}
   \end{equation}        
\end{widetext}

For more stable numerical performance, we use the following equation for $a_t$:
\begin{equation}
    \begin{aligned}
    \partial_z(ika_y-\partial_za_t)+i\sum_I(\Phi_I^*\partial_zu_I+v_I\partial_z\Phi_I\\-u_I\partial_z\Phi_I^*-\Phi_I\partial_zv_I)=0\,,
    \end{aligned}
\end{equation}
which comes from the constraint equation~\eqref{constraint}. Moreover, we require~\eqref{at} to be satisfied at the AdS boundary $z=0$, yielding
\begin{equation}
(k\partial_za_y+\omega \partial_za_t=0)|_{z=0}\,. 
\end{equation}
Then, by considering~\eqref{relation}, \eqref{at} is also satisfied in the whole bulk.  Regarding other perturbed fields, we impose the source free boundary condition at the AdS boundary. Finally, the system results in a generalized eigenvalue problem:
\begin{equation}
\label{eig}
    M_ku_k=i\omega_k Bu_k, \quad u_k=\{u_1,v_1,u_2,v_2,a_t,a_y\}_k^\mathrm{T}\,.
\end{equation}

The corresponding QNMs are extracted by solving the above generalized eigenvalue problem. Then we can  numerically obtain $\omega$ for each $k$ and velocity $v_y$. Due to dissipation into the normal component, the quasi-normal frequencies generically take a complex value. Since $\delta\Phi_I\sim e^{-i\omega t}$, the stationary configuration will become dynamically unstable whenever $\mathrm{Im}(\omega)>0$. The larger the positive imaginary part is, the more unstable the system becomes.

There are some general features of this generalized eigenvalue problem.
Two types of symmetries lie in the EoMs. The first one links EoMs for $k$ and $-k$. If $u_k$ and $\omega_k$ is a solution for \eqref{eig}, we would also have
\begin{equation}
\label{eig-k}
    M_{-k}{u'}_{k}^{*}=-i\omega_k^*{u'}_{k}^{*}\,,\quad {u'}_k=\{v_1,u_1,v_2,u_2,a_t,a_y\}_k^\mathrm{T}\,,
\end{equation}
which means $u_{-k}={u'}_k^*$, $\omega_{-k}=-\omega_k^*$. The second symmetry originates from the symmetry between the two components. For $(v_1)_y=\pm (v_2)_y$, we have
\begin{equation}
\label{eigv}
\begin{aligned}
        M_{\pm k}(\Phi_{01}\leftrightarrow\Phi_{02},A_y\rightarrow \pm A_y)u^\pm_k=i\omega_ku^\pm_k,\\
    u^\pm_k=\{u_2,v_2,u_1,v_1,a_t,\pm a_y\}_k^\mathrm{T}\,.
\end{aligned}
\end{equation}
Taking into account $\Phi_{01}=\Phi_{02}$ and $A_y=0$ for $(v_1)_y=-(v_2)_y$, we get $M_{\pm k}(\Phi_{01}\leftrightarrow\Phi_{02},A_y\rightarrow \pm A_y)=M_{\pm k}$. Then, combining~(\ref{eig-k}) and~(\ref{eigv}), for counterflow case one gets
\begin{equation}
\label{eigm}
\begin{aligned}
       & M_{k}\Bar{u}^-_k=-i\omega_k^*\Bar{u}^-_k, \\
   & \Bar{u}^-_k=\{v_2,u_2,v_1,u_1,a_t,-a_y\}_k^\mathrm{\dagger}\,.
\end{aligned}
\end{equation}
Therefore for the counterflow case, $\omega_k$ and $-\omega_k^*$ come as pairs for given $k$. If $\omega_k=-\omega_k^*$, \emph{i.e.} $\omega_k$ purely imaginary, we would also have $(u_1,v_1)=(v_2,u_2)^*e^{i\alpha}$, where $\alpha$ is a global phase. For the coflow case, we have $u_k=u^+_k$, which gives $(u_1,v_1)=(u_2,v_2)e^{i\alpha}=(u_1,v_1)e^{2i\alpha}\Rightarrow e^{i\alpha}=\pm 1$. Therefore, there are two sets of solutions $(u_1,v_1)=\pm(u_2,v_2)$ for the coflow case. In the main text, we see that one of the solutions shows the Landau instability and the other shows the coflow instability.

\section{Sound velocity from QNMs and susceptibilities}\label{app:sound}

In Figure~\ref{vs_F} and Figure~\ref{thermo}, we plot the sound velocity extracted from QNMs directly by $v_s=\lim_{k\rightarrow 0}\omega/k$. According to~\eqref{soundspeed}, one can also extract the sound velocity from the susceptibilities:
\begin{equation}
    (v_s)_\pm=\frac{-\chi_{n+}\pm\sqrt{\chi_{n+}^2+\chi_{++}\chi_{nn}}}{\chi_{nn}}\,.
\end{equation}
 The sound velocity extracted from these two methods should be in agreement. As a self-consistency check, below in Figure~\ref{dv}, we plot the discrepancy between sound velocities from the two methods $\Delta v_s/(v_s)_{\mathrm{QNM}}=((v_s)_{\mathrm{QNM}}-(v_s)_{\mathrm{susceptibility}})/(v_s)_{\mathrm{QNM}}$ for the larger sound velocity of the counterflow and coflow cases separately. We find the results agree well within numerical error.

\begin{figure}[htpb]
    \centering
        \includegraphics[width=0.95\linewidth]{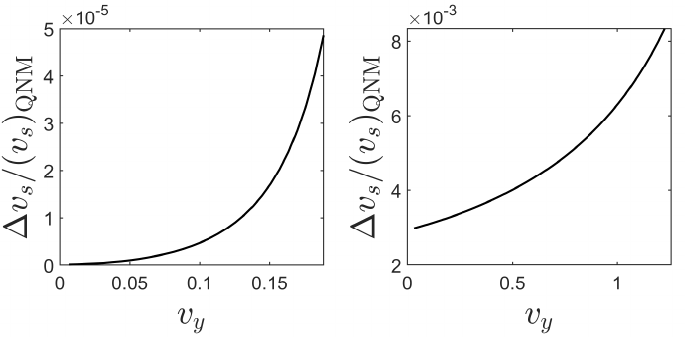}
        \caption{Discrepancy between sound velocities extracted from QNMs and susceptibilities, $\Delta v_s/(v_s)_{\mathrm{QNM}}=((v_s)_{\mathrm{QNM}}-(v_s)_{\mathrm{susceptibility}})/(v_s)_{\mathrm{QNM}}$, for different superfluid velocities. \textbf{Left:} discrepancy of the larger sound velocity of the counterflow case. \textbf{Right:} discrepancy of the larger sound velocity of the coflow case. Relevant parameters are $T/T_c=0.677$ and $\nu=-0.2$.}
    \label{dv}
\end{figure}

\bibliography{biblio}

\begin{thebibliography}{101}%
\makeatletter
\providecommand \@ifxundefined [1]{%
 \@ifx{#1\undefined}
}%
\providecommand \@ifnum [1]{%
 \ifnum #1\expandafter \@firstoftwo
 \else \expandafter \@secondoftwo
 \fi
}%
\providecommand \@ifx [1]{%
 \ifx #1\expandafter \@firstoftwo
 \else \expandafter \@secondoftwo
 \fi
}%
\providecommand \natexlab [1]{#1}%
\providecommand \enquote  [1]{``#1''}%
\providecommand \bibnamefont  [1]{#1}%
\providecommand \bibfnamefont [1]{#1}%
\providecommand \citenamefont [1]{#1}%
\providecommand \href@noop [0]{\@secondoftwo}%
\providecommand \href [0]{\begingroup \@sanitize@url \@href}%
\providecommand \@href[1]{\@@startlink{#1}\@@href}%
\providecommand \@@href[1]{\endgroup#1\@@endlink}%
\providecommand \@sanitize@url [0]{\catcode `\\12\catcode `\$12\catcode `\&12\catcode `\#12\catcode `\^12\catcode `\_12\catcode `\%12\relax}%
\providecommand \@@startlink[1]{}%
\providecommand \@@endlink[0]{}%
\providecommand \url  [0]{\begingroup\@sanitize@url \@url }%
\providecommand \@url [1]{\endgroup\@href {#1}{\urlprefix }}%
\providecommand \urlprefix  [0]{URL }%
\providecommand \Eprint [0]{\href }%
\providecommand \doibase [0]{http://dx.doi.org/}%
\providecommand \selectlanguage [0]{\@gobble}%
\providecommand \bibinfo  [0]{\@secondoftwo}%
\providecommand \bibfield  [0]{\@secondoftwo}%
\providecommand \translation [1]{[#1]}%
\providecommand \BibitemOpen [0]{}%
\providecommand \bibitemStop [0]{}%
\providecommand \bibitemNoStop [0]{.\EOS\space}%
\providecommand \EOS [0]{\spacefactor3000\relax}%
\providecommand \BibitemShut  [1]{\csname bibitem#1\endcsname}%
\let\auto@bib@innerbib\@empty
\bibitem [{\citenamefont {Helmholtz}(1868)}]{KH1}%
  \BibitemOpen
  \bibfield  {author} {\bibinfo {author} {\bibnamefont {Helmholtz}},\ }\href {\doibase 10.1080/14786446808640073} {\bibfield  {journal} {\bibinfo  {journal} {The London, Edinburgh, and Dublin Philosophical Magazine and Journal of Science}\ }\textbf {\bibinfo {volume} {36}},\ \bibinfo {pages} {337} (\bibinfo {year} {1868})}\BibitemShut {NoStop}%
\bibitem [{\citenamefont {Thomson}(1871)}]{KH2}%
  \BibitemOpen
  \bibfield  {author} {\bibinfo {author} {\bibfnamefont {W.}~\bibnamefont {Thomson}},\ }\href {\doibase 10.1080/14786447108640585} {\bibfield  {journal} {\bibinfo  {journal} {The London, Edinburgh, and Dublin Philosophical Magazine and Journal of Science}\ }\textbf {\bibinfo {volume} {42}},\ \bibinfo {pages} {362} (\bibinfo {year} {1871})}\BibitemShut {NoStop}%
\bibitem [{\citenamefont {Takeuchi}\ \emph {et~al.}(2010{\natexlab{a}})\citenamefont {Takeuchi}, \citenamefont {Suzuki}, \citenamefont {Kasamatsu}, \citenamefont {Saito},\ and\ \citenamefont {Tsubota}}]{QKH2}%
  \BibitemOpen
  \bibfield  {author} {\bibinfo {author} {\bibfnamefont {H.}~\bibnamefont {Takeuchi}}, \bibinfo {author} {\bibfnamefont {N.}~\bibnamefont {Suzuki}}, \bibinfo {author} {\bibfnamefont {K.}~\bibnamefont {Kasamatsu}}, \bibinfo {author} {\bibfnamefont {H.}~\bibnamefont {Saito}}, \ and\ \bibinfo {author} {\bibfnamefont {M.}~\bibnamefont {Tsubota}},\ }\href {\doibase 10.1103/PhysRevB.81.094517} {\bibfield  {journal} {\bibinfo  {journal} {Physical Review B}\ }\textbf {\bibinfo {volume} {81}},\ \bibinfo {pages} {094517} (\bibinfo {year} {2010}{\natexlab{a}})}\BibitemShut {NoStop}%
\bibitem [{\citenamefont {Volovik}(2002)}]{QKH3}%
  \BibitemOpen
  \bibfield  {author} {\bibinfo {author} {\bibfnamefont {G.~E.}\ \bibnamefont {Volovik}},\ }\href {\doibase 10.1134/1.1490014} {\bibfield  {journal} {\bibinfo  {journal} {Journal of Experimental and Theoretical Physics Letters}\ }\textbf {\bibinfo {volume} {75}},\ \bibinfo {pages} {418} (\bibinfo {year} {2002})}\BibitemShut {NoStop}%
\bibitem [{\citenamefont {Blaauwgeers}\ \emph {et~al.}(2002)\citenamefont {Blaauwgeers}, \citenamefont {Eltsov}, \citenamefont {Eska}, \citenamefont {Finne}, \citenamefont {Haley}, \citenamefont {Krusius}, \citenamefont {Ruohio}, \citenamefont {Skrbek},\ and\ \citenamefont {Volovik}}]{QKH4}%
  \BibitemOpen
  \bibfield  {author} {\bibinfo {author} {\bibfnamefont {R.}~\bibnamefont {Blaauwgeers}}, \bibinfo {author} {\bibfnamefont {V.~B.}\ \bibnamefont {Eltsov}}, \bibinfo {author} {\bibfnamefont {G.}~\bibnamefont {Eska}}, \bibinfo {author} {\bibfnamefont {A.~P.}\ \bibnamefont {Finne}}, \bibinfo {author} {\bibfnamefont {R.~P.}\ \bibnamefont {Haley}}, \bibinfo {author} {\bibfnamefont {M.}~\bibnamefont {Krusius}}, \bibinfo {author} {\bibfnamefont {J.~J.}\ \bibnamefont {Ruohio}}, \bibinfo {author} {\bibfnamefont {L.}~\bibnamefont {Skrbek}}, \ and\ \bibinfo {author} {\bibfnamefont {G.~E.}\ \bibnamefont {Volovik}},\ }\href {\doibase 10.1103/PhysRevLett.89.155301} {\bibfield  {journal} {\bibinfo  {journal} {Phys. Rev. Lett.}\ }\textbf {\bibinfo {volume} {89}},\ \bibinfo {pages} {155301} (\bibinfo {year} {2002})}\BibitemShut {NoStop}%
\bibitem [{\citenamefont {Finne}\ \emph {et~al.}(2006)\citenamefont {Finne}, \citenamefont {Eltsov}, \citenamefont {Hänninen}, \citenamefont {Kopnin}, \citenamefont {Kopu}, \citenamefont {Krusius}, \citenamefont {Tsubota},\ and\ \citenamefont {Volovik}}]{QKH5}%
  \BibitemOpen
  \bibfield  {author} {\bibinfo {author} {\bibfnamefont {A.~P.}\ \bibnamefont {Finne}}, \bibinfo {author} {\bibfnamefont {V.~B.}\ \bibnamefont {Eltsov}}, \bibinfo {author} {\bibfnamefont {R.}~\bibnamefont {Hänninen}}, \bibinfo {author} {\bibfnamefont {N.~B.}\ \bibnamefont {Kopnin}}, \bibinfo {author} {\bibfnamefont {J.}~\bibnamefont {Kopu}}, \bibinfo {author} {\bibfnamefont {M.}~\bibnamefont {Krusius}}, \bibinfo {author} {\bibfnamefont {M.}~\bibnamefont {Tsubota}}, \ and\ \bibinfo {author} {\bibfnamefont {G.~E.}\ \bibnamefont {Volovik}},\ }\href {\doibase 10.1088/0034-4885/69/12/R03} {\bibfield  {journal} {\bibinfo  {journal} {Reports on Progress in Physics}\ }\textbf {\bibinfo {volume} {69}},\ \bibinfo {pages} {3157} (\bibinfo {year} {2006})}\BibitemShut {NoStop}%
\bibitem [{\citenamefont {Eltsov}\ \emph {et~al.}(2019)\citenamefont {Eltsov}, \citenamefont {Gordeev},\ and\ \citenamefont {Krusius}}]{QKH6}%
  \BibitemOpen
  \bibfield  {author} {\bibinfo {author} {\bibfnamefont {V.~B.}\ \bibnamefont {Eltsov}}, \bibinfo {author} {\bibfnamefont {A.}~\bibnamefont {Gordeev}}, \ and\ \bibinfo {author} {\bibfnamefont {M.}~\bibnamefont {Krusius}},\ }\href {\doibase 10.1103/PhysRevB.99.054104} {\bibfield  {journal} {\bibinfo  {journal} {Phys. Rev. B}\ }\textbf {\bibinfo {volume} {99}},\ \bibinfo {pages} {054104} (\bibinfo {year} {2019})}\BibitemShut {NoStop}%
\bibitem [{\citenamefont {Abad}\ \emph {et~al.}(2015)\citenamefont {Abad}, \citenamefont {Recati}, \citenamefont {Stringari},\ and\ \citenamefont {Chevy}}]{counterflow1}%
  \BibitemOpen
  \bibfield  {author} {\bibinfo {author} {\bibfnamefont {M.}~\bibnamefont {Abad}}, \bibinfo {author} {\bibfnamefont {A.}~\bibnamefont {Recati}}, \bibinfo {author} {\bibfnamefont {S.}~\bibnamefont {Stringari}}, \ and\ \bibinfo {author} {\bibfnamefont {F.}~\bibnamefont {Chevy}},\ }\href {\doibase 10.1140/epjd/e2015-50851-y} {\bibfield  {journal} {\bibinfo  {journal} {The European Physical Journal D}\ }\textbf {\bibinfo {volume} {69}},\ \bibinfo {pages} {126} (\bibinfo {year} {2015})}\BibitemShut {NoStop}%
\bibitem [{\citenamefont {Law}\ \emph {et~al.}(2001)\citenamefont {Law}, \citenamefont {Chan}, \citenamefont {Leung},\ and\ \citenamefont {Chu}}]{counterflow2}%
  \BibitemOpen
  \bibfield  {author} {\bibinfo {author} {\bibfnamefont {C.~K.}\ \bibnamefont {Law}}, \bibinfo {author} {\bibfnamefont {C.~M.}\ \bibnamefont {Chan}}, \bibinfo {author} {\bibfnamefont {P.~T.}\ \bibnamefont {Leung}}, \ and\ \bibinfo {author} {\bibfnamefont {M.~C.}\ \bibnamefont {Chu}},\ }\href {\doibase 10.1103/PhysRevA.63.063612} {\bibfield  {journal} {\bibinfo  {journal} {Physical Review A}\ }\textbf {\bibinfo {volume} {63}},\ \bibinfo {pages} {063612} (\bibinfo {year} {2001})}\BibitemShut {NoStop}%
\bibitem [{\citenamefont {Ishino}\ \emph {et~al.}(2011)\citenamefont {Ishino}, \citenamefont {Tsubota},\ and\ \citenamefont {Takeuchi}}]{counterflow3}%
  \BibitemOpen
  \bibfield  {author} {\bibinfo {author} {\bibfnamefont {S.}~\bibnamefont {Ishino}}, \bibinfo {author} {\bibfnamefont {M.}~\bibnamefont {Tsubota}}, \ and\ \bibinfo {author} {\bibfnamefont {H.}~\bibnamefont {Takeuchi}},\ }\href {\doibase 10.1103/PhysRevA.83.063602} {\bibfield  {journal} {\bibinfo  {journal} {Phys. Rev. A}\ }\textbf {\bibinfo {volume} {83}},\ \bibinfo {pages} {063602} (\bibinfo {year} {2011})}\BibitemShut {NoStop}%
\bibitem [{\citenamefont {Takeuchi}\ \emph {et~al.}(2010{\natexlab{b}})\citenamefont {Takeuchi}, \citenamefont {Ishino},\ and\ \citenamefont {Tsubota}}]{counterflow4}%
  \BibitemOpen
  \bibfield  {author} {\bibinfo {author} {\bibfnamefont {H.}~\bibnamefont {Takeuchi}}, \bibinfo {author} {\bibfnamefont {S.}~\bibnamefont {Ishino}}, \ and\ \bibinfo {author} {\bibfnamefont {M.}~\bibnamefont {Tsubota}},\ }\href {\doibase 10.1103/PhysRevLett.105.205301} {\bibfield  {journal} {\bibinfo  {journal} {Physical Review Letters}\ }\textbf {\bibinfo {volume} {105}},\ \bibinfo {pages} {205301} (\bibinfo {year} {2010}{\natexlab{b}})}\BibitemShut {NoStop}%
\bibitem [{\citenamefont {Schmitt}(2014)}]{two-stream1}%
  \BibitemOpen
  \bibfield  {author} {\bibinfo {author} {\bibfnamefont {A.}~\bibnamefont {Schmitt}},\ }\href {\doibase 10.1103/PhysRevD.89.065024} {\bibfield  {journal} {\bibinfo  {journal} {Phys. Rev. D}\ }\textbf {\bibinfo {volume} {89}},\ \bibinfo {pages} {065024} (\bibinfo {year} {2014})}\BibitemShut {NoStop}%
\bibitem [{\citenamefont {Delehaye}\ \emph {et~al.}(2015)\citenamefont {Delehaye}, \citenamefont {Laurent}, \citenamefont {Ferrier-Barbut}, \citenamefont {Jin}, \citenamefont {Chevy},\ and\ \citenamefont {Salomon}}]{two-stream2}%
  \BibitemOpen
  \bibfield  {author} {\bibinfo {author} {\bibfnamefont {M.}~\bibnamefont {Delehaye}}, \bibinfo {author} {\bibfnamefont {S.}~\bibnamefont {Laurent}}, \bibinfo {author} {\bibfnamefont {I.}~\bibnamefont {Ferrier-Barbut}}, \bibinfo {author} {\bibfnamefont {S.}~\bibnamefont {Jin}}, \bibinfo {author} {\bibfnamefont {F.}~\bibnamefont {Chevy}}, \ and\ \bibinfo {author} {\bibfnamefont {C.}~\bibnamefont {Salomon}},\ }\href {\doibase 10.1103/PhysRevLett.115.265303} {\bibfield  {journal} {\bibinfo  {journal} {Phys. Rev. Lett.}\ }\textbf {\bibinfo {volume} {115}},\ \bibinfo {pages} {265303} (\bibinfo {year} {2015})}\BibitemShut {NoStop}%
\bibitem [{\citenamefont {Andersson}\ \emph {et~al.}(2004)\citenamefont {Andersson}, \citenamefont {Comer},\ and\ \citenamefont {Prix}}]{two-stream3}%
  \BibitemOpen
  \bibfield  {author} {\bibinfo {author} {\bibfnamefont {N.}~\bibnamefont {Andersson}}, \bibinfo {author} {\bibfnamefont {G.~L.}\ \bibnamefont {Comer}}, \ and\ \bibinfo {author} {\bibfnamefont {R.}~\bibnamefont {Prix}},\ }\href {\doibase 10.1111/j.1365-2966.2004.08166.x} {\bibfield  {journal} {\bibinfo  {journal} {Monthly Notices of the Royal Astronomical Society}\ }\textbf {\bibinfo {volume} {354}},\ \bibinfo {pages} {101} (\bibinfo {year} {2004})}\BibitemShut {NoStop}%
\bibitem [{\citenamefont {Haber}\ \emph {et~al.}(2016)\citenamefont {Haber}, \citenamefont {Schmitt},\ and\ \citenamefont {Stetina}}]{two-stream4}%
  \BibitemOpen
  \bibfield  {author} {\bibinfo {author} {\bibfnamefont {A.}~\bibnamefont {Haber}}, \bibinfo {author} {\bibfnamefont {A.}~\bibnamefont {Schmitt}}, \ and\ \bibinfo {author} {\bibfnamefont {S.}~\bibnamefont {Stetina}},\ }\href {\doibase 10.1103/PhysRevD.93.025011} {\bibfield  {journal} {\bibinfo  {journal} {Physical Review D}\ }\textbf {\bibinfo {volume} {93}},\ \bibinfo {pages} {025011} (\bibinfo {year} {2016})}\BibitemShut {NoStop}%
\bibitem [{\citenamefont {Khalatnikov}(1973)}]{khalatnikov1973sound}%
  \BibitemOpen
  \bibfield  {author} {\bibinfo {author} {\bibfnamefont {I.}~\bibnamefont {Khalatnikov}},\ }\href@noop {} {\bibfield  {journal} {\bibinfo  {journal} {Zh. Eksp. Teor. Fiz}\ }\textbf {\bibinfo {volume} {17}},\ \bibinfo {pages} {534} (\bibinfo {year} {1973})}\BibitemShut {NoStop}%
\bibitem [{\citenamefont {Nepomnyashchii}(1974)}]{Nepomnyashchii1974}%
  \BibitemOpen
  \bibfield  {author} {\bibinfo {author} {\bibfnamefont {Y.~A.}\ \bibnamefont {Nepomnyashchii}},\ }\href {\doibase 10.1007/BF01040171} {\bibfield  {journal} {\bibinfo  {journal} {"Theoretical and Mathematical Physics"}\ }\textbf {\bibinfo {volume} {20}},\ \bibinfo {pages} {904} (\bibinfo {year} {1974})}\BibitemShut {NoStop}%
\bibitem [{\citenamefont {Nepomnyashchii}(1976)}]{nepomnyashchil1976microscopic}%
  \BibitemOpen
  \bibfield  {author} {\bibinfo {author} {\bibfnamefont {Y.~A.}\ \bibnamefont {Nepomnyashchii}},\ }\href@noop {} {\bibfield  {journal} {\bibinfo  {journal} {Zh. Eksp. Teor. Fiz}\ }\textbf {\bibinfo {volume} {70}},\ \bibinfo {pages} {1071} (\bibinfo {year} {1976})}\BibitemShut {NoStop}%
\bibitem [{\citenamefont {Yukalov}(1980)}]{Yukalov1980-03-01}%
  \BibitemOpen
  \bibfield  {author} {\bibinfo {author} {\bibfnamefont {V.~I.}\ \bibnamefont {Yukalov}},\ }\href@noop {} {\bibfield  {journal} {\bibinfo  {journal} {Acta Phys. Pol., A}\ }\textbf {\bibinfo {volume} {57:3}} (\bibinfo {year} {1980})}\BibitemShut {NoStop}%
\bibitem [{\citenamefont {Hamner}\ \emph {et~al.}(2011)\citenamefont {Hamner}, \citenamefont {Chang}, \citenamefont {Engels},\ and\ \citenamefont {Hoefer}}]{PhysRevLett.106.065302}%
  \BibitemOpen
  \bibfield  {author} {\bibinfo {author} {\bibfnamefont {C.}~\bibnamefont {Hamner}}, \bibinfo {author} {\bibfnamefont {J.~J.}\ \bibnamefont {Chang}}, \bibinfo {author} {\bibfnamefont {P.}~\bibnamefont {Engels}}, \ and\ \bibinfo {author} {\bibfnamefont {M.~A.}\ \bibnamefont {Hoefer}},\ }\href {\doibase 10.1103/PhysRevLett.106.065302} {\bibfield  {journal} {\bibinfo  {journal} {Phys. Rev. Lett.}\ }\textbf {\bibinfo {volume} {106}},\ \bibinfo {pages} {065302} (\bibinfo {year} {2011})}\BibitemShut {NoStop}%
\bibitem [{\citenamefont {Beattie}\ \emph {et~al.}(2013{\natexlab{a}})\citenamefont {Beattie}, \citenamefont {Moulder}, \citenamefont {Fletcher},\ and\ \citenamefont {Hadzibabic}}]{PhysRevLett.110.025301}%
  \BibitemOpen
  \bibfield  {author} {\bibinfo {author} {\bibfnamefont {S.}~\bibnamefont {Beattie}}, \bibinfo {author} {\bibfnamefont {S.}~\bibnamefont {Moulder}}, \bibinfo {author} {\bibfnamefont {R.~J.}\ \bibnamefont {Fletcher}}, \ and\ \bibinfo {author} {\bibfnamefont {Z.}~\bibnamefont {Hadzibabic}},\ }\href {\doibase 10.1103/PhysRevLett.110.025301} {\bibfield  {journal} {\bibinfo  {journal} {Phys. Rev. Lett.}\ }\textbf {\bibinfo {volume} {110}},\ \bibinfo {pages} {025301} (\bibinfo {year} {2013}{\natexlab{a}})}\BibitemShut {NoStop}%
\bibitem [{\citenamefont {Kim}\ \emph {et~al.}(2017)\citenamefont {Kim}, \citenamefont {Seo},\ and\ \citenamefont {Shin}}]{PhysRevLett.119.185302}%
  \BibitemOpen
  \bibfield  {author} {\bibinfo {author} {\bibfnamefont {J.~H.}\ \bibnamefont {Kim}}, \bibinfo {author} {\bibfnamefont {S.~W.}\ \bibnamefont {Seo}}, \ and\ \bibinfo {author} {\bibfnamefont {Y.}~\bibnamefont {Shin}},\ }\href {\doibase 10.1103/PhysRevLett.119.185302} {\bibfield  {journal} {\bibinfo  {journal} {Phys. Rev. Lett.}\ }\textbf {\bibinfo {volume} {119}},\ \bibinfo {pages} {185302} (\bibinfo {year} {2017})}\BibitemShut {NoStop}%
\bibitem [{\citenamefont {Pethick}\ and\ \citenamefont {Smith}(2008)}]{pethick2008bose}%
  \BibitemOpen
  \bibfield  {author} {\bibinfo {author} {\bibfnamefont {C.~J.}\ \bibnamefont {Pethick}}\ and\ \bibinfo {author} {\bibfnamefont {H.}~\bibnamefont {Smith}},\ }\href@noop {} {\emph {\bibinfo {title} {Bose--Einstein condensation in dilute gases}}}\ (\bibinfo  {publisher} {Cambridge university press},\ \bibinfo {year} {2008})\BibitemShut {NoStop}%
\bibitem [{\citenamefont {Pitaevskii}\ and\ \citenamefont {Stringari}(2003)}]{pitaevskii2003bose}%
  \BibitemOpen
  \bibfield  {author} {\bibinfo {author} {\bibfnamefont {L.}~\bibnamefont {Pitaevskii}}\ and\ \bibinfo {author} {\bibfnamefont {S.}~\bibnamefont {Stringari}},\ }\href {https://books.google.co.il/books?id=rIobbOxC4j4C} {\emph {\bibinfo {title} {Bose-Einstein Condensation}}},\ International Series of Monographs on Physics\ (\bibinfo  {publisher} {Clarendon Press},\ \bibinfo {year} {2003})\BibitemShut {NoStop}%
\bibitem [{\citenamefont {Landau}\ and\ \citenamefont {Lifshitz}(1987)}]{landaubookfluids}%
  \BibitemOpen
  \bibfield  {author} {\bibinfo {author} {\bibfnamefont {L.}~\bibnamefont {Landau}}\ and\ \bibinfo {author} {\bibfnamefont {E.}~\bibnamefont {Lifshitz}},\ }\href {\doibase https://doi.org/10.1016/B978-0-08-033933-7.50006-4} {\emph {\bibinfo {title} {Fluid Mechanics (Second Edition)}}},\ \bibinfo {edition} {second edition}\ ed.,\ edited by\ \bibinfo {editor} {\bibfnamefont {L.}~\bibnamefont {Landau}}\ and\ \bibinfo {editor} {\bibfnamefont {E.}~\bibnamefont {Lifshitz}}\ (\bibinfo  {publisher} {Pergamon},\ \bibinfo {year} {1987})\BibitemShut {NoStop}%
\bibitem [{\citenamefont {Chaikin}\ and\ \citenamefont {Lubensky}(1995)}]{chaikinlubensky1995}%
  \BibitemOpen
  \bibfield  {author} {\bibinfo {author} {\bibfnamefont {P.~M.}\ \bibnamefont {Chaikin}}\ and\ \bibinfo {author} {\bibfnamefont {T.~C.}\ \bibnamefont {Lubensky}},\ }\href {\doibase 10.1017/CBO9780511813467} {\emph {\bibinfo {title} {Principles of Condensed Matter Physics}}}\ (\bibinfo  {publisher} {Cambridge University Press},\ \bibinfo {year} {1995})\BibitemShut {NoStop}%
\bibitem [{\citenamefont {Kovtun}(2012)}]{Kovtun:2012rj}%
  \BibitemOpen
  \bibfield  {author} {\bibinfo {author} {\bibfnamefont {P.}~\bibnamefont {Kovtun}},\ }\href {\doibase 10.1088/1751-8113/45/47/473001} {\bibfield  {journal} {\bibinfo  {journal} {J. Phys. A}\ }\textbf {\bibinfo {volume} {45}},\ \bibinfo {pages} {473001} (\bibinfo {year} {2012})},\ \Eprint {http://arxiv.org/abs/1205.5040} {arXiv:1205.5040 [hep-th]} \BibitemShut {NoStop}%
\bibitem [{\citenamefont {Forster}(2018)}]{forster2018hydrodynamic}%
  \BibitemOpen
  \bibfield  {author} {\bibinfo {author} {\bibfnamefont {D.}~\bibnamefont {Forster}},\ }\href@noop {} {\emph {\bibinfo {title} {Hydrodynamic fluctuations, broken symmetry, and correlation functions}}}\ (\bibinfo  {publisher} {CRC Press},\ \bibinfo {year} {2018})\BibitemShut {NoStop}%
\bibitem [{\citenamefont {Lucas}\ and\ \citenamefont {Fong}(2018)}]{Lucas:2017idv}%
  \BibitemOpen
  \bibfield  {author} {\bibinfo {author} {\bibfnamefont {A.}~\bibnamefont {Lucas}}\ and\ \bibinfo {author} {\bibfnamefont {K.~C.}\ \bibnamefont {Fong}},\ }\href {\doibase 10.1088/1361-648X/aaa274} {\bibfield  {journal} {\bibinfo  {journal} {J. Phys. Condens. Matter}\ }\textbf {\bibinfo {volume} {30}},\ \bibinfo {pages} {053001} (\bibinfo {year} {2018})},\ \Eprint {http://arxiv.org/abs/1710.08425} {arXiv:1710.08425 [cond-mat.str-el]} \BibitemShut {NoStop}%
\bibitem [{\citenamefont {Fritz}\ and\ \citenamefont {Scaffidi}(2024)}]{fritz2024hydrodynamic}%
  \BibitemOpen
  \bibfield  {author} {\bibinfo {author} {\bibfnamefont {L.}~\bibnamefont {Fritz}}\ and\ \bibinfo {author} {\bibfnamefont {T.}~\bibnamefont {Scaffidi}},\ }\href@noop {} {\bibfield  {journal} {\bibinfo  {journal} {Annual Review of Condensed Matter Physics}\ }\textbf {\bibinfo {volume} {15}},\ \bibinfo {pages} {17} (\bibinfo {year} {2024})},\ \Eprint {http://arxiv.org/abs/2303.14205} {arXiv:2303.14205 [cond-mat.str-el]} \BibitemShut {NoStop}%
\bibitem [{\citenamefont {Luzum}\ and\ \citenamefont {Romatschke}(2008)}]{Luzum:2008cw}%
  \BibitemOpen
  \bibfield  {author} {\bibinfo {author} {\bibfnamefont {M.}~\bibnamefont {Luzum}}\ and\ \bibinfo {author} {\bibfnamefont {P.}~\bibnamefont {Romatschke}},\ }\href {\doibase 10.1103/PhysRevC.78.034915} {\bibfield  {journal} {\bibinfo  {journal} {Phys. Rev. C}\ }\textbf {\bibinfo {volume} {78}},\ \bibinfo {pages} {034915} (\bibinfo {year} {2008})},\ \bibinfo {note} {[Erratum: Phys.Rev.C 79, 039903 (2009)]},\ \Eprint {http://arxiv.org/abs/0804.4015} {arXiv:0804.4015 [nucl-th]} \BibitemShut {NoStop}%
\bibitem [{\citenamefont {Ammon}\ and\ \citenamefont {Erdmenger}(2015)}]{Ammon:2015wua}%
  \BibitemOpen
  \bibfield  {author} {\bibinfo {author} {\bibfnamefont {M.}~\bibnamefont {Ammon}}\ and\ \bibinfo {author} {\bibfnamefont {J.}~\bibnamefont {Erdmenger}},\ }\href {\doibase 10.1017/CBO9780511846373} {\emph {\bibinfo {title} {{Gauge/gravity duality}: {Foundations and applications}}}}\ (\bibinfo  {publisher} {Cambridge University Press},\ \bibinfo {address} {Cambridge},\ \bibinfo {year} {2015})\BibitemShut {NoStop}%
\bibitem [{\citenamefont {Zaanen}\ \emph {et~al.}(2015)\citenamefont {Zaanen}, \citenamefont {Liu}, \citenamefont {Sun},\ and\ \citenamefont {Schalm}}]{RN499}%
  \BibitemOpen
  \bibfield  {author} {\bibinfo {author} {\bibfnamefont {J.}~\bibnamefont {Zaanen}}, \bibinfo {author} {\bibfnamefont {Y.}~\bibnamefont {Liu}}, \bibinfo {author} {\bibfnamefont {Y.-W.}\ \bibnamefont {Sun}}, \ and\ \bibinfo {author} {\bibfnamefont {K.}~\bibnamefont {Schalm}},\ }\href {\doibase DOI: 10.1017/CBO9781139942492} {\emph {\bibinfo {title} {Holographic Duality in Condensed Matter Physics}}}\ (\bibinfo  {publisher} {Cambridge University Press},\ \bibinfo {address} {Cambridge},\ \bibinfo {year} {2015})\BibitemShut {NoStop}%
\bibitem [{\citenamefont {{Hartnoll}}\ \emph {et~al.}()\citenamefont {{Hartnoll}}, \citenamefont {{Lucas}},\ and\ \citenamefont {{Sachdev}}}]{holoreview}%
  \BibitemOpen
  \bibfield  {author} {\bibinfo {author} {\bibfnamefont {S.~A.}\ \bibnamefont {{Hartnoll}}}, \bibinfo {author} {\bibfnamefont {A.}~\bibnamefont {{Lucas}}}, \ and\ \bibinfo {author} {\bibfnamefont {S.}~\bibnamefont {{Sachdev}}},\ }\href@noop {} {\ }\Eprint {http://arxiv.org/abs/1612.07324} {arXiv:1612.07324 [hep-th]} \BibitemShut {NoStop}%
\bibitem [{\citenamefont {Maldacena}(1998)}]{Maldacena:1997re}%
  \BibitemOpen
  \bibfield  {author} {\bibinfo {author} {\bibfnamefont {J.~M.}\ \bibnamefont {Maldacena}},\ }\href {\doibase 10.4310/ATMP.1998.v2.n2.a1} {\bibfield  {journal} {\bibinfo  {journal} {Adv. Theor. Math. Phys.}\ }\textbf {\bibinfo {volume} {2}},\ \bibinfo {pages} {231} (\bibinfo {year} {1998})},\ \Eprint {http://arxiv.org/abs/hep-th/9711200} {arXiv:hep-th/9711200} \BibitemShut {NoStop}%
\bibitem [{\citenamefont {Witten}(1998)}]{Witten:1998qj}%
  \BibitemOpen
  \bibfield  {author} {\bibinfo {author} {\bibfnamefont {E.}~\bibnamefont {Witten}},\ }\href {\doibase 10.4310/ATMP.1998.v2.n2.a2} {\bibfield  {journal} {\bibinfo  {journal} {Adv. Theor. Math. Phys.}\ }\textbf {\bibinfo {volume} {2}},\ \bibinfo {pages} {253} (\bibinfo {year} {1998})},\ \Eprint {http://arxiv.org/abs/hep-th/9802150} {arXiv:hep-th/9802150} \BibitemShut {NoStop}%
\bibitem [{\citenamefont {Gubser}\ \emph {et~al.}(1998)\citenamefont {Gubser}, \citenamefont {Klebanov},\ and\ \citenamefont {Polyakov}}]{Gubser:1998bc}%
  \BibitemOpen
  \bibfield  {author} {\bibinfo {author} {\bibfnamefont {S.~S.}\ \bibnamefont {Gubser}}, \bibinfo {author} {\bibfnamefont {I.~R.}\ \bibnamefont {Klebanov}}, \ and\ \bibinfo {author} {\bibfnamefont {A.~M.}\ \bibnamefont {Polyakov}},\ }\href {\doibase 10.1016/S0370-2693(98)00377-3} {\bibfield  {journal} {\bibinfo  {journal} {Phys. Lett. B}\ }\textbf {\bibinfo {volume} {428}},\ \bibinfo {pages} {105} (\bibinfo {year} {1998})},\ \Eprint {http://arxiv.org/abs/hep-th/9802109} {arXiv:hep-th/9802109} \BibitemShut {NoStop}%
\bibitem [{\citenamefont {Kovtun}\ \emph {et~al.}(2005)\citenamefont {Kovtun}, \citenamefont {Son},\ and\ \citenamefont {Starinets}}]{Kovtun:2004de}%
  \BibitemOpen
  \bibfield  {author} {\bibinfo {author} {\bibfnamefont {P.}~\bibnamefont {Kovtun}}, \bibinfo {author} {\bibfnamefont {D.~T.}\ \bibnamefont {Son}}, \ and\ \bibinfo {author} {\bibfnamefont {A.~O.}\ \bibnamefont {Starinets}},\ }\href {\doibase 10.1103/PhysRevLett.94.111601} {\bibfield  {journal} {\bibinfo  {journal} {Phys. Rev. Lett.}\ }\textbf {\bibinfo {volume} {94}},\ \bibinfo {pages} {111601} (\bibinfo {year} {2005})},\ \Eprint {http://arxiv.org/abs/hep-th/0405231} {arXiv:hep-th/0405231} \BibitemShut {NoStop}%
\bibitem [{\citenamefont {Sch\"afer}\ and\ \citenamefont {Teaney}(2009)}]{Schafer:2009dj}%
  \BibitemOpen
  \bibfield  {author} {\bibinfo {author} {\bibfnamefont {T.}~\bibnamefont {Sch\"afer}}\ and\ \bibinfo {author} {\bibfnamefont {D.}~\bibnamefont {Teaney}},\ }\href {\doibase 10.1088/0034-4885/72/12/126001} {\bibfield  {journal} {\bibinfo  {journal} {Rept. Prog. Phys.}\ }\textbf {\bibinfo {volume} {72}},\ \bibinfo {pages} {126001} (\bibinfo {year} {2009})},\ \Eprint {http://arxiv.org/abs/0904.3107} {arXiv:0904.3107 [hep-ph]} \BibitemShut {NoStop}%
\bibitem [{\citenamefont {Hartnoll}(2015)}]{Hartnoll:2014lpa}%
  \BibitemOpen
  \bibfield  {author} {\bibinfo {author} {\bibfnamefont {S.~A.}\ \bibnamefont {Hartnoll}},\ }\href {\doibase 10.1038/nphys3174} {\bibfield  {journal} {\bibinfo  {journal} {Nature Phys.}\ }\textbf {\bibinfo {volume} {11}},\ \bibinfo {pages} {54} (\bibinfo {year} {2015})},\ \Eprint {http://arxiv.org/abs/1405.3651} {arXiv:1405.3651 [cond-mat.str-el]} \BibitemShut {NoStop}%
\bibitem [{\citenamefont {Brown}\ \emph {et~al.}(2019)\citenamefont {Brown} \emph {et~al.}}]{Brown:2019vef}%
  \BibitemOpen
  \bibfield  {author} {\bibinfo {author} {\bibfnamefont {P.~T.}\ \bibnamefont {Brown}} \emph {et~al.},\ }\href {\doibase 10.1126/science.aat4134} {\bibfield  {journal} {\bibinfo  {journal} {Science}\ }\textbf {\bibinfo {volume} {363}},\ \bibinfo {pages} {aat4134} (\bibinfo {year} {2019})},\ \Eprint {http://arxiv.org/abs/1802.09456} {arXiv:1802.09456 [cond-mat.quant-gas]} \BibitemShut {NoStop}%
\bibitem [{\citenamefont {Zhang}\ \emph {et~al.}(2017)\citenamefont {Zhang}, \citenamefont {Levenson-Falk}, \citenamefont {Ramshaw}, \citenamefont {Bonn}, \citenamefont {Liang}, \citenamefont {Hardy}, \citenamefont {Hartnoll},\ and\ \citenamefont {Kapitulnik}}]{Zhang:2016ofh}%
  \BibitemOpen
  \bibfield  {author} {\bibinfo {author} {\bibfnamefont {J.~C.}\ \bibnamefont {Zhang}}, \bibinfo {author} {\bibfnamefont {E.~M.}\ \bibnamefont {Levenson-Falk}}, \bibinfo {author} {\bibfnamefont {B.~J.}\ \bibnamefont {Ramshaw}}, \bibinfo {author} {\bibfnamefont {D.~A.}\ \bibnamefont {Bonn}}, \bibinfo {author} {\bibfnamefont {R.}~\bibnamefont {Liang}}, \bibinfo {author} {\bibfnamefont {W.~N.}\ \bibnamefont {Hardy}}, \bibinfo {author} {\bibfnamefont {S.~A.}\ \bibnamefont {Hartnoll}}, \ and\ \bibinfo {author} {\bibfnamefont {A.}~\bibnamefont {Kapitulnik}},\ }\href {\doibase 10.1073/pnas.1703416114} {\bibfield  {journal} {\bibinfo  {journal} {Proc. Nat. Acad. Sci.}\ }\textbf {\bibinfo {volume} {114}},\ \bibinfo {pages} {5378} (\bibinfo {year} {2017})},\ \Eprint {http://arxiv.org/abs/1610.05845} {arXiv:1610.05845 [cond-mat.supr-con]} \BibitemShut {NoStop}%
\bibitem [{\citenamefont {Landau}(1941)}]{Landau:1941}%
  \BibitemOpen
  \bibfield  {author} {\bibinfo {author} {\bibfnamefont {L.}~\bibnamefont {Landau}},\ }\href {\doibase 10.1103/PhysRev.60.356} {\bibfield  {journal} {\bibinfo  {journal} {Phys. Rev.}\ }\textbf {\bibinfo {volume} {60}},\ \bibinfo {pages} {356} (\bibinfo {year} {1941})}\BibitemShut {NoStop}%
\bibitem [{\citenamefont {Schmitt}(2015)}]{Schmitt:2014eka}%
  \BibitemOpen
  \bibfield  {author} {\bibinfo {author} {\bibfnamefont {A.}~\bibnamefont {Schmitt}},\ }\href {\doibase 10.1007/978-3-319-07947-9} {\emph {\bibinfo {title} {{Introduction to Superfluidity}: {Field-theoretical approach and applications}}}},\ Vol.\ \bibinfo {volume} {888}\ (\bibinfo {year} {2015})\ \Eprint {http://arxiv.org/abs/1404.1284} {arXiv:1404.1284 [hep-ph]} \BibitemShut {NoStop}%
\bibitem [{\citenamefont {Pitaevskii}\ and\ \citenamefont {Stringari}(2016)}]{PitStringbook}%
  \BibitemOpen
  \bibfield  {author} {\bibinfo {author} {\bibfnamefont {L.}~\bibnamefont {Pitaevskii}}\ and\ \bibinfo {author} {\bibfnamefont {S.}~\bibnamefont {Stringari}},\ }\href {\doibase 10.1093/acprof:oso/9780198758884.001.0001} {\emph {\bibinfo {title} {Bose-Einstein Condensation and Superfluidity}}}\ (\bibinfo  {publisher} {Oxford University Press},\ \bibinfo {year} {2016})\BibitemShut {NoStop}%
\bibitem [{\citenamefont {Sonin}(2016)}]{Sonin_2016}%
  \BibitemOpen
  \bibfield  {author} {\bibinfo {author} {\bibfnamefont {E.~B.}\ \bibnamefont {Sonin}},\ }\href@noop {} {\emph {\bibinfo {title} {Dynamics of Quantised Vortices in Superfluids}}}\ (\bibinfo  {publisher} {Cambridge University Press},\ \bibinfo {year} {2016})\BibitemShut {NoStop}%
\bibitem [{\citenamefont {Gubser}(2008)}]{Gubser:2008px}%
  \BibitemOpen
  \bibfield  {author} {\bibinfo {author} {\bibfnamefont {S.~S.}\ \bibnamefont {Gubser}},\ }\href {\doibase 10.1103/PhysRevD.78.065034} {\bibfield  {journal} {\bibinfo  {journal} {Phys. Rev. D}\ }\textbf {\bibinfo {volume} {78}},\ \bibinfo {pages} {065034} (\bibinfo {year} {2008})},\ \Eprint {http://arxiv.org/abs/0801.2977} {arXiv:0801.2977 [hep-th]} \BibitemShut {NoStop}%
\bibitem [{\citenamefont {Hartnoll}\ \emph {et~al.}(2008{\natexlab{a}})\citenamefont {Hartnoll}, \citenamefont {Herzog},\ and\ \citenamefont {Horowitz}}]{Hartnoll:2008kx}%
  \BibitemOpen
  \bibfield  {author} {\bibinfo {author} {\bibfnamefont {S.~A.}\ \bibnamefont {Hartnoll}}, \bibinfo {author} {\bibfnamefont {C.~P.}\ \bibnamefont {Herzog}}, \ and\ \bibinfo {author} {\bibfnamefont {G.~T.}\ \bibnamefont {Horowitz}},\ }\href {\doibase 10.1088/1126-6708/2008/12/015} {\bibfield  {journal} {\bibinfo  {journal} {JHEP}\ }\textbf {\bibinfo {volume} {12}},\ \bibinfo {pages} {015} (\bibinfo {year} {2008}{\natexlab{a}})},\ \Eprint {http://arxiv.org/abs/0810.1563} {arXiv:0810.1563 [hep-th]} \BibitemShut {NoStop}%
\bibitem [{\citenamefont {Hartnoll}\ \emph {et~al.}(2008{\natexlab{b}})\citenamefont {Hartnoll}, \citenamefont {Herzog},\ and\ \citenamefont {Horowitz}}]{Hartnoll:2008vx}%
  \BibitemOpen
  \bibfield  {author} {\bibinfo {author} {\bibfnamefont {S.~A.}\ \bibnamefont {Hartnoll}}, \bibinfo {author} {\bibfnamefont {C.~P.}\ \bibnamefont {Herzog}}, \ and\ \bibinfo {author} {\bibfnamefont {G.~T.}\ \bibnamefont {Horowitz}},\ }\href {\doibase 10.1103/PhysRevLett.101.031601} {\bibfield  {journal} {\bibinfo  {journal} {Phys. Rev. Lett.}\ }\textbf {\bibinfo {volume} {101}},\ \bibinfo {pages} {031601} (\bibinfo {year} {2008}{\natexlab{b}})},\ \Eprint {http://arxiv.org/abs/0803.3295} {arXiv:0803.3295 [hep-th]} \BibitemShut {NoStop}%
\bibitem [{\citenamefont {Amado}\ \emph {et~al.}(2009)\citenamefont {Amado}, \citenamefont {Kaminski},\ and\ \citenamefont {Landsteiner}}]{Amado:2009ts}%
  \BibitemOpen
  \bibfield  {author} {\bibinfo {author} {\bibfnamefont {I.}~\bibnamefont {Amado}}, \bibinfo {author} {\bibfnamefont {M.}~\bibnamefont {Kaminski}}, \ and\ \bibinfo {author} {\bibfnamefont {K.}~\bibnamefont {Landsteiner}},\ }\href {\doibase 10.1088/1126-6708/2009/05/021} {\bibfield  {journal} {\bibinfo  {journal} {JHEP}\ }\textbf {\bibinfo {volume} {05}},\ \bibinfo {pages} {021} (\bibinfo {year} {2009})},\ \Eprint {http://arxiv.org/abs/0903.2209} {arXiv:0903.2209 [hep-th]} \BibitemShut {NoStop}%
\bibitem [{\citenamefont {Herzog}\ and\ \citenamefont {Yarom}(2009)}]{Herzog:2009md}%
  \BibitemOpen
  \bibfield  {author} {\bibinfo {author} {\bibfnamefont {C.~P.}\ \bibnamefont {Herzog}}\ and\ \bibinfo {author} {\bibfnamefont {A.}~\bibnamefont {Yarom}},\ }\href {\doibase 10.1103/PhysRevD.80.106002} {\bibfield  {journal} {\bibinfo  {journal} {Phys. Rev. D}\ }\textbf {\bibinfo {volume} {80}},\ \bibinfo {pages} {106002} (\bibinfo {year} {2009})},\ \Eprint {http://arxiv.org/abs/0906.4810} {arXiv:0906.4810 [hep-th]} \BibitemShut {NoStop}%
\bibitem [{\citenamefont {Sonner}\ and\ \citenamefont {Withers}(2010)}]{Sonner:2010yx}%
  \BibitemOpen
  \bibfield  {author} {\bibinfo {author} {\bibfnamefont {J.}~\bibnamefont {Sonner}}\ and\ \bibinfo {author} {\bibfnamefont {B.}~\bibnamefont {Withers}},\ }\href {\doibase 10.1103/PhysRevD.82.026001} {\bibfield  {journal} {\bibinfo  {journal} {Phys. Rev.}\ }\textbf {\bibinfo {volume} {D82}},\ \bibinfo {pages} {026001} (\bibinfo {year} {2010})},\ \Eprint {http://arxiv.org/abs/1004.2707} {arXiv:1004.2707 [hep-th]} \BibitemShut {NoStop}%
\bibitem [{\citenamefont {Herzog}\ \emph {et~al.}(2011)\citenamefont {Herzog}, \citenamefont {Lisker}, \citenamefont {Surowka},\ and\ \citenamefont {Yarom}}]{Herzog:2011ec}%
  \BibitemOpen
  \bibfield  {author} {\bibinfo {author} {\bibfnamefont {C.~P.}\ \bibnamefont {Herzog}}, \bibinfo {author} {\bibfnamefont {N.}~\bibnamefont {Lisker}}, \bibinfo {author} {\bibfnamefont {P.}~\bibnamefont {Surowka}}, \ and\ \bibinfo {author} {\bibfnamefont {A.}~\bibnamefont {Yarom}},\ }\href {\doibase 10.1007/JHEP08(2011)052} {\bibfield  {journal} {\bibinfo  {journal} {JHEP}\ }\textbf {\bibinfo {volume} {08}},\ \bibinfo {pages} {052} (\bibinfo {year} {2011})},\ \Eprint {http://arxiv.org/abs/1101.3330} {arXiv:1101.3330 [hep-th]} \BibitemShut {NoStop}%
\bibitem [{\citenamefont {Bhattacharya}\ \emph {et~al.}(2011)\citenamefont {Bhattacharya}, \citenamefont {Bhattacharyya},\ and\ \citenamefont {Minwalla}}]{Bhattacharya:2011eea}%
  \BibitemOpen
  \bibfield  {author} {\bibinfo {author} {\bibfnamefont {J.}~\bibnamefont {Bhattacharya}}, \bibinfo {author} {\bibfnamefont {S.}~\bibnamefont {Bhattacharyya}}, \ and\ \bibinfo {author} {\bibfnamefont {S.}~\bibnamefont {Minwalla}},\ }\href {\doibase 10.1007/JHEP04(2011)125} {\bibfield  {journal} {\bibinfo  {journal} {JHEP}\ }\textbf {\bibinfo {volume} {04}},\ \bibinfo {pages} {125} (\bibinfo {year} {2011})},\ \Eprint {http://arxiv.org/abs/1101.3332} {arXiv:1101.3332 [hep-th]} \BibitemShut {NoStop}%
\bibitem [{\citenamefont {Gout\'eraux}\ and\ \citenamefont {Mefford}(2020{\natexlab{a}})}]{Gouteraux:2019kuy}%
  \BibitemOpen
  \bibfield  {author} {\bibinfo {author} {\bibfnamefont {B.}~\bibnamefont {Gout\'eraux}}\ and\ \bibinfo {author} {\bibfnamefont {E.}~\bibnamefont {Mefford}},\ }\href {\doibase 10.1103/PhysRevLett.124.161604} {\bibfield  {journal} {\bibinfo  {journal} {Phys. Rev. Lett.}\ }\textbf {\bibinfo {volume} {124}},\ \bibinfo {pages} {161604} (\bibinfo {year} {2020}{\natexlab{a}})},\ \Eprint {http://arxiv.org/abs/1912.08849} {arXiv:1912.08849 [hep-th]} \BibitemShut {NoStop}%
\bibitem [{\citenamefont {Gout\'eraux}\ and\ \citenamefont {Mefford}(2020{\natexlab{b}})}]{Gouteraux:2020asq}%
  \BibitemOpen
  \bibfield  {author} {\bibinfo {author} {\bibfnamefont {B.}~\bibnamefont {Gout\'eraux}}\ and\ \bibinfo {author} {\bibfnamefont {E.}~\bibnamefont {Mefford}},\ }\href {\doibase 10.1007/JHEP11(2020)091} {\bibfield  {journal} {\bibinfo  {journal} {JHEP}\ }\textbf {\bibinfo {volume} {11}},\ \bibinfo {pages} {091} (\bibinfo {year} {2020}{\natexlab{b}})},\ \Eprint {http://arxiv.org/abs/2008.02289} {arXiv:2008.02289 [hep-th]} \BibitemShut {NoStop}%
\bibitem [{\citenamefont {Arean}\ \emph {et~al.}(2021)\citenamefont {Arean}, \citenamefont {Baggioli}, \citenamefont {Grieninger},\ and\ \citenamefont {Landsteiner}}]{Arean:2021tks}%
  \BibitemOpen
  \bibfield  {author} {\bibinfo {author} {\bibfnamefont {D.}~\bibnamefont {Arean}}, \bibinfo {author} {\bibfnamefont {M.}~\bibnamefont {Baggioli}}, \bibinfo {author} {\bibfnamefont {S.}~\bibnamefont {Grieninger}}, \ and\ \bibinfo {author} {\bibfnamefont {K.}~\bibnamefont {Landsteiner}},\ }\href {\doibase 10.1007/JHEP11(2021)206} {\bibfield  {journal} {\bibinfo  {journal} {JHEP}\ }\textbf {\bibinfo {volume} {11}},\ \bibinfo {pages} {206} (\bibinfo {year} {2021})},\ \Eprint {http://arxiv.org/abs/2107.08802} {arXiv:2107.08802 [hep-th]} \BibitemShut {NoStop}%
\bibitem [{\citenamefont {Goutéraux}\ \emph {et~al.}(2023)\citenamefont {Goutéraux}, \citenamefont {Sottovia},\ and\ \citenamefont {Mefford}}]{thermoinstability}%
  \BibitemOpen
  \bibfield  {author} {\bibinfo {author} {\bibfnamefont {B.}~\bibnamefont {Goutéraux}}, \bibinfo {author} {\bibfnamefont {F.}~\bibnamefont {Sottovia}}, \ and\ \bibinfo {author} {\bibfnamefont {E.}~\bibnamefont {Mefford}},\ }\href {\doibase 10.1103/PhysRevD.108.L081903} {\bibfield  {journal} {\bibinfo  {journal} {Physical Review D}\ }\textbf {\bibinfo {volume} {108}},\ \bibinfo {pages} {L081903} (\bibinfo {year} {2023})}\BibitemShut {NoStop}%
\bibitem [{\citenamefont {Are\'an}\ \emph {et~al.}(2024)\citenamefont {Are\'an}, \citenamefont {Gout\'eraux}, \citenamefont {Mefford},\ and\ \citenamefont {Sottovia}}]{Arean:2023nnn}%
  \BibitemOpen
  \bibfield  {author} {\bibinfo {author} {\bibfnamefont {D.}~\bibnamefont {Are\'an}}, \bibinfo {author} {\bibfnamefont {B.}~\bibnamefont {Gout\'eraux}}, \bibinfo {author} {\bibfnamefont {E.}~\bibnamefont {Mefford}}, \ and\ \bibinfo {author} {\bibfnamefont {F.}~\bibnamefont {Sottovia}},\ }\href {\doibase 10.1007/JHEP05(2024)272} {\bibfield  {journal} {\bibinfo  {journal} {JHEP}\ }\textbf {\bibinfo {volume} {05}},\ \bibinfo {pages} {272} (\bibinfo {year} {2024})},\ \Eprint {http://arxiv.org/abs/2312.08243} {arXiv:2312.08243 [hep-th]} \BibitemShut {NoStop}%
\bibitem [{\citenamefont {An}\ \emph {et~al.}(2024{\natexlab{a}})\citenamefont {An}, \citenamefont {Li}, \citenamefont {Xia},\ and\ \citenamefont {Zeng}}]{An:2024ebg}%
  \BibitemOpen
  \bibfield  {author} {\bibinfo {author} {\bibfnamefont {Y.-P.}\ \bibnamefont {An}}, \bibinfo {author} {\bibfnamefont {L.}~\bibnamefont {Li}}, \bibinfo {author} {\bibfnamefont {C.-Y.}\ \bibnamefont {Xia}}, \ and\ \bibinfo {author} {\bibfnamefont {H.-B.}\ \bibnamefont {Zeng}},\ }\href {\doibase 10.1103/PhysRevD.109.106022} {\bibfield  {journal} {\bibinfo  {journal} {Phys. Rev. D}\ }\textbf {\bibinfo {volume} {109}},\ \bibinfo {pages} {106022} (\bibinfo {year} {2024}{\natexlab{a}})}\BibitemShut {NoStop}%
\bibitem [{\citenamefont {An}\ \emph {et~al.}(2024{\natexlab{b}})\citenamefont {An}, \citenamefont {Li},\ and\ \citenamefont {Zeng}}]{An:2024dkn}%
  \BibitemOpen
  \bibfield  {author} {\bibinfo {author} {\bibfnamefont {Y.}~\bibnamefont {An}}, \bibinfo {author} {\bibfnamefont {L.}~\bibnamefont {Li}}, \ and\ \bibinfo {author} {\bibfnamefont {H.}~\bibnamefont {Zeng}},\ }\href@noop {} {\  (\bibinfo {year} {2024}{\natexlab{b}})},\ \Eprint {http://arxiv.org/abs/2406.13965} {arXiv:2406.13965 [cond-mat.quant-gas]} \BibitemShut {NoStop}%
\bibitem [{\citenamefont {Landau}\ \emph {et~al.}(1980)\citenamefont {Landau}, \citenamefont {Lifshitz},\ and\ \citenamefont {Pitaevskii}}]{landau1980course9}%
  \BibitemOpen
  \bibfield  {author} {\bibinfo {author} {\bibfnamefont {L.}~\bibnamefont {Landau}}, \bibinfo {author} {\bibfnamefont {E.}~\bibnamefont {Lifshitz}}, \ and\ \bibinfo {author} {\bibfnamefont {L.}~\bibnamefont {Pitaevskii}},\ }\href@noop {} {\emph {\bibinfo {title} {Course of Theoretical Physics: Statistical Physics, Part 2 : by E.M. Lifshitz and L.P. Pitaevskii}}},\ \bibinfo {number} {vol.~9}\ (\bibinfo  {publisher} {Pergamon},\ \bibinfo {year} {1980})\BibitemShut {NoStop}%
\bibitem [{\citenamefont {Bardeen}(1962)}]{RevModPhys.34.667}%
  \BibitemOpen
  \bibfield  {author} {\bibinfo {author} {\bibfnamefont {J.}~\bibnamefont {Bardeen}},\ }\href {\doibase 10.1103/RevModPhys.34.667} {\bibfield  {journal} {\bibinfo  {journal} {Rev. Mod. Phys.}\ }\textbf {\bibinfo {volume} {34}},\ \bibinfo {pages} {667} (\bibinfo {year} {1962})}\BibitemShut {NoStop}%
\bibitem [{\citenamefont {Gout\'eraux}\ and\ \citenamefont {Mefford}(2024)}]{Gouteraux:2024adm}%
  \BibitemOpen
  \bibfield  {author} {\bibinfo {author} {\bibfnamefont {B.}~\bibnamefont {Gout\'eraux}}\ and\ \bibinfo {author} {\bibfnamefont {E.}~\bibnamefont {Mefford}},\ }\href@noop {} {\  (\bibinfo {year} {2024})},\ \Eprint {http://arxiv.org/abs/2407.07939} {arXiv:2407.07939 [hep-th]} \BibitemShut {NoStop}%
\bibitem [{\citenamefont {Khalatnikov}(1957)}]{khalatnikov1957hydrodynamics}%
  \BibitemOpen
  \bibfield  {author} {\bibinfo {author} {\bibfnamefont {I.}~\bibnamefont {Khalatnikov}},\ }\href@noop {} {\bibfield  {journal} {\bibinfo  {journal} {SOVIET PHYSICS JETP-USSR}\ }\textbf {\bibinfo {volume} {5}},\ \bibinfo {pages} {542} (\bibinfo {year} {1957})}\BibitemShut {NoStop}%
\bibitem [{\citenamefont {Andreev}\ and\ \citenamefont {Bashkin}(1975)}]{andreev1975three}%
  \BibitemOpen
  \bibfield  {author} {\bibinfo {author} {\bibfnamefont {A.}~\bibnamefont {Andreev}}\ and\ \bibinfo {author} {\bibfnamefont {E.}~\bibnamefont {Bashkin}},\ }\href@noop {} {\bibfield  {journal} {\bibinfo  {journal} {Soviet Journal of Experimental and Theoretical Physics}\ }\textbf {\bibinfo {volume} {42}},\ \bibinfo {pages} {164} (\bibinfo {year} {1975})}\BibitemShut {NoStop}%
\bibitem [{\citenamefont {Novak}\ \emph {et~al.}(2020)\citenamefont {Novak}, \citenamefont {Sonner},\ and\ \citenamefont {Withers}}]{Novak:2019wqg}%
  \BibitemOpen
  \bibfield  {author} {\bibinfo {author} {\bibfnamefont {I.}~\bibnamefont {Novak}}, \bibinfo {author} {\bibfnamefont {J.}~\bibnamefont {Sonner}}, \ and\ \bibinfo {author} {\bibfnamefont {B.}~\bibnamefont {Withers}},\ }\href {\doibase 10.1007/JHEP07(2020)165} {\bibfield  {journal} {\bibinfo  {journal} {JHEP}\ }\textbf {\bibinfo {volume} {07}},\ \bibinfo {pages} {165} (\bibinfo {year} {2020})},\ \Eprint {http://arxiv.org/abs/1911.02578} {arXiv:1911.02578 [hep-th]} \BibitemShut {NoStop}%
\bibitem [{\citenamefont {de~Boer}\ \emph {et~al.}(2020)\citenamefont {de~Boer}, \citenamefont {Hartong}, \citenamefont {Have}, \citenamefont {Obers},\ and\ \citenamefont {Sybesma}}]{deBoer:2020xlc}%
  \BibitemOpen
  \bibfield  {author} {\bibinfo {author} {\bibfnamefont {J.}~\bibnamefont {de~Boer}}, \bibinfo {author} {\bibfnamefont {J.}~\bibnamefont {Hartong}}, \bibinfo {author} {\bibfnamefont {E.}~\bibnamefont {Have}}, \bibinfo {author} {\bibfnamefont {N.~A.}\ \bibnamefont {Obers}}, \ and\ \bibinfo {author} {\bibfnamefont {W.}~\bibnamefont {Sybesma}},\ }\href {\doibase 10.21468/SciPostPhys.9.2.018} {\bibfield  {journal} {\bibinfo  {journal} {SciPost Phys.}\ }\textbf {\bibinfo {volume} {9}},\ \bibinfo {pages} {018} (\bibinfo {year} {2020})},\ \Eprint {http://arxiv.org/abs/2004.10759} {arXiv:2004.10759 [hep-th]} \BibitemShut {NoStop}%
\bibitem [{\citenamefont {Armas}\ and\ \citenamefont {Jain}(2021)}]{Armas:2020mpr}%
  \BibitemOpen
  \bibfield  {author} {\bibinfo {author} {\bibfnamefont {J.}~\bibnamefont {Armas}}\ and\ \bibinfo {author} {\bibfnamefont {A.}~\bibnamefont {Jain}},\ }\href {\doibase 10.21468/SciPostPhys.11.3.054} {\bibfield  {journal} {\bibinfo  {journal} {SciPost Phys.}\ }\textbf {\bibinfo {volume} {11}},\ \bibinfo {pages} {054} (\bibinfo {year} {2021})},\ \Eprint {http://arxiv.org/abs/2010.15782} {arXiv:2010.15782 [hep-th]} \BibitemShut {NoStop}%
\bibitem [{\citenamefont {Armas}\ and\ \citenamefont {Have}(2024)}]{Armas:2023ouk}%
  \BibitemOpen
  \bibfield  {author} {\bibinfo {author} {\bibfnamefont {J.}~\bibnamefont {Armas}}\ and\ \bibinfo {author} {\bibfnamefont {E.}~\bibnamefont {Have}},\ }\href {\doibase 10.21468/SciPostPhys.16.1.039} {\bibfield  {journal} {\bibinfo  {journal} {SciPost Phys.}\ }\textbf {\bibinfo {volume} {16}},\ \bibinfo {pages} {039} (\bibinfo {year} {2024})},\ \Eprint {http://arxiv.org/abs/2304.09596} {arXiv:2304.09596 [hep-th]} \BibitemShut {NoStop}%
\bibitem [{\citenamefont {Jensen}\ \emph {et~al.}(2012)\citenamefont {Jensen}, \citenamefont {Kaminski}, \citenamefont {Kovtun}, \citenamefont {Meyer}, \citenamefont {Ritz},\ and\ \citenamefont {Yarom}}]{Jensen:2012jh}%
  \BibitemOpen
  \bibfield  {author} {\bibinfo {author} {\bibfnamefont {K.}~\bibnamefont {Jensen}}, \bibinfo {author} {\bibfnamefont {M.}~\bibnamefont {Kaminski}}, \bibinfo {author} {\bibfnamefont {P.}~\bibnamefont {Kovtun}}, \bibinfo {author} {\bibfnamefont {R.}~\bibnamefont {Meyer}}, \bibinfo {author} {\bibfnamefont {A.}~\bibnamefont {Ritz}}, \ and\ \bibinfo {author} {\bibfnamefont {A.}~\bibnamefont {Yarom}},\ }\href {\doibase 10.1103/PhysRevLett.109.101601} {\bibfield  {journal} {\bibinfo  {journal} {Phys. Rev. Lett.}\ }\textbf {\bibinfo {volume} {109}},\ \bibinfo {pages} {101601} (\bibinfo {year} {2012})},\ \Eprint {http://arxiv.org/abs/1203.3556} {arXiv:1203.3556 [hep-th]} \BibitemShut {NoStop}%
\bibitem [{\citenamefont {Amado}\ \emph {et~al.}(2014)\citenamefont {Amado}, \citenamefont {Are{\'a}n}, \citenamefont {Jim{\'e}nez-Alba}, \citenamefont {Landsteiner}, \citenamefont {Melgar},\ and\ \citenamefont {Salazar~Landea}}]{Amado:2013aea}%
  \BibitemOpen
  \bibfield  {author} {\bibinfo {author} {\bibfnamefont {I.}~\bibnamefont {Amado}}, \bibinfo {author} {\bibfnamefont {D.}~\bibnamefont {Are{\'a}n}}, \bibinfo {author} {\bibfnamefont {A.}~\bibnamefont {Jim{\'e}nez-Alba}}, \bibinfo {author} {\bibfnamefont {K.}~\bibnamefont {Landsteiner}}, \bibinfo {author} {\bibfnamefont {L.}~\bibnamefont {Melgar}}, \ and\ \bibinfo {author} {\bibfnamefont {I.}~\bibnamefont {Salazar~Landea}},\ }\href {\doibase 10.1007/JHEP02(2014)063} {\bibfield  {journal} {\bibinfo  {journal} {JHEP}\ }\textbf {\bibinfo {volume} {02}},\ \bibinfo {pages} {063} (\bibinfo {year} {2014})},\ \Eprint {http://arxiv.org/abs/1307.8100} {arXiv:1307.8100 [hep-th]} \BibitemShut {NoStop}%
\bibitem [{\citenamefont {Beattie}\ \emph {et~al.}(2013{\natexlab{b}})\citenamefont {Beattie}, \citenamefont {Moulder}, \citenamefont {Fletcher},\ and\ \citenamefont {Hadzibabic}}]{beattie2013persistent}%
  \BibitemOpen
  \bibfield  {author} {\bibinfo {author} {\bibfnamefont {S.}~\bibnamefont {Beattie}}, \bibinfo {author} {\bibfnamefont {S.}~\bibnamefont {Moulder}}, \bibinfo {author} {\bibfnamefont {R.~J.}\ \bibnamefont {Fletcher}}, \ and\ \bibinfo {author} {\bibfnamefont {Z.}~\bibnamefont {Hadzibabic}},\ }\href@noop {} {\bibfield  {journal} {\bibinfo  {journal} {Physical review letters}\ }\textbf {\bibinfo {volume} {110}},\ \bibinfo {pages} {025301} (\bibinfo {year} {2013}{\natexlab{b}})}\BibitemShut {NoStop}%
\bibitem [{\citenamefont {Yakimenko}\ \emph {et~al.}(2013)\citenamefont {Yakimenko}, \citenamefont {Isaieva}, \citenamefont {Vilchinskii},\ and\ \citenamefont {Weyrauch}}]{yakimenko2013stability}%
  \BibitemOpen
  \bibfield  {author} {\bibinfo {author} {\bibfnamefont {A.}~\bibnamefont {Yakimenko}}, \bibinfo {author} {\bibfnamefont {K.}~\bibnamefont {Isaieva}}, \bibinfo {author} {\bibfnamefont {S.}~\bibnamefont {Vilchinskii}}, \ and\ \bibinfo {author} {\bibfnamefont {M.}~\bibnamefont {Weyrauch}},\ }\href@noop {} {\bibfield  {journal} {\bibinfo  {journal} {Physical Review A—Atomic, Molecular, and Optical Physics}\ }\textbf {\bibinfo {volume} {88}},\ \bibinfo {pages} {051602} (\bibinfo {year} {2013})}\BibitemShut {NoStop}%
\bibitem [{\citenamefont {Abad}\ \emph {et~al.}(2014)\citenamefont {Abad}, \citenamefont {Sartori}, \citenamefont {Finazzi},\ and\ \citenamefont {Recati}}]{abad2014persistent}%
  \BibitemOpen
  \bibfield  {author} {\bibinfo {author} {\bibfnamefont {M.}~\bibnamefont {Abad}}, \bibinfo {author} {\bibfnamefont {A.}~\bibnamefont {Sartori}}, \bibinfo {author} {\bibfnamefont {S.}~\bibnamefont {Finazzi}}, \ and\ \bibinfo {author} {\bibfnamefont {A.}~\bibnamefont {Recati}},\ }\href@noop {} {\bibfield  {journal} {\bibinfo  {journal} {Physical Review A}\ }\textbf {\bibinfo {volume} {89}},\ \bibinfo {pages} {053602} (\bibinfo {year} {2014})}\BibitemShut {NoStop}%
\bibitem [{\citenamefont {An}\ and\ \citenamefont {Li}(2024)}]{An:2024mbx}%
  \BibitemOpen
  \bibfield  {author} {\bibinfo {author} {\bibfnamefont {Y.}~\bibnamefont {An}}\ and\ \bibinfo {author} {\bibfnamefont {L.}~\bibnamefont {Li}},\ }\href@noop {} {\  (\bibinfo {year} {2024})},\ \Eprint {http://arxiv.org/abs/2409.08310} {arXiv:2409.08310 [hep-th]} \BibitemShut {NoStop}%
\bibitem [{\citenamefont {Amado}\ \emph {et~al.}(2013)\citenamefont {Amado}, \citenamefont {Areán}, \citenamefont {Jimenez-Alba}, \citenamefont {Landsteiner}, \citenamefont {Melgar},\ and\ \citenamefont {Landea}}]{type2G}%
  \BibitemOpen
  \bibfield  {author} {\bibinfo {author} {\bibfnamefont {I.}~\bibnamefont {Amado}}, \bibinfo {author} {\bibfnamefont {D.}~\bibnamefont {Areán}}, \bibinfo {author} {\bibfnamefont {A.}~\bibnamefont {Jimenez-Alba}}, \bibinfo {author} {\bibfnamefont {K.}~\bibnamefont {Landsteiner}}, \bibinfo {author} {\bibfnamefont {L.}~\bibnamefont {Melgar}}, \ and\ \bibinfo {author} {\bibfnamefont {I.~S.}\ \bibnamefont {Landea}},\ }\href {\doibase 10.1007/JHEP07(2013)108} {\bibfield  {journal} {\bibinfo  {journal} {Journal of High Energy Physics}\ }\textbf {\bibinfo {volume} {2013}},\ \bibinfo {pages} {108} (\bibinfo {year} {2013})}\BibitemShut {NoStop}%
\bibitem [{\citenamefont {Manakov}(1974)}]{manakov1974theory}%
  \BibitemOpen
  \bibfield  {author} {\bibinfo {author} {\bibfnamefont {S.~V.}\ \bibnamefont {Manakov}},\ }\href@noop {} {\bibfield  {journal} {\bibinfo  {journal} {Soviet Physics-JETP}\ }\textbf {\bibinfo {volume} {38}},\ \bibinfo {pages} {248} (\bibinfo {year} {1974})}\BibitemShut {NoStop}%
\bibitem [{\citenamefont {Hidaka}(2013)}]{countingrule}%
  \BibitemOpen
  \bibfield  {author} {\bibinfo {author} {\bibfnamefont {Y.}~\bibnamefont {Hidaka}},\ }\href {\doibase 10.1103/PhysRevLett.110.091601} {\bibfield  {journal} {\bibinfo  {journal} {Phys. Rev. Lett.}\ }\textbf {\bibinfo {volume} {110}},\ \bibinfo {pages} {091601} (\bibinfo {year} {2013})}\BibitemShut {NoStop}%
\bibitem [{\citenamefont {Basu}\ \emph {et~al.}(2010)\citenamefont {Basu}, \citenamefont {He}, \citenamefont {Mukherjee}, \citenamefont {Rozali},\ and\ \citenamefont {Shieh}}]{Basu:2010fa}%
  \BibitemOpen
  \bibfield  {author} {\bibinfo {author} {\bibfnamefont {P.}~\bibnamefont {Basu}}, \bibinfo {author} {\bibfnamefont {J.}~\bibnamefont {He}}, \bibinfo {author} {\bibfnamefont {A.}~\bibnamefont {Mukherjee}}, \bibinfo {author} {\bibfnamefont {M.}~\bibnamefont {Rozali}}, \ and\ \bibinfo {author} {\bibfnamefont {H.-H.}\ \bibnamefont {Shieh}},\ }\href {\doibase 10.1007/JHEP10(2010)092} {\bibfield  {journal} {\bibinfo  {journal} {JHEP}\ }\textbf {\bibinfo {volume} {10}},\ \bibinfo {pages} {092} (\bibinfo {year} {2010})},\ \Eprint {http://arxiv.org/abs/1007.3480} {arXiv:1007.3480 [hep-th]} \BibitemShut {NoStop}%
\bibitem [{\citenamefont {Cai}\ \emph {et~al.}(2013)\citenamefont {Cai}, \citenamefont {Li}, \citenamefont {Li},\ and\ \citenamefont {Wang}}]{Cai:2013wma}%
  \BibitemOpen
  \bibfield  {author} {\bibinfo {author} {\bibfnamefont {R.-G.}\ \bibnamefont {Cai}}, \bibinfo {author} {\bibfnamefont {L.}~\bibnamefont {Li}}, \bibinfo {author} {\bibfnamefont {L.-F.}\ \bibnamefont {Li}}, \ and\ \bibinfo {author} {\bibfnamefont {Y.-Q.}\ \bibnamefont {Wang}},\ }\href {\doibase 10.1007/JHEP09(2013)074} {\bibfield  {journal} {\bibinfo  {journal} {JHEP}\ }\textbf {\bibinfo {volume} {09}},\ \bibinfo {pages} {074} (\bibinfo {year} {2013})},\ \Eprint {http://arxiv.org/abs/1307.2768} {arXiv:1307.2768 [hep-th]} \BibitemShut {NoStop}%
\bibitem [{\citenamefont {{Lan}}\ \emph {et~al.}()\citenamefont {{Lan}}, \citenamefont {{Liu}}, \citenamefont {{Tian}},\ and\ \citenamefont {{Zhang}}}]{Landauinstability}%
  \BibitemOpen
  \bibfield  {author} {\bibinfo {author} {\bibfnamefont {S.}~\bibnamefont {{Lan}}}, \bibinfo {author} {\bibfnamefont {H.}~\bibnamefont {{Liu}}}, \bibinfo {author} {\bibfnamefont {Y.}~\bibnamefont {{Tian}}}, \ and\ \bibinfo {author} {\bibfnamefont {H.}~\bibnamefont {{Zhang}}},\ }\href@noop {} {\ }\Eprint {http://arxiv.org/abs/2010.06232} {arXiv:2010.06232 [hep-th]} \BibitemShut {NoStop}%
\bibitem [{\citenamefont {An}\ and\ \citenamefont {Li}(2025)}]{An:2025gid}%
  \BibitemOpen
  \bibfield  {author} {\bibinfo {author} {\bibfnamefont {Y.}~\bibnamefont {An}}\ and\ \bibinfo {author} {\bibfnamefont {L.}~\bibnamefont {Li}},\ }\href@noop {} {\  (\bibinfo {year} {2025})},\ \Eprint {http://arxiv.org/abs/2501.03561} {arXiv:2501.03561 [hep-th]} \BibitemShut {NoStop}%
\bibitem [{\citenamefont {Ishino}\ \emph {et~al.}(2013{\natexlab{a}})\citenamefont {Ishino}, \citenamefont {Tsubota},\ and\ \citenamefont {Takeuchi}}]{vortex1}%
  \BibitemOpen
  \bibfield  {author} {\bibinfo {author} {\bibfnamefont {S.}~\bibnamefont {Ishino}}, \bibinfo {author} {\bibfnamefont {M.}~\bibnamefont {Tsubota}}, \ and\ \bibinfo {author} {\bibfnamefont {H.}~\bibnamefont {Takeuchi}},\ }\href {\doibase 10.1103/PhysRevA.88.063617} {\bibfield  {journal} {\bibinfo  {journal} {Physical Review A}\ }\textbf {\bibinfo {volume} {88}},\ \bibinfo {pages} {063617} (\bibinfo {year} {2013}{\natexlab{a}})}\BibitemShut {NoStop}%
\bibitem [{\citenamefont {Ishino}\ \emph {et~al.}(2013{\natexlab{b}})\citenamefont {Ishino}, \citenamefont {Tsubota},\ and\ \citenamefont {Takeuchi}}]{vortex2}%
  \BibitemOpen
  \bibfield  {author} {\bibinfo {author} {\bibfnamefont {S.}~\bibnamefont {Ishino}}, \bibinfo {author} {\bibfnamefont {M.}~\bibnamefont {Tsubota}}, \ and\ \bibinfo {author} {\bibfnamefont {H.}~\bibnamefont {Takeuchi}},\ }\href {\doibase 10.1007/s10909-012-0798-x} {\bibfield  {journal} {\bibinfo  {journal} {Journal of Low Temperature Physics}\ }\textbf {\bibinfo {volume} {171}},\ \bibinfo {pages} {429} (\bibinfo {year} {2013}{\natexlab{b}})}\BibitemShut {NoStop}%
\bibitem [{\citenamefont {Kuopanportti}\ \emph {et~al.}(2019)\citenamefont {Kuopanportti}, \citenamefont {Bandyopadhyay}, \citenamefont {Roy},\ and\ \citenamefont {Angom}}]{vortex3}%
  \BibitemOpen
  \bibfield  {author} {\bibinfo {author} {\bibfnamefont {P.}~\bibnamefont {Kuopanportti}}, \bibinfo {author} {\bibfnamefont {S.}~\bibnamefont {Bandyopadhyay}}, \bibinfo {author} {\bibfnamefont {A.}~\bibnamefont {Roy}}, \ and\ \bibinfo {author} {\bibfnamefont {D.}~\bibnamefont {Angom}},\ }\href {\doibase 10.1103/PhysRevA.100.033615} {\bibfield  {journal} {\bibinfo  {journal} {Physical Review A}\ }\textbf {\bibinfo {volume} {100}},\ \bibinfo {pages} {033615} (\bibinfo {year} {2019})}\BibitemShut {NoStop}%
\bibitem [{\citenamefont {Armas}\ \emph {et~al.}(2017)\citenamefont {Armas}, \citenamefont {Bhattacharya}, \citenamefont {Jain},\ and\ \citenamefont {Kundu}}]{Armas:2016xxg}%
  \BibitemOpen
  \bibfield  {author} {\bibinfo {author} {\bibfnamefont {J.}~\bibnamefont {Armas}}, \bibinfo {author} {\bibfnamefont {J.}~\bibnamefont {Bhattacharya}}, \bibinfo {author} {\bibfnamefont {A.}~\bibnamefont {Jain}}, \ and\ \bibinfo {author} {\bibfnamefont {N.}~\bibnamefont {Kundu}},\ }\href {\doibase 10.1007/JHEP06(2017)090} {\bibfield  {journal} {\bibinfo  {journal} {JHEP}\ }\textbf {\bibinfo {volume} {06}},\ \bibinfo {pages} {090} (\bibinfo {year} {2017})},\ \Eprint {http://arxiv.org/abs/1612.08088} {arXiv:1612.08088 [hep-th]} \BibitemShut {NoStop}%
\bibitem [{\citenamefont {Crossno}\ \emph {et~al.}(2016)\citenamefont {Crossno} \emph {et~al.}}]{Crossno:2016fvs}%
  \BibitemOpen
  \bibfield  {author} {\bibinfo {author} {\bibfnamefont {J.}~\bibnamefont {Crossno}} \emph {et~al.},\ }\href {\doibase 10.1126/science.aad0343} {\bibfield  {journal} {\bibinfo  {journal} {Science}\ }\textbf {\bibinfo {volume} {351}},\ \bibinfo {pages} {aad0343} (\bibinfo {year} {2016})}\BibitemShut {NoStop}%
\bibitem [{\citenamefont {Lucas}\ \emph {et~al.}(2016)\citenamefont {Lucas}, \citenamefont {Crossno}, \citenamefont {Fong}, \citenamefont {Kim},\ and\ \citenamefont {Sachdev}}]{Lucas:2015sya}%
  \BibitemOpen
  \bibfield  {author} {\bibinfo {author} {\bibfnamefont {A.}~\bibnamefont {Lucas}}, \bibinfo {author} {\bibfnamefont {J.}~\bibnamefont {Crossno}}, \bibinfo {author} {\bibfnamefont {K.~C.}\ \bibnamefont {Fong}}, \bibinfo {author} {\bibfnamefont {P.}~\bibnamefont {Kim}}, \ and\ \bibinfo {author} {\bibfnamefont {S.}~\bibnamefont {Sachdev}},\ }\href {\doibase 10.1103/PhysRevB.93.075426} {\bibfield  {journal} {\bibinfo  {journal} {Phys. Rev. B}\ }\textbf {\bibinfo {volume} {93}},\ \bibinfo {pages} {075426} (\bibinfo {year} {2016})},\ \Eprint {http://arxiv.org/abs/1510.01738} {arXiv:1510.01738 [cond-mat.str-el]} \BibitemShut {NoStop}%
\bibitem [{\citenamefont {LaSalle}\ and\ \citenamefont {Lefschetz}(1961)}]{LiapunovBook}%
  \BibitemOpen
  \bibfield  {author} {\bibinfo {author} {\bibfnamefont {J.}~\bibnamefont {LaSalle}}\ and\ \bibinfo {author} {\bibfnamefont {S.}~\bibnamefont {Lefschetz}},\ }\href@noop {} {\emph {\bibinfo {title} {Stability by Liapunov's Direct Method: With Applications}}}\ (\bibinfo  {publisher} {Academic Press},\ \bibinfo {year} {1961})\BibitemShut {NoStop}%
\bibitem [{\citenamefont {Nicolis}(1979)}]{Nicolis1979}%
  \BibitemOpen
  \bibfield  {author} {\bibinfo {author} {\bibfnamefont {G.}~\bibnamefont {Nicolis}},\ }\href {\doibase 10.1088/0034-4885/42/2/001} {\bibfield  {journal} {\bibinfo  {journal} {Reports on Progress in Physics}\ }\textbf {\bibinfo {volume} {42}},\ \bibinfo {pages} {225} (\bibinfo {year} {1979})}\BibitemShut {NoStop}%
\bibitem [{\citenamefont {Policastro}\ \emph {et~al.}(2001)\citenamefont {Policastro}, \citenamefont {Son},\ and\ \citenamefont {Starinets}}]{Policastro:2001yc}%
  \BibitemOpen
  \bibfield  {author} {\bibinfo {author} {\bibfnamefont {G.}~\bibnamefont {Policastro}}, \bibinfo {author} {\bibfnamefont {D.~T.}\ \bibnamefont {Son}}, \ and\ \bibinfo {author} {\bibfnamefont {A.~O.}\ \bibnamefont {Starinets}},\ }\href {\doibase 10.1103/PhysRevLett.87.081601} {\bibfield  {journal} {\bibinfo  {journal} {Phys. Rev. Lett.}\ }\textbf {\bibinfo {volume} {87}},\ \bibinfo {pages} {081601} (\bibinfo {year} {2001})},\ \Eprint {http://arxiv.org/abs/hep-th/0104066} {arXiv:hep-th/0104066} \BibitemShut {NoStop}%
\bibitem [{\citenamefont {Policastro}\ \emph {et~al.}(2002{\natexlab{a}})\citenamefont {Policastro}, \citenamefont {Son},\ and\ \citenamefont {Starinets}}]{Policastro:2002se}%
  \BibitemOpen
  \bibfield  {author} {\bibinfo {author} {\bibfnamefont {G.}~\bibnamefont {Policastro}}, \bibinfo {author} {\bibfnamefont {D.~T.}\ \bibnamefont {Son}}, \ and\ \bibinfo {author} {\bibfnamefont {A.~O.}\ \bibnamefont {Starinets}},\ }\href {\doibase 10.1088/1126-6708/2002/09/043} {\bibfield  {journal} {\bibinfo  {journal} {JHEP}\ }\textbf {\bibinfo {volume} {09}},\ \bibinfo {pages} {043} (\bibinfo {year} {2002}{\natexlab{a}})},\ \Eprint {http://arxiv.org/abs/hep-th/0205052} {arXiv:hep-th/0205052} \BibitemShut {NoStop}%
\bibitem [{\citenamefont {Policastro}\ \emph {et~al.}(2002{\natexlab{b}})\citenamefont {Policastro}, \citenamefont {Son},\ and\ \citenamefont {Starinets}}]{Policastro:2002tn}%
  \BibitemOpen
  \bibfield  {author} {\bibinfo {author} {\bibfnamefont {G.}~\bibnamefont {Policastro}}, \bibinfo {author} {\bibfnamefont {D.~T.}\ \bibnamefont {Son}}, \ and\ \bibinfo {author} {\bibfnamefont {A.~O.}\ \bibnamefont {Starinets}},\ }\href {\doibase 10.1088/1126-6708/2002/12/054} {\bibfield  {journal} {\bibinfo  {journal} {JHEP}\ }\textbf {\bibinfo {volume} {12}},\ \bibinfo {pages} {054} (\bibinfo {year} {2002}{\natexlab{b}})},\ \Eprint {http://arxiv.org/abs/hep-th/0210220} {arXiv:hep-th/0210220} \BibitemShut {NoStop}%
\bibitem [{\citenamefont {Herzog}(2003)}]{Herzog:2003ke}%
  \BibitemOpen
  \bibfield  {author} {\bibinfo {author} {\bibfnamefont {C.~P.}\ \bibnamefont {Herzog}},\ }\href {\doibase 10.1103/PhysRevD.68.024013} {\bibfield  {journal} {\bibinfo  {journal} {Phys. Rev. D}\ }\textbf {\bibinfo {volume} {68}},\ \bibinfo {pages} {024013} (\bibinfo {year} {2003})},\ \Eprint {http://arxiv.org/abs/hep-th/0302086} {arXiv:hep-th/0302086} \BibitemShut {NoStop}%
\bibitem [{\citenamefont {Bhattacharyya}\ \emph {et~al.}(2008)\citenamefont {Bhattacharyya}, \citenamefont {Hubeny}, \citenamefont {Minwalla},\ and\ \citenamefont {Rangamani}}]{Bhattacharyya:2007vjd}%
  \BibitemOpen
  \bibfield  {author} {\bibinfo {author} {\bibfnamefont {S.}~\bibnamefont {Bhattacharyya}}, \bibinfo {author} {\bibfnamefont {V.~E.}\ \bibnamefont {Hubeny}}, \bibinfo {author} {\bibfnamefont {S.}~\bibnamefont {Minwalla}}, \ and\ \bibinfo {author} {\bibfnamefont {M.}~\bibnamefont {Rangamani}},\ }\href {\doibase 10.1088/1126-6708/2008/02/045} {\bibfield  {journal} {\bibinfo  {journal} {JHEP}\ }\textbf {\bibinfo {volume} {02}},\ \bibinfo {pages} {045} (\bibinfo {year} {2008})},\ \Eprint {http://arxiv.org/abs/0712.2456} {arXiv:0712.2456 [hep-th]} \BibitemShut {NoStop}%
\bibitem [{\citenamefont {Banerjee}\ \emph {et~al.}(2011)\citenamefont {Banerjee}, \citenamefont {Bhattacharya}, \citenamefont {Bhattacharyya}, \citenamefont {Dutta}, \citenamefont {Loganayagam},\ and\ \citenamefont {Surowka}}]{Banerjee:2008th}%
  \BibitemOpen
  \bibfield  {author} {\bibinfo {author} {\bibfnamefont {N.}~\bibnamefont {Banerjee}}, \bibinfo {author} {\bibfnamefont {J.}~\bibnamefont {Bhattacharya}}, \bibinfo {author} {\bibfnamefont {S.}~\bibnamefont {Bhattacharyya}}, \bibinfo {author} {\bibfnamefont {S.}~\bibnamefont {Dutta}}, \bibinfo {author} {\bibfnamefont {R.}~\bibnamefont {Loganayagam}}, \ and\ \bibinfo {author} {\bibfnamefont {P.}~\bibnamefont {Surowka}},\ }\href {\doibase 10.1007/JHEP01(2011)094} {\bibfield  {journal} {\bibinfo  {journal} {JHEP}\ }\textbf {\bibinfo {volume} {01}},\ \bibinfo {pages} {094} (\bibinfo {year} {2011})},\ \Eprint {http://arxiv.org/abs/0809.2596} {arXiv:0809.2596 [hep-th]} \BibitemShut {NoStop}%
\bibitem [{\citenamefont {Erdmenger}\ \emph {et~al.}(2009)\citenamefont {Erdmenger}, \citenamefont {Haack}, \citenamefont {Kaminski},\ and\ \citenamefont {Yarom}}]{Erdmenger:2008rm}%
  \BibitemOpen
  \bibfield  {author} {\bibinfo {author} {\bibfnamefont {J.}~\bibnamefont {Erdmenger}}, \bibinfo {author} {\bibfnamefont {M.}~\bibnamefont {Haack}}, \bibinfo {author} {\bibfnamefont {M.}~\bibnamefont {Kaminski}}, \ and\ \bibinfo {author} {\bibfnamefont {A.}~\bibnamefont {Yarom}},\ }\href {\doibase 10.1088/1126-6708/2009/01/055} {\bibfield  {journal} {\bibinfo  {journal} {JHEP}\ }\textbf {\bibinfo {volume} {01}},\ \bibinfo {pages} {055} (\bibinfo {year} {2009})},\ \Eprint {http://arxiv.org/abs/0809.2488} {arXiv:0809.2488 [hep-th]} \BibitemShut {NoStop}%
\bibitem [{\citenamefont {Kovtun}\ and\ \citenamefont {Starinets}(2005)}]{Kovtun:2005ev}%
  \BibitemOpen
  \bibfield  {author} {\bibinfo {author} {\bibfnamefont {P.~K.}\ \bibnamefont {Kovtun}}\ and\ \bibinfo {author} {\bibfnamefont {A.~O.}\ \bibnamefont {Starinets}},\ }\href {\doibase 10.1103/PhysRevD.72.086009} {\bibfield  {journal} {\bibinfo  {journal} {Phys. Rev. D}\ }\textbf {\bibinfo {volume} {72}},\ \bibinfo {pages} {086009} (\bibinfo {year} {2005})},\ \Eprint {http://arxiv.org/abs/hep-th/0506184} {arXiv:hep-th/0506184} \BibitemShut {NoStop}%
\bibitem [{\citenamefont {Son}\ and\ \citenamefont {Starinets}(2002)}]{Son:2002sd}%
  \BibitemOpen
  \bibfield  {author} {\bibinfo {author} {\bibfnamefont {D.~T.}\ \bibnamefont {Son}}\ and\ \bibinfo {author} {\bibfnamefont {A.~O.}\ \bibnamefont {Starinets}},\ }\href {\doibase 10.1088/1126-6708/2002/09/042} {\bibfield  {journal} {\bibinfo  {journal} {JHEP}\ }\textbf {\bibinfo {volume} {09}},\ \bibinfo {pages} {042} (\bibinfo {year} {2002})},\ \Eprint {http://arxiv.org/abs/hep-th/0205051} {arXiv:hep-th/0205051} \BibitemShut {NoStop}%
\bibitem [{\citenamefont {Zhao}\ \emph {et~al.}(2023)\citenamefont {Zhao}, \citenamefont {Zhang},\ and\ \citenamefont {Nie}}]{0order}%
  \BibitemOpen
  \bibfield  {author} {\bibinfo {author} {\bibfnamefont {Z.-Q.}\ \bibnamefont {Zhao}}, \bibinfo {author} {\bibfnamefont {X.-K.}\ \bibnamefont {Zhang}}, \ and\ \bibinfo {author} {\bibfnamefont {Z.-Y.}\ \bibnamefont {Nie}},\ }\href {\doibase 10.1007/JHEP02(2023)023} {\bibfield  {journal} {\bibinfo  {journal} {Journal of High Energy Physics}\ }\textbf {\bibinfo {volume} {2023}},\ \bibinfo {pages} {23} (\bibinfo {year} {2023})}\BibitemShut {NoStop}%
\end{thebibliography}%
\end{document}